\documentclass[showpacs,preprintnumbers,amsmath,amssymb]{revtex4}

\usepackage{epsfig}% Include figure files
\usepackage{graphicx}% Include figure files
\usepackage{dcolumn}% Align table columns on decimal point
\usepackage{bm}% bold math
\usepackage{epstopdf}
\usepackage{subfigure}

\begin{document}

%%%%%%%%%%%%%%%%%%%%%%%%%%%%%%%%%%%%%%%%%%%%%%%%%%%%%%%%%%%%%%%%%%%%%%%%%%
%                               Title                                    
%%%%%%%%%%%%%%%%%%%%%%%%%%%%%%%%%%%%%%%%%%%%%%%%%%%%%%%%%%%%%%%%%%%%%%%%%%

\title{Pseudofermion scattering theory}
\author{J. M. P. Carmelo and D. Bozi}
\affiliation{GCEP - Center of Physics, University of
Minho, Campus Gualtar, P-4710-057 Braga, Portugal}
\affiliation{Department of Physics, Massachusetts Institute of Technology, Cambridge,
Massachusetts 02139-4307, USA} 
\author{P. D. Sacramento}
\affiliation{Departamento de F\'{\i}sica and CFIF, Instituto Superior T\'ecnico,
P-1049-001 Lisboa, Portugal}
%\date{21 March 2006}
%\date{\today}

%%%%%%%%%%%%%%%%%%%%%%%%%%%%%%%%%%%%%%%%%%%%%%%%%%%%%%%%%%%%%%%%%%%%%%%%%%
%                              abstract                                  
%%%%%%%%%%%%%%%%%%%%%%%%%%%%%%%%%%%%%%%%%%%%%%%%%%%%%%%%%%%%%%%%%%%%%%%%%%

\begin{abstract}
In this paper we study the scattering theory associated with the 
pseudofermion dynamical theory for the Hubbard chain. While for the electronic
basis the problem is non perturbative and strongly correlated, in terms
of pseudofermions the spectral properties are controlled by zero-momentum
forward scattering only. Indeed, we find that each ground-state --
excited-energy-eigenstate transition corresponds to a well defined set of elementary
two-pseudofermion zero-momentum forward-scattering events. 
An important point of the theory is that independent
$\eta$-spin $1/2$ holons and spin $1/2$ spinons are neither scatterers nor scattering
centers. Instead, the scatterers and scattering centers are spin-less and
$\eta$-spin-less $c0$ pseudofermions, $\eta$-spin-zero $2\nu$-holon composite $c\nu$
pseudofermions, spin-zero $2\nu$-spinon composite $s\nu$ pseudofermions such that
$\nu=1,2,3,...$, and the corresponding pseudofermion holes. Similarly to chromodynamics,
where all quark-composite physical particles are color-neutral, for the pseudofermion
dynamical theory all $2\nu$-holon (and $2\nu$-spinon) composite pseudofermion 
scatterers and scattering centers are $\eta$-spin-neutral (and spin-neutral). Thus, 
the pseudofermion $S$ matrix is a mere phase factor, which is behind the simple form 
of the pseudofermion anti-commutators and the simplification of the study of the 
finite-energy spectral and dynamical properties. 
The pseudofermion $S$ matrix is expressed as a commutative product
of $S$ matrices, each corresponding to an elementary two-pseudofermion scattering event.
This commutative factorization is stronger than the usual factorization associated with
Yang-Baxter Equation for the original spin $1/2$ electron bare $S$ matrix. Our results
reveal the scattering mechanisms which control the exotic finite-energy spectral properties 
of the low-dimensional complex materials and correlated systems of cold fermionic atoms 
on an optical lattice. Importantly, the exotic scatterers and scattering centers predicted
by the theory were observed by angle-resolved photoelectron spectroscopy in 
low-dimensional organic metals.
\end{abstract}

\pacs{70}

\maketitle
%%%%%%%%%%%%%%%%%%%%%%%%%%%%%%%%%%%%%%%%%%%%%%%%%%%%%%%%%%%%%%%%%%%%%%%%%%
%                              body of paper                             
%%%%%%%%%%%%%%%%%%%%%%%%%%%%%%%%%%%%%%%%%%%%%%%%%%%%%%%%%%%%%%%%%%%%%%%%%%
\section{INTRODUCTION}

The one-dimensional (1D) Hubbard model is solvable by coordinate Bethe ansatz (BA)
\cite{Lieb,Takahashi}. The solution of the model can also be achieved by the
inverse-scattering algebraic BA \cite{Martins98}. In terms of electronic scattering the model describes
a very complex non-perturbative strongly correlated problem. The main goal
of this paper is to show that in terms of the pseudofermions associated with
the holons and spinons introduced in Refs. \cite{I,II,IIIb} the scattering
problem considerably simplifies and involves two-pseudofermion zero-momentum 
forward scattering only. In the last twenty years the low-energy behavior of the
model correlation functions has been the subject of many studies 
\cite{Woy,Ogata,Kawakami,Frahm,Brech,93-94,Karlo}. Recently, the use of
the pseudofermion representation extended such studies to finite
energy \cite{V,V-1,LE}.

The generalization of the electron - rotated-electron unitary transformation 
introduced in Ref. \cite{Harris} for all values of the on-site repulsion $U$ 
plays an important role in the construction of the holon and spinon
representation of Ref. \cite{I}. Such a representation is faithful for
the whole Hilbert space, including the subspace generated by application
onto the BA-solution states of the off-diagonal generators of the
spin and $\eta$-spin $SU(2)$ algebras \cite{HL,Yang89}. In turn,
there is one pseudofermion representation for each initial
ground state whose subspace is spanned by the excited
energy eigenstates contained in the excitations generated
by application of one- and two-electron operators onto the
former state \cite{IIIb,V-1}. The pseudofermion dynamical 
theory (PDT) \cite{V,V-1,LE} is a suitable starting point for the study of the dynamical and spectral 
properties of the model for all values of the momentum, energy,
on-site repulsion $U$, and electronic density $n$. However, the relation of
the PDT to the scattering mechanisms remains an open question. 
The theory is a generalization for all values of the on-site repulsion $U$ of the
$U/t>>1$ method of Ref. \cite{Penc}. Here $t$ is the first-neighbor transfer
integral. 

The finite-energy PDT reproduces the well-known behavior of spectral and 
correlation functions in the limit of low energy \cite{LE}, which was previously 
obtained \cite{Woy,Ogata,Kawakami,Frahm,Brech,93-94,Karlo} by use of methods 
such as conformal-field theory \cite{CFT} and bosonization \cite{Bozo}. The
theory was successfully applied to the
description of the unusual finite-energy spectral properties of low-dimensional complex
materials \cite{spectral0,spectral,super}: For the one-electron removal spectral function
the singular spectral features predicted by the PDT show quantitative agreement for
the whole energy band width with the peak dispersions observed by angle-resolved
photoelectron spectroscopy in the quasi-1D organic conductor TTF-TCNQ
\cite{spectral0,spectral}. (Results for the TTF-TCNQ spectrum consistent with those 
of the PDT were obtained by the dynamical density matrix renormalization group method
\cite{Eric}.) Moreover, the theory was also used in the description 
of the phase diagram of other low-dimensional complex materials \cite{super}
and is of interest for the study of the spectral properties 
of the new quantum systems described by cold
fermionic atoms on an optical lattice \cite{Zoller}. (New experiments involving cold
fermionic atoms [such as $^6Li$] on an optical lattice formed by interfering laser fields
are in progress \cite{Chen}.)

We are able to calculate explicitly the pseudofermion and pseudofermion hole $S$  
matrices and overall phase shifts and to clarify how these quantities control the 
finite-energy spectral properties. Interestingly, the unusual independent charge 
and spin spectral features observed by angle-resolved photoelectron 
spectroscopy in low-dimensional organic metals \cite{spectral0,spectral}
correspond to the exotic independent charge and spin pseudofermion scatters
and scattering centers introduced here. This paper contains a detailed presentation 
of the preliminary results presented in short form elsewhere \cite{Car04}.

The relation of the holon, spinon, and pseudofermion description of 
Refs. \cite{I,II,IIIb,V,V-1,LE} used here to the conventional
holon and spinon representation and scattering theory of Refs. \cite{Natan,S0,S} 
was very recently clarified in Ref. \cite{relation}.  Such an investigation confirms
that both representations are faithful and thus that
there is no inconsistency between the two corresponding definitions of quantum
objects. Moreover, that study confirms that the holon, spinon, and pseudofermion
description of Refs.  \cite{I,II,IIIb,V,V-1,LE} is the most suitable for the study of the
finite-energy spectral and dynamical properties.

The paper is organized as follows: In Sec. II we introduce 
the 1D Hubbard model and summarize the basic information
about the pseudofermion description needed for our studies. The general pseudofermion
scattering theory and the pseudofermion $S$ matrix are introduced in Sec. III. In that
section we also discuss the relation between the pseudofermion scattering theory and the
spectral and dynamical properties. In Sec. IV we introduce and study the pseudofermion
phase shifts. Finally, Sec. V contains the concluding remarks.

%%%%%%%%%%%%%%%%%%%%%%%%%%%%%%%%%%%%%%%%%%%%%%%%%%%%%%%%%%%%%%%%%%%%%%%%%%
\section{THE 1D HUBBARD MODEL AND THE PSEUDOFERMION DESCRIPTION}

In this section we introduce the 1D Hubbard model and summarize the concepts and results
concerning rotated electrons \cite{Harris,I} and the pseudofermion description
\cite{IIIb} that are needed for our studies.

%%%%%%%%%%%%%%%%%%%%%%%%%%%%%%%%%%%%%%%%%%%%%%%%%%%%%%%%%%%%%%%%%%%%%%%%%%
\subsection{THE 1D HUBBARD MODEL AND ROTATED ELECTRONS}

The exotic quantum objects associated with the BA solution of the 1D Hubbard model are
related to the electrons through the rotated electrons \cite{I,II,IIIb}. Let us start by
introducing the model.

%%%%%%%%%%%%%%%%%%%%%%%%%%%%%%%%%%%%%%%%%%%%%%%%%%%%%%%%%%%%%%%%%%%%%%%%%%
\subsubsection{THE 1D HUBBARD MODEL}

In a chemical potential $\mu $ and magnetic field $H$ the 1D Hubbard Hamiltonian can be
written as,
\begin{equation}
\hat{H}={\hat{H}}_{SO(4)} + \sum_{\alpha =c,\,s}\mu_{\alpha}\, {\hat{S}}_{\alpha}^z \, ,
\label{H}
\end{equation}
where
\begin{equation}
{\hat{H}}_{SO(4)} = {\hat{H}}_{H} - {U\over 2}\Bigl[\,\hat{N} - {N_a\over 2}\Bigr] \, ;
\hspace{1cm}{\hat{H}}_{H} = \hat{T}+U\,\hat{D} \, . \label{HH}
\end{equation}
Here ${\hat{H}}_{H}$ is the ``simple" Hubbard model, $\hat{T}=-t\sum_{\sigma=\uparrow
,\,\downarrow }\sum_{j=1}^{N_a}\Bigl[c_{j,\,\sigma}^{\dag}\,c_{j+1,\,\sigma} + h.
c.\Bigr]$ is the {\it kinetic-energy} operator, $\hat{D} =
\sum_{j=1}^{N_a}c_{j,\,\uparrow}^{\dag}\,c_{j,\,\uparrow}\,
c_{j,\,\downarrow}^{\dag}\,c_{j,\,\downarrow} =
\sum_{j=1}^{N_a}\hat{n}_{j,\,\uparrow}\,\hat{n}_{j,\,\downarrow}$ is the electron
double-occupation operator, and the operator
$\hat{n}_{j,\,\sigma} = c_{j,\,\sigma }^{\dagger }\,c_{j,\,\sigma }$ counts the number of
spin-projection $\sigma$ electrons at lattice site $j$. The operator
$c_{j,\,\sigma}^{\dagger}$ (and $c_{j,\,\sigma}$) that appears in the above equations
creates (and annihilates) a spin-projection $\sigma $ electron at lattice site
$j=1,2,...,N_a$. We consider that the number of lattice sites $N_a$ is large
and even. On the right-hand side of Eq. (\ref{H}), $\mu_c=2\mu$, $\mu_s=2\mu_0 H$,
$\mu_0$ is the Bohr magneton, and the number operators ${\hat{S }}_c^z=-{1\over
2}[N_a-\hat{N}]$ and ${\hat{S }}_s^z= -{1\over 2}[{\hat{N}}_{\uparrow}-
{\hat{N}}_{\downarrow}]$ are the diagonal generators of the $\eta$-spin and spin $SU(2)$
algebras \cite{HL,Yang89}, respectively. Here the electronic number operators read
${\hat{N}}=\sum_{\sigma=\uparrow ,\,\downarrow }\,\hat{N}_{\sigma}$ and
${\hat{N}}_{\sigma}=\sum_{j=1}^{N_a}\hat{n}_{j,\,\sigma}$.
The momentum operator is given by $\hat{P} = \sum_{\sigma=\uparrow
,\,\downarrow }\sum_{k}\, \hat{n}_{\sigma} (k)\, k$, where the spin-projection $\sigma$
momentum distribution operator reads $\hat{n}_{\sigma} (k) = c_{k,\,\sigma }^{\dagger
}\,c_{k,\,\sigma }$ and the operator $c_{k,\,\sigma}^{\dagger}$ (and $c_{k,\,\sigma}$)
creates (and annihilates) a spin-projection $\sigma $ electron of momentum $k$.

Throughout this paper we use units of both Planck constant $\hbar$ and lattice constant
$a$ one. We denote the electronic charge by $-e$, the lattice length by $L=N_a\,a=N_a$,
and the $\eta$-spin value $\eta$ (and spin value $S$) and $\eta$-spin projection $\eta_z$
(and spin projection $S_z$) of the energy eigenstates by $S_c$ and $S_c^z$ (and $S_s$ and
$S_s^z$), respectively. The Hamiltonian $\hat{H}_{SO(4)}$ given in Eq. (\ref{HH}) commutes with the six
generators of the $\eta$-spin and spin $SU(2)$ algebras and has $SO(4)$ symmetry
\cite{HL,Yang89}. While the expressions of the two corresponding diagonal generators were
given above, the off-diagonal generators of these two $SU(2)$ algebras read
${\hat{S}}_c^{\dagger}=\sum_{j}(-1)^j\,c_{j,\,\downarrow}^{\dagger}\,c_{j,\,\uparrow}^{\dagger}$ 
and ${\hat{S}}_c =\sum_{j}(-1)^j\,c_{j,\,\uparrow}\,c_{j,\,\downarrow}$ for $\eta$ spin and
${\hat{S}}_s^{\dagger}=\sum_{j}\,c_{j,\,\downarrow}^{\dagger}\,c_{j,\,\uparrow}$ and 
${\hat{S}}_s=\sum_{j}c_{j,\,\uparrow}^{\dagger}\,
c_{j,\,\downarrow}$ for spin. The BA solvability of the 1D Hubbard model (\ref{H}) is
restricted to the Hilbert subspace spanned by the lowest-weight states (LWSs)
\cite{Lieb,Takahashi} or highest-weight states (HWSs) \cite{Martins98} of the $\eta$-spin
and spin algebras, that is by the states whose $S_{\alpha}$ and $S_{\alpha}^z$ numbers
are such that $S_{\alpha}= -S_{\alpha}^z$ or $S_{\alpha}=S_{\alpha}^z$, respectively,
where $\alpha =c$ for $\eta$-spin and $\alpha =s$ for spin. Such states have
electronic densities $n=N/L$ and spin densities $m=[N_{\uparrow}-N_{\downarrow}]/L$ 
in the domains $0\leq n \leq 1$ and $0\leq m \leq n$, respectively. The description 
of the states corresponding to the extended domains $0\leq n \leq 1$\, ;
$1\leq n \leq 2$ and $-n\leq m \leq n$\, ; $-(2-n)\leq m \leq (2-n)$, respectively, is
achieved by application onto the latter states of off-diagonal generators of the
$\eta$-spin and spin $SU(2)$ algebras \cite{I}. The scattering processes studied
in this paper result from ground-state - excited-state transitions. For simplicity,
here we consider initial ground states with densities in the domains $0\leq n \leq 1$ 
and $0\leq m \leq n$, respectively. (Some of our results correspond to initial ground 
states with densities in the ranges $0< n < 1$ and $0< m < n$.)

%%%%%%%%%%%%%%%%%%%%%%%%%%%%%%%%%%%%%%%%%%%%%%%%%%%%%%%%%%%%%%%%%%%%%%%%%%
\subsubsection{ROTATED ELECTRONS}

Each lattice site $j=1,2,...,N_a$ of the model (\ref{H}) can either be doubly occupied,
unoccupied, or singly occupied by a spin-down or spin-up electron. The maximum number of
electrons is $2N_a$ and corresponds to density $n=2$. Besides the $N$ electrons, it is
useful to consider $N^h=[2N_a-N]$ {\it electronic holes}. (Here we use the designation
{\it electronic hole} instead of {\it hole}, in order to distinguish this type of hole
from the pseudofermion hole.) Our definition of electronic hole is such that when a
lattice site is unoccupied, we say that it is occupied by two electronic holes. If a
lattice site is singly occupied, we say that it is occupied by an electron and an
electronic hole. If a lattice site is doubly occupied, it is unoccupied by electronic
holes. The same definition applies to the rotated-electronic holes.

The electron - rotated-electron unitary transformation maps the electrons onto rotated
electrons such that rotated-electron double occupation, non-occupation, and spin-up and
spin-down single occupation are good quantum numbers for all values of $U/t$ \cite{I,II}.
The lattice occupied by rotated electrons is identical to the original electronic
lattice. We call $c_{j,\,\sigma}^{\dag}$ the electrons that occur in the 1D Hubbard model
(\ref{H}) and (\ref{HH}), while the operator ${\tilde{c}}_{j,\,\sigma}^{\dag}$ such that
${\tilde{c}}_{j,\,\sigma}^{\dag} =
{\hat{V}}^{\dag}(U/t)\,c_{j,\,\sigma}^{\dag}\,{\hat{V}}(U/t)$ represents the rotated
electrons, where ${\hat{V}}(U/t)$ denotes the electron - rotated-electron unitary
operator. Similarly, $c_{j,\,\sigma}^{\dag} =
{\hat{V}}(U/t)\,{\tilde{c}}_{j,\,\sigma}^{\dag}\,{\hat{V}}^{\dag}(U/t)$. Note that
for $m=0$ $c_{j,\,\sigma}^{\dag}$ and ${\tilde{c}}_{j,\,\sigma}^{\dag}$ are only identical in the
$U/t\rightarrow\infty$ limit where electron double occupation becomes a good quantum
number. The operators ${\hat{V}}^{\dag}(U/t)$ and ${\hat{V}}(U/t)$ are uniquely defined
for all values of $U/t$ by Eqs. (21)-(23) of Ref. \cite{I}. The electron -
rotated-electron unitary transformation was introduced in Ref. \cite{Harris}. The
rotated-electron double occupation operator $\tilde{D}$ given in Eq. (20) of Ref.
\cite{I} commutes with the 1D Hubbard model. We denote the rotated-electron double
occupation by $D_r$. 

%%%%%%%%%%%%%%%%%%%%%%%%%%%%%%%%%%%%%%%%%%%%%%%%%%%%%%%%%%%%%%%%
\subsection{THE PSEUDOFERMION DESCRIPTION}

Here we summarize the pseudofermion properties that are needed for the studies of this
paper. The pseudoparticles studied in Refs. \cite{I,II} and the pseudofermions used in
the investigations of Refs. \cite{V-1,LE} are closely related. While the pseudoparticles have discrete
bare-momentum values $q_j$ such that $q_{j+1}-q_j =2\pi/L$, the corresponding
pseudofermions have {\it canonical-momentum} values ${\bar{q}}_j = q_j +
Q^{\Phi}_{\alpha\nu} (q_j)/L$. (The designation "bare-momentum" follows from
the discrete values $q_j$ in units of $2\pi/L$ corresponding to quantum numbers,
as given in Eq. (B.1) of Ref. \cite{I}; The designation "canonical-momentum"
stems from the analogy of the momentum shift $Q^{\Phi}_{\alpha\nu} (q_j)/L$
with the shift within the canonical momentum of electrons in the presence of a 
vector potential.) Here $Q^{\Phi}_{\alpha\nu} (q_j)/L$ is the functional given in Eq. (14) of
Ref. \cite{IIIb} and $\alpha\nu$ labels the pseudofermion branch, as discussed below.
Although that functional is of the order $1/L$, the discrete canonical-momentum are such
that ${\bar{q}}_{j+1}-{\bar{q}}_j =2\pi/L + O(1/L^2)$. Except for the slightly different
discrete canonical-momentum values ${\bar{q}}_j$ and discrete bare-momentum values $q_j$,
the pseudofermions have the same properties as the corresponding pseudoparticles. For
instance, they have the same values of charge, $\eta$-spin, or spin and for the branches
other than the $c0$ branch, also the same holon or spinon contents. The pseudofermion
description refers to a Hilbert subspace called in Ref. \cite{IIIb} {\it pseudofermion
subspace} (PS). (All one-,two-, and any other finite-number-electron excitations are
contained in the PS.) In the PS the energy eigenstates are described by the same
pseudoparticle \cite{I} and pseudofermion \cite{IIIb} occupancy configurations and the
$\alpha\nu$ pseudoparticles and $\alpha\nu$ pseudofermions are related by a unitary
transformation \cite{IIIb}. Thus, the basic pseudofermion properties summarized below
have many similarities with the corresponding pseudoparticle properties studied in Refs.
\cite{I,II}.

%%%%%%%%%%%%%%%%%%%%%%%%%%%%%%%%%%%%%%%%%%%%%%%%%%%%%%%%%%%%%%%%
\subsubsection{$c0$ PSEUDOFERMIONS, COMPOSITE PSEUDOFERMIONS, YANG HOLONS, AND HL SPINONS}

A key result needed for our study is that the energy eigenstates that span the PS can be
described in terms of occupancy configurations of holons, spinons, and $c0$
pseudofermions \cite{I,IIIb}. We recall that the holons and spinons considered
here are different from those of the conventional spinon-holon representation 
used in the studies of Refs. \cite{Natan,S0,S} and that the 
relation between the two alternative holon and spinon representations
is clarified in Ref. \cite{relation}. For the simplest excited energy eigenstates, the 
holon (and spinon) of the conventional representation involves mixing of the $c0$
pseudofermion hole and Yang holon (and $s1$ pseudofermion hole and HL
spinon) considered below. The holons (and spinons) introduced in Ref.
\cite{I} have $\eta$-spin $1/2$, $\eta$-spin projection $\pm 1/2$, charge 
$\pm 2e$, and spin zero (spin $1/2$, spin
projection $\pm 1/2$, and no charge degrees of freedom). We use the notation $\pm 1/2$
holons (and $\pm 1/2$ spinons) according to the value of $\eta$-spin projection (and spin
projection). The rotated-electron double occupation $D_r$ equals the number of $-1/2$
holons. Within the description of charge transport in terms of electrons (and electronic
holes), the $c0$ pseudofermions carry charge $-e$ (and $+e$) and have no spin or
$\eta$-spin degrees of freedom. Moreover, the $c\nu$ pseudofermions (and $s\nu$
pseudofermions) are $\eta$-spin zero (and spin zero) composite objects of an equal number
$\nu=1,2,...$ of $-1/2$ holons and $+1/2$ holons (and $-1/2$ spinons and $+1/2$ spinons).
Within the description of charge transport in terms of electrons (and electronic holes),
the $c\nu$ pseudofermions carry charge $-2\nu e$ (and $+2\nu e$) where $\nu =1,2,...$. In
this paper we use the notation $\alpha\nu$ pseudofermion, where $\alpha=c,\,s$ and $\nu
=0,1,2,...$ for the $c\nu$ branches and $\nu =1,2,...$ for the $s\nu$ branches. The $\pm
1/2$ holons (and $\pm 1/2$ spinons) which are not part of $2\nu$-holon composite $c\nu$
pseudofermions (and $2\nu$-spinon composite $s\nu$ pseudofermions) are called $\pm 1/2$
Yang holons (and $\pm 1/2$ HL spinons). In the designations {\it HL spinon} and {\it Yang
holon}, HL stands for Heilmann and Lieb and Yang refers to C. N. Yang, respectively, who
are the authors of Refs. \cite{HL,Yang89}. We denote the number of $\alpha\nu$
pseudofermions by $N_{\alpha\nu}$ and the number $\pm 1/2$ Yang holons ($\alpha =c$) and
$\pm 1/2$ HL spinons ($\alpha =s$) by $L_{\alpha,\,\pm 1/2}$. Note that $N_{c0}$ equals
the number of rotated-electron singly occupied sites, $[N_a -N_{c0}]$ equals the number
of rotated-electron doubly occupied plus unoccupied sites, and $L_{\alpha,\,\pm
1/2}=S_{\alpha} \mp S_{\alpha}^z$. We call $M_{\alpha,\,\pm 1/2}$ the number of $\pm 1/2$
holons ($\alpha =c$) and $\pm 1/2$ spinons ($\alpha =s$) such that $M_{\alpha,\,\pm
1/2}=L_{\alpha,\,\pm 1/2} +\sum_{\nu =1}^{\infty}\nu\,N_{\alpha\nu}$. These numbers are
given by $M_{c,\,-1/2}=[N-N_{c0}]/2$, $M_{c,\,+1/2}=[N^h-N_{c0}]/2$,
$M_{s,\,-1/2}=[N_{c0}-N_{\uparrow}+N_{\downarrow}]/2$, and
$M_{s,\,+1/2}=[N_{c0}+N_{\uparrow}-N_{\downarrow}]/2$. Furthermore,
$M_{\alpha}=[M_{\alpha,\,-1/2}+M_{\alpha,\,+1/2}]$ denotes the number of holons ($\alpha
=c$) or spinons ($\alpha =s$) such that $M_c =[N_a -N_{c0}]$ and $M_s =N_{c0}$ and
$L_{\alpha}=[L_{\alpha,\,-1/2}+L_{\alpha,\,+1/2}]$ denotes the number of Yang holons
($\alpha =c$) or HL spinons ($\alpha =s$) such that $L_c =2S_c = 2\eta$ and $L_s =2S_s =
2S$. An important point is that for the ground state and densities such that $0\leq n\leq
1$ and $0\leq m\leq n$ one finds that $N_{c0}=N$, $N_{s1}=N_{\downarrow}$,
$M_{c,\,+1/2}=L_{c,\,+1/2}=[N_a -N]$, $M_{s,\,-1/2}=N_{\downarrow}$,
$M_{s,\,+1/2}=N_{\uparrow}$, $L_{s,\,+1/2}=[N_{\uparrow}-N_{\downarrow}]$, and
$N_{\alpha\nu}=M_{c,\,-1/2}=L_{\alpha,\,-1/2}=0$ for $\alpha\nu\neq c0,\,s1$ and $\alpha
=c,\,s$.

Often in this paper we use the notation $\alpha\nu\neq c0,\,s1$ branches, which refers to
all $\alpha\nu$ branches except the $c0$ and $s1$ branches. Moreover, the summations (and
products) $\sum_{\alpha\nu}$, $\sum_{\alpha\nu =c0,\,s1}$, and $\sum_{\alpha\nu\neq
c0,\,s1}$ (and $\prod_{\alpha\nu}$, $\prod_{\alpha\nu =c0,\,s1}$, and
$\prod_{\alpha\nu\neq c0,\,s1}$) run over all $\alpha\nu$ branches with finite
$\alpha\nu$ pseudofermion occupancy in the corresponding state or subspace, the $c0$ and
$s1$ branches only, and all $\alpha\nu$ branches with finite $\alpha\nu$ pseudofermion
occupancy in the corresponding state or subspace except the $c0$ and $s1$ branches,
respectively.

%%%%%%%%%%%%%%%%%%%%%%%%%%%%%%%%%%%%%%%%%%%%%%%%%%%%%%%%%%%%%%%%
\subsubsection{THE PSEUDOFERMION CANONICAL MOMENTUM AND ASSOCIATED FUNCTIONALS}

As mentioned above, the $\alpha\nu$ pseudofermion discrete canonical-momentum values
${\bar{q}}_j$ are of the following form,
\begin{equation}
{\bar{q}}_j = {\bar{q}} (q_j) = q_j + {Q^{\Phi}_{\alpha\nu} (q_j)\over L} = {2\pi\over
L}I^{\alpha\nu}_j + {Q^{\Phi}_{\alpha\nu} (q_j)\over L} \, ; \hspace{0.5cm}
j=1,2,...,N_{\alpha\nu}^* \, . \label{barqan}
\end{equation}
where $I^{\alpha\nu}_j$ are integers or half-odd integers \cite{I},
$N^*_{\alpha\nu}=N_{\alpha\nu}+N_{\alpha\nu}^h$, and $N_{\alpha\nu}^h$ denotes the number
of $\alpha\nu$ pseudofermion holes. The latter number equals the corresponding number of
$\alpha\nu$ pseudoparticle holes given in Eqs. (B7) and (B8) of Ref. \cite{I}. Note that
besides equaling the number of discrete canonical-momentum values in the $\alpha\nu$
canonical-momentum band, $N^*_{\alpha\nu}=N_{\alpha\nu}+N_{\alpha\nu}^h$ also equals the
number of sites of the $\alpha\nu$ effective lattice \cite{IIIb}, which plays an
important role in the pseudofermion description. In addition to the $\alpha\nu$
pseudofermions of canonical momentum ${\bar{q}}$, there are local $\alpha\nu$
pseudofermions, whose creation and annihilation operators correspond to the sites of the
effective $\alpha\nu$ lattice. Such a lattice has spatial coordinates $x_j
=a_{\alpha\nu}\,j$ where $j=1,2,...,N^*_{\alpha\nu}$ and $N^*_{\alpha\nu}$ is the number
of sites defined in Eqs. (B.6)-(B.8) and (B.11) of Ref. \cite{I} and $a_{\alpha\nu} =
L/N^*_{\alpha\nu}$ is the effective $\alpha\nu$ lattice constant. Each $\alpha\nu$
pseudofermion band is associated with an effective $\alpha\nu$ lattice whose length
$L=N^*_{\alpha\nu}\,a_{\alpha\nu}$ is the same as that of the original real-space
lattice. \cite{IIIb}. The relation between the momentum and local pseudofermion operators
is given in Eq. (34) of Ref. \cite{IIIb}.

The discrete bare-momentum $q_j$ is a good quantum number whose allowed occupancies are
one and zero only. (Also the corresponding discrete canonical-momentum ${\bar{q}}_j$ has
allowed occupancies one and zero only.) Thus, for the bare-momentum occupancy
configuration describing a given energy eigenstate the bare-momentum distribution
function $N_{\alpha\nu} (q_j)$ is such that $N_{\alpha\nu} (q_j)=1$ for occupied
bare-momentum values and $N_{\alpha\nu} (q_j)=0$ for unoccupied bare-momentum values. We
denote the ground-state bare-momentum distribution function by $N^{0}_{\alpha\nu} (q_j)$.
It is given in Eqs. (C.1)-(C.3) of Ref. \cite{I}. Although the $\alpha\nu$ pseudoparticles
carry bare-momentum $q_j$, one can also label the corresponding $\alpha\nu$
pseudofermions by such a bare-momentum. This is because there is a one-to-one
correspondence between the bare momentum $q_j$ and the pseudofermion canonical momentum
${\bar{q}}_j = q_j + Q^{\Phi}_{\alpha\nu} (q_j)/L$. Thus, when one refers to the
pseudofermion bare-momentum $q_j$, one means that $q_j$ is the bare-momentum value that
corresponds to the pseudofermion canonical momentum ${\bar{q}}_j = q_j +
Q^{\Phi}_{\alpha\nu} (q_j)/L$. The pseudofermion canonical-momentum shift functional
$Q^{\Phi}_{\alpha\nu} (q_j)/L$ is given by,
\begin{equation}
{Q^{\Phi}_{\alpha\nu} (q_j)\over L} = {2\pi\over L} \sum_{\alpha'\nu'}\,\,
\sum_{j'=1}^{N^*_{\alpha'\nu'}}\,\Phi_{\alpha\nu,\,\alpha'\nu'}(q_j,q_{j'})\, \Delta
N_{\alpha'\nu'}(q_{j'}) \, , \label{qcan1j}
\end{equation}
where,
\begin{equation}
\Delta N_{\alpha\nu} (q_j) \equiv N_{\alpha\nu} (q_j) - N^{0}_{\alpha\nu} (q_j) \, ,
\label{DNq}
\end{equation}
is the $\alpha\nu$ bare-momentum distribution-function deviation. A PS excited energy
eigenstate is uniquely defined by the values of the set of deviations $\{\Delta
N_{\alpha\nu} (q_j)\}$ for all values of $q_j$ corresponding to the $\alpha\nu$
branches with finite pseudofermion occupancy in the state and by the values $L_{c
,\,-1/2}$ and $L_{s ,\,-1/2}$. Moreover, the quantity
$\Phi_{\alpha\nu,\,\alpha'\nu'}(q,q')$ on the right-hand side of Eq. (\ref{qcan1j}) is a
function of both the bare-momentum values $q$ and $q'$ given by,
\begin{equation}
\Phi_{\alpha\nu,\,\alpha'\nu'}(q,q') = \bar{\Phi }_{\alpha\nu,\,\alpha'\nu'}
\left({4t\,\Lambda^{0}_{\alpha\nu}(q)\over U}, {4t\,\Lambda^{0}_{\alpha'\nu'}(q')\over
U}\right) \, , \label{Phi-barPhi}
\end{equation}
where the function $\bar{\Phi }_{\alpha\nu,\,\alpha'\nu'} (r ,\,r')$ is the unique
solution of the integral equations (A1)-(A13) of Ref. \cite{IIIb}. The ground-state
rapidity functions $\Lambda_{\alpha\nu}^0 (q)$ appearing in Eq. (\ref{Phi-barPhi}), where
$\Lambda^0_{c0}(q)\equiv\sin k^0 (q)$ for $\alpha\nu=c0$, are defined in terms of the
inverse functions of $k^0 (q)$ and $\Lambda_{\alpha\nu}^0 (q)$ for $\nu
>0$ in Eqs. (A.1) and (A.2) of Ref. \cite{V-1}. 

It is found below that $\pi\,\Phi_{\alpha\nu,\,\alpha'\nu'}(q,q')$ [or
$-\pi\,\Phi_{\alpha\nu,\,\alpha'\nu'}(q,q')$] is an elementary {\it two-pseudofermion
phase shift} such that $q$ is the bare-momentum value of a $\alpha\nu$ pseudofermion or
$\alpha\nu$ pseudofermion hole scattered by a $\alpha'\nu'$ pseudofermion [or
$\alpha'\nu'$ pseudofermion hole] of bare-momentum $q'$ created under a ground-state -
excited-energy-eigenstate transition. As discussed in Sec. IV-D, there are no $c\nu\neq c0$ 
(and $s\nu\neq s1$) bare-momentum bands for $n=1$ (and $m=0$) ground states. 
Indeed, $N^*_{c\nu}=0$ (and $N^*_{s\nu}=0$) for such states and then
the corresponding ground-state rapidity functions $\Lambda^{0}_{c\nu}(q)$
(and $\Lambda^{0}_{s\nu}(q)$) cannot be defined. Fortunately, expression
(\ref{Phi-barPhi}) remains valid in that case provided that the ground-state
rapidity functions are suitably replaced by those of the excited states.
(Indeed, we find below that the functions (\ref{Phi-barPhi}) are phase
shifts originated by well-defined ground-state - excited-state transitions;
Thus, in the particular case of the  $n=1$ and/or $m=0$ ground states
the quantities (\ref{Phi-barPhi}) are functionals rather than functions,
with the rapidity functions for the $c\nu\neq c0$ and/or $s\nu\neq s1$
branches being those of the excited state under consideration.)
 
The form of the functional (\ref{qcan1j}) reveals that for the initial ground state the
discrete canonical-momentum value ${\bar{q}}_j$ and corresponding discrete 
bare-momentum value $q_j$ are such that ${\bar{q}}_j = q_j$. The ground-state 
continuum bare-momentum and canonical-momentum values belong to the domain $q\in
[-q^0_{\alpha\nu},\,+q^0_{\alpha\nu}]$, where the limiting value $q^0_{\alpha\nu}$ reads,
\begin{equation}
q^0_{c0} = \pi \, ; \hspace{0.5cm} q^0_{s1} = k_{F\uparrow} \, ; \hspace{0.5cm}
q^0_{c\nu} = [\pi -2k_F] \, , \hspace{0.3cm} \nu
>0 \, ; \hspace{0.5cm} q^0_{s\nu} =
[k_{F\uparrow}-k_{F\downarrow}] \, , \hspace{0.3cm} \nu >1 \, . \label{qcanGS}
\end{equation}
The ground-state is described by a compact $c0$ and $s1$ pseudofermion finite occupancy
for $q\in [-q^0_{F\alpha\nu},\,+q^0_{F\alpha\nu}]$, while the remaining branches have
vanishing ground-state occupancy. Here the $c0$ and $s1$ {\it Fermi} points are given by,
\begin{equation}
q^0_{Fc0} = 2k_F \, ; \hspace{0.5cm} q^0_{Fs1} = k_{F\downarrow} \, . \label{q0Fcs}
\end{equation}
Both the limiting values of Eq. (\ref{qcanGS}) and the ground-state {\it Fermi} values of
Eq. (\ref{q0Fcs}) are given except for corrections of order $1/L$. The limiting bare-momentum 
values and ground-state {\it Fermi} bare-momentum including the $1/L$ corrections 
are provided in Eqs.  (B.14)-(B.17) and (C.4)-(C.11), respectively, of Ref. \cite{I}. 

%%%%%%%%%%%%%%%%%%%%%%%%%%%%%%%%%%%%%%%%%%%%%%%%%%%%%%%%%%%%%%%%
\subsubsection{THE PSEUDOFERMION ENERGY SPECTRUM}

The PS general energy spectrum can be expressed in terms of the pseudofermion
canonical-momentum distribution-function deviations as follows \cite{IIIb},
\begin{eqnarray}
\Delta E & = & \sum_{{\bar{q}}=-\pi}^{+\pi} \,\epsilon_{c0} ({\bar{q}})\Delta
{\cal{N}}_{c0} ({\bar{q}}) +
\sum_{{\bar{q}}=-k_{F\uparrow}}^{+k_{F\uparrow}}\epsilon_{s1}({\bar{q}})\,\Delta
{\cal{N}}_{s1}({\bar{q}}) \nonumber \\
& + & E_{h} + \sum_{\nu =2}^{\infty}\,\sum_{{\bar{q}}=-[k_{F\uparrow}-
k_{F\downarrow}]}^{+[k_{F\uparrow}-k_{F\downarrow}]}\epsilon^0_{s\nu}({\bar{q}})\,\Delta
{\cal{N}}_{s\nu}({\bar{q}}) + E_{uhb} + \sum_{\nu =1}^{\infty}\,\sum_{{\bar{q}}=-[\pi
-2k_F]}^{+[\pi -2k_F]}\epsilon^0_{c\nu}({\bar{q}})\,\Delta {\cal{N}}_{c\nu}({\bar{q}}) \,
. \label{DE}
\end{eqnarray}
Here $\Delta {\cal{N}}_{\alpha\nu}({\bar{q}})=\Delta N_{\alpha\nu}(q)$ and the
$\alpha\nu$ energy bands are defined in Eqs. (C.15)-(C.21) of Ref. \cite{I} and plotted in
Figs. 6 to 9 of Ref. \cite{II} for $m=0$. The zero-energy level of these energy bands is
such that,
\begin{equation}
\epsilon_{c0} (\pm 2k_F) =\epsilon_{s1} (\pm k_{F\downarrow})= \epsilon_{c\nu}^0 (\pm
[\pi -2k_F])=\epsilon_{s\nu}^0 (\pm [k_{F\uparrow}-k_{F\downarrow}])=0 \, .
\label{eplev0}
\end{equation}

The rotated-electron double occupation $D_r$ and the number $S_{r}$ of spin-down
rotated-electron singly occupied sites whose rotated electrons are not associated with
the $s1$ pseudofermions play an important role in the finite-energy physics and are given
by,
\begin{equation}
D_r\equiv M_{c,\,-1/2} = [L_{c,\,-1/2} +\sum_{\nu =1}^{\infty}\nu N_{c\nu}] \, ;
\hspace{0.5cm} S_r\equiv [M_{s,\,-1/2}-N_{s1}] = [L_{s,\,-1/2} +\sum_{\nu =2}^{\infty}\nu
N_{s\nu}] \, . \label{DrS-r}
\end{equation}
These quantities fully determine the value for the upper-Hubbard band (UHB) energy gap
$E_{uhb}$ and spin gap $E_h$ of the PS energy spectrum (\ref{DE}) as follows,
\begin{equation}
E_{uhb} = 2\mu\,D_r \, ; \hspace{1cm} E_h = 2\mu_0\,H\,S_r \, . \label{om0}
\end{equation}
Note that $E_{uhb}=E_h =0$ for the initial ground state. In equation (\ref{DrS-r})
$L_{\alpha ,\,-1/2}$ are the numbers of $-1/2$ Yang holons ($\alpha =c$) and $-1/2$ HL
spinons ($\alpha =s$) of the excited energy eigenstate, and $N_{\alpha\nu}$ is the number
of $\alpha\nu$ pseudofermions of the same state for the $\alpha\nu\neq c0,\,s1$ branches.

%%%%%%%%%%%%%%%%%%%%%%%%%%%%%%%%%%%%%%%%%%%%%%%%%%%%%%%%%%%%%%%%
\section{THE PSEUDOFERMION SCATTERING THEORY: THE PSEUDOFERMION S MATRIX}

In this section we introduce the pseudofermion scattering theory. We derive the
pseudofermion and pseudofermion-hole $S$ matrices and discuss their relation to the
spectral properties. Our analysis of the problem follows the standard quantum
non-relativistic scattering theory of spin-less particles \cite{Taylor}. A ground-state -
excited-energy-eigenstate transition involves a set of elementary two-pseudofermion
scattering events. Such a transition is divided below into three steps. The
first and second steps have a scatter-less character and lead to the excitation momentum and energy
associated with the transition. Through these two steps the ground-state -
excited-energy-eigenstate transition reaches the many-pseudofermion ``in" state,
which contains the one-pseudofermion ``in" asymptote states of the pseudofermion
scattering theory. The third step corresponds to a well defined set of elementary
two-pseudofermion scattering events which give rise to the ``out" state and
conserve both the momentum and the energy. The latter many-pseudofermion state 
is a excited energy eigenstate and contains the one-pseudofermion ``out" asymptote states of the
pseudofermion scattering theory. An important point for applications to the study of
the finite-energy spectral and dynamical properties is that all ``in" and ``out" states of the 
theory are energy eigenstates

We start our analysis by a discussion of the PS subspaces and the pseudofermion creation
and annihilation operator anti-commutation relations.

%%%%%%%%%%%%%%%%%%%%%%%%%%%%%%%%%%%%%%%%%%%%%%%%%%%%%%%%%%%%%%%%
\subsection{PS SUBSPACES AND THE PSEUDOFERMION OPERATOR ANTICOMMUTATORS}

Several PS subspaces play an important role in the pseudofermion theory. An {\it
electronic ensemble space} is a subspace spanned by all energy eigenstates with the same
values for the electronic numbers $N_{\uparrow}$ and $N_{\downarrow}$. A {\it CPHS
ensemble space} is a subspace spanned by all energy eigenstates with the same values for
the numbers $\{M_{\alpha,\,\pm 1/2}\}$ of $\pm 1/2$ holons ($\alpha =c$) and $\pm 1/2$
spinons ($\alpha =c$) \cite{I,II}. (In CPHS ensemble space, CPHS refers to $c0$
pseudofermion, holon, and spinon.) A {\it CPHS ensemble subspace} is spanned by all
energy eigenstates with the same values for the sets of numbers $N_{c0}$, $\{N_{c\nu}\}$,
$\{N_{s\nu}\}$, $L_{c,\,-1/2}$, and $L_{s,\,-1/2}$ such that $\nu =1,2,...$.

At zero absolute temperature the pseudofermion description corresponds to a ground-state 
normal-ordered theory \cite{IIIb}. Thus, there is a pseudofermion theory for each initial ground 
state. The minimum excitation energy value of the energy eigenstates that span a given 
CPHS ensemble subspace involves the gap parameters of Eq. (\ref{om0}) and is given by,
\begin{equation}
\omega_0 = \omega_0 (D_r,\,S_r) = E_{uhb} + E_h = 2\mu\,D_r + 2\mu_0\,H\,S_{r} \, .
\label{omega0}
\end{equation}
For the ground state $D_r=S_r=0$ and thus $\omega_0=0$. The application onto the latter 
state of an one-,two-, or any other finite-number-electron operator ${\cal{O}}^{\dag}$ 
generates an excitation which can be described as a suitable superposition of PS excited 
energy eigenstates. The PDT provides the matrix elements associated with the coefficients 
of such a superposition \cite{V,V-1,LE}. Therefore, the elementary 
ground-state - excited-energy-eigenstate transition plays a central role in the 
pseudofermion theory.

An excitation ${\cal{O}}^{\dag}\vert GS\rangle$ associated with
small values for the deviations $\Delta N_{\uparrow}$ and $\Delta N_{\downarrow}$ is
contained in a well defined direct sum of CPHS ensemble subspaces,
\begin{equation}
{\cal{S}}^1_{cphs} \oplus {\cal{S}}^2_{cphs} \oplus {\cal{S}}^3_{cphs} \oplus
{\cal{S}}^4_{cphs} \oplus ... \, . \label{cphses}
\end{equation}
Here ${\cal{S}}^i_{cphs}$ with $i=1,\,2,\,3,...$ corresponds to different CPHS ensemble
subspaces. The pseudofermion, Yang holon, and HL spinon number deviations of all the CPHS
ensemble subspaces of such a direct sum obey the sum rules (18) and (19) of Ref. \cite{V}
and the selection rules given in Eq. (21) of the same reference.

Let us consider a $\alpha\nu$ pseudofermion of canonical momentum ${\bar{q}}$ and a
$\alpha'\nu'$ pseudofermion of canonical momentum ${\bar{q}'}$ such that the
canonical-momentum values ${\bar{q}}$ and ${\bar{q}'}$ correspond to an excited energy
eigenstate and the initial ground state, respectively. Following the results of Refs.
\cite{V,V-1,LE}, the anticommutators involving the creation and/or annihilation operators of
these two pseudofermions play a key role in the study of the finite-energy spectral and
dynamical properties. Such anticommutators read \cite{IIIb},
\begin{equation}
\{f^{\dag }_{{\bar{q}},\,\alpha\nu},\,f_{{\bar{q}}',\,\alpha'\nu'}\} =
{\delta_{\alpha\nu,\,\alpha'\nu'}\over N^*_{\alpha\nu}}\,e^{-i({\bar{q}}-{\bar{q}}')/
2}\,e^{iQ_{\alpha\nu}(q)/2}\,{\sin\Bigl(Q_{\alpha\nu} (q)/
2\Bigr)\over\sin ([{\bar{q}}-{\bar{q}}']/2)} \, ; \hspace{0.25cm} \{f^{\dag
}_{{\bar{q}},\,\alpha\nu},\,f^{\dag
}_{{\bar{q}}',\,\alpha'\nu'}\}=\{f_{{\bar{q}},\,\alpha\nu},\,f_{{\bar{q}}',\,\alpha'\nu'}\}=0
\, . \label{pfacrGS}
\end{equation}
Here $Q_{\alpha\nu}(q)/2$ is the value of the following functional for the above excited
energy eigenstate,
\begin{equation}
Q_{\alpha\nu}(q_j)/2 = Q_{\alpha\nu}^0/2 + Q^{\Phi}_{\alpha\nu} (q_j)/2 \, ,
\label{Qcan1j}
\end{equation}
where $Q^{\Phi}_{\alpha\nu} (q_j)/2$ is the functional given in Eq. (\ref{qcan1j}) whose
bare-momentum distribution function deviations correspond to that state and \cite{IIIb},
\begin{eqnarray}
Q_{c0}^0 & = & 0 \, ; \hspace{0.5cm} \sum_{\alpha =c,\,s}\,\sum_{\nu=1}^{\infty} \Delta
N_{\alpha\nu} \hspace{0.25cm} {\rm even} \, ;  \hspace{1.0cm} Q_{c0}^0=\pm\pi \, ;
\hspace{0.5cm} \sum_{\alpha =c,\,s}\,\sum_{\nu=1}^{\infty} \Delta
N_{\alpha\nu} \hspace{0.25cm} {\rm odd} \, ; \nonumber \\
Q_{\alpha\nu}^0 & = & 0 \, ; \hspace{0.5cm} \Delta N_{c0}+\Delta N_{\alpha\nu}
\hspace{0.25cm} {\rm even} \, ; \hspace{1.0cm} Q_{\alpha\nu}^0=\pm\pi \, ; \hspace{0.5cm}
\Delta N_{c0}+\Delta N_{\alpha\nu} \hspace{0.25cm} {\rm odd} \, ; \hspace{0.5cm} \alpha =
c,\,s \, , \hspace{0.25cm} \nu > 0 \, . \label{pic0an}
\end{eqnarray}
When $Q_{\alpha\nu}^0=\pm\pi$ for the $\alpha\nu\neq c0$ bands, the uniquely
chosen and only permitted value $Q_{\alpha\nu}^0=\pi$ or $Q_{\alpha\nu}^0=-\pi$ is that which
leads to symmetrical limiting discrete bare-momentum values $\pm [\pi/L][N^*_{\alpha\nu}-1]$
for the excited-state bare-momentum band. (See Eq. (B.14) of Ref. \cite{I}.)
In turn, for the $c0$ branch the bare-momentum band width is $2\pi$. Thus, in
this case $Q_{c0}^0=\pi$ and $Q_{c0}^0=-\pi$ lead to allowed occupancy configurations
of alternative excited energy eigenstates. (In the particular case that the $c0$
band is full for the excited energy eigenstate, the two values $Q_{c0}^0=\pi$ and 
$Q_{c0}^0=-\pi$ refer to two equivalent representations of that state.)

The quantity $Q_{\alpha\nu}^0/L$ is the shift in the discrete bare-momentum value $q_j =
[2\pi/L] I^{\alpha\nu}_j$ of Eq. (\ref{barqan}) that arises due to the transition from
the ground state to the excited energy eigenstate. Furthermore, $Q_{\alpha\nu}(q_j)/L$ is
the corresponding shift in the discrete canonical-momentum
values that occurs as a result of the same transition. It follows from Eq.
(\ref{pfacrGS}) that the functional $Q_{\alpha\nu}(q)/2$ fully controls the pseudofermion
anticommutators associated with the ground-state - excited-energy-eigenstate transition. The one-
and two-electron matrix elements between the initial ground state and the excited energy
eigenstates can be expressed in terms of the anticommutators (\ref{pfacrGS}) \cite{V}.
This justifies the importance of the functional $Q_{\alpha\nu}(q)/2$ given in Eq.
(\ref{Qcan1j}), once it controls the quantum overlaps associated with the one- and
two-electron finite-energy spectral properties \cite{V,V-1,LE}.

Let us introduce the $\alpha\nu$ pseudofermion scattering theory. Within such a theory
the functional $Q_{\alpha\nu}(q)/2$ is an overall $\alpha\nu$ pseudofermion or hole phase
shift.

%%%%%%%%%%%%%%%%%%%%%%%%%%%%%%%%%%%%%%%%%%%%%%%%%%%%%%%%%%%%%%%%
\subsection{THE GROUND-STATE - VIRTUAL-STATE TRANSITION}

From now on and until section IV-D our analysis refers to initial ground states with 
density values in the ranges $0<n<1$ and $0<n<m$. The specific properties of the 
scattering theory for initial ground states corresponding to $n=1$ and/or $m=0$ are
considered in that section. The preliminary analysis of the pseudofermion scattering
problem presented in Ref. \cite{Car04} divided each transition from the initial ground 
state to a PS excited energy eigenstate into two main steps. However, it is useful for 
our study to divide the first step considered in that reference into two processes. 
The first process considered here is a scatter-less 
finite-energy and finite-momentum excitation which transforms the ground state onto 
a well defined virtual state. For $\nu>0$ branches, that excitation can involve a change 
in the number of discrete bare-momentum values given by,
\begin{equation}
\Delta N^*_{s1} = \Delta N_{c0} - \Delta N_{s1} - 2\sum_{\nu =2}^{\infty} \Delta N_{s\nu}
\, ; \hspace{0.5cm} \Delta N^*_{\alpha\nu} = \Delta L_{\alpha} + 2\sum_{\nu'=\nu
+1}^{\infty} (\nu' -\nu) \Delta N_{\alpha\nu'} \, ; \hspace{0.5cm} \alpha\nu\neq c0,\, s1
\, . \label{DN*s1an}
\end{equation}
For the initial ground state these numbers read,
\begin{equation}
N^{0,*}_{c\nu}=(N_a -N) \, ; \hspace{1cm} N^{0,*}_{s1}=N_{\uparrow} \, ; \hspace{1cm}
N^{0,*}_{s\nu}=(N_{\uparrow} -N_{\downarrow}) \, , \hspace{0.3cm} \nu > 1 \, ,
\label{N*csnu}
\end{equation}
and $N^{0,*}_{c0}=N^*_{c0}$ is given by $N^*_{c0}=N_a$ for the whole Hilbert space.  
Although the $\alpha\nu\neq c0,\, s1$ branches have no finite pseudofermion occupancy 
in the initial ground state, for densities $0<n<1$ and $0<m<n$ one can define the values
$N^*_{\alpha\nu}=N^h_{\alpha\nu}$ for the corresponding empty bands. For the
$\alpha\nu\neq c0,1,s1$ branches, those are the numbers $N^{0,*}_{c\nu}$ and
$N^{0,*}_{s\nu}$ given in Eq. (\ref{N*csnu}). Thus, for $\alpha\nu\neq c0,\, s1$ branches
with finite pseudofermion occupancy in the virtual state the deviations (\ref{DN*s1an})
and discrete bare-momentum shifts (\ref{pic0an}) are relative to the values of these
empty bands. 

In addition and following the change in the number of discrete bare-momentum values,
this excitation also involves the pseudofermion creation and annihilation processes and 
pseudofermion particle-hole processes associated with PS excited states. The 
momentum and energy of this ground-state - virtual-state transition is given by,
\begin{eqnarray}
\Delta P & = & \sum_{q=-\pi}^{+\pi} \,q\,\Delta N_{c0} (q) +
\sum_{q=-k_{F\uparrow}}^{+k_{F\uparrow}} \,q\,\Delta {\cal{N}}_{s1} (q) +
\sum_{\nu =2}^{\infty}\,\sum_{q=-[k_{F\uparrow}-
k_{F\downarrow}]}^{+[k_{F\uparrow}-k_{F\downarrow}]}\,q\,\Delta
N_{s\nu}(q) \nonumber \\
& + &  \pi\,[L_{c,\,-1/2} +\sum_{\nu =1}^{\infty}\nu N_{c\nu}] + \sum_{\nu
=1}^{\infty}\,\sum_{q=-[\pi -2k_F]}^{+[\pi -2k_F]}\,[\pi -q]\,\Delta N_{c\nu}(q)
\, , \label{DP}
\end{eqnarray}
and
\begin{eqnarray}
\Delta E & = & \sum_{q=-\pi}^{+\pi} \,\epsilon_{c0} (q)\Delta N_{c0} (q) +
\sum_{q=-k_{F\uparrow}}^{+k_{F\uparrow}}\epsilon_{s1}(q)\,\Delta
N_{s1}(q) \nonumber \\
& + & E_{h} + \sum_{\nu =2}^{\infty}\,\sum_{q=-[k_{F\uparrow}-
k_{F\downarrow}]}^{+[k_{F\uparrow}-k_{F\downarrow}]}\epsilon^0_{s\nu}(q)\,\Delta
N_{s\nu}(q) + E_{uhb} + \sum_{\nu =1}^{\infty}\,\sum_{q=-[\pi -2k_F]}^{+[\pi
-2k_F]}\epsilon^0_{c\nu}(q)\,\Delta N_{c\nu}(q) \, , \label{DE-0}
\end{eqnarray}
respectively, where all quantities were defined above. Except for $1/L$ energy corrections, 
the energy spectra (\ref{DE}) and (\ref{DE-0}) are identical. In this first scatter-less 
step the pseudofermions acquire the excitation momentum and energy needed for the 
second and third steps.

%%%%%%%%%%%%%%%%%%%%%%%%%%%%%%%%%%%%%%%%%%%%%%%%%%%%%%%%%%%%%%%%
\subsection{PSEUDOFERMION SCATTERING PROCESSES, S MATRICES, AND PHASE SHIFTS}

In order to study the second and third processes of the ground-state
- excited-energy-eigenstate transition, it is useful to express the
many-pseudofermion states and operators in terms of one-pseudofermion
states and operators, respectively.
The PS energy and momentum eigenstates can be written as direct products of  
states spanned by the occupancy configurations of each of the 
$\alpha\nu$ branches with finite pseudofermion occupancy in the state
under consideration. Moreover, the many-pseudofermion states spanned
by occupancy configurations of each $\alpha\nu$ branch can be expressed 
as a direct product of $N^*_{\alpha\nu}$ one-pseudofermion states,
each referring to one discrete bare-momentum value $q_j$, where
$j=1,2,...,N^*_{\alpha\nu}$.

Within the pseudofermion description, the 1D Hubbard model in normal order relative 
to the initial ground state has no residual-interaction energy terms. Thus,
when acting in the PS it has a uniquely defined expression of the general 
form \cite{IIIb},
\begin{equation}
:\hat{H}: = \sum_{\alpha\nu}\sum_{j=1}^{N^*_{\alpha\nu}}\hat{H}_{\alpha\nu,q_j} +
\sum_{\alpha}\hat{H}_{\alpha} \, ,
\label{Hexp}
\end{equation}
where we denoted the ground-state normal ordered Hamiltonian by $:\hat{H}:$,
$\hat{H}_{\alpha\nu,q_j}$ is the one-pseudofermion Hamiltonian which
describes the $\alpha\nu$ pseudofermion or hole of bare-momentum $q_j$,
and $\hat{H}_{\alpha}$ refers to the Yang holons ($\alpha =c$) and HL spinons 
($\alpha =s$), which are scatter-less objects. Thus, for each 
many-pseudofermion PS virtual state reached in the first step of the
transition from the ground state to the excited energy eigenstate,
the number of Hamiltonians $\hat{H}_{\alpha\nu,q_j}$ equals that of 
one-pseudofermion states of the virtual state given by,
\begin{equation}
N^*_{c0} + N^*_{s1} + \sum_{\alpha\nu\neq c0 ,\,s1} \theta (\vert\Delta
N_{\alpha\nu}\vert)\, N^*_{\alpha\nu} \, . \label{DimSm}
\end{equation}
Here $\theta (x)=1$ for $x>0$ and $\theta (x)=0$ for $x= 0$ and 
$N^*_{c0} =N_{c0}+N^h_{c0}$, $N^*_{s1}=N_{s1}+N^h_{s1}$, and
$N^*_{\alpha\nu}=N_{\alpha\nu}+N^h_{\alpha\nu}$ refer to
the virtual state and corresponding excited energy eigenstate 
under consideration. The pseudofermion-hole numbers of these
expressions read \cite{I},
\begin{equation}
N^h_{c0} = N_a - N_{c0} \, ; \hspace{0.5cm} N^h_{c\nu} = N^h_{c0} -
\sum_{\nu'=1}^{\infty} \Bigl(\nu + \nu' - \vert\nu - \nu'\vert\Bigl) N_{c\nu'} \, ;
\hspace{0.5cm} N^h_{s\nu} = N_{c0} - \sum_{\nu'=1}^{\infty} \Bigl(\nu + \nu' - \vert\nu -
\nu'\vert\Bigl) N_{s\nu'} \, . \label{Nhag}
\end{equation}

For the pseudofermion description only momentum and energy contributions 
of order zero and one in $1/L$ are physical and thus our analysis refers 
to periodic boundary conditions and large values of $L$ such that $L>>1$. 
The second scatter-less process generates the ``in" state. 
Indeed, the one-pseudofermion states belonging 
to the many-pseudofermion ``in" state are the ``in" asymptote states of the 
pseudofermion scattering theory. The generator of the virtual-state - ``in"-state 
transition is of the form,
\begin{equation}
{\hat{S}}^{0} = \prod_{\alpha\nu}\prod_{j=1}^{N^*_{\alpha\nu}}
{\hat{S}}^{0}_{\alpha\nu ,q_j} \, ,
\label{S0}
\end{equation}
where ${\hat{S}}^{0}_{\alpha\nu ,q_j}$ is a well-defined one-pseudofermion 
unitary operator. Application of ${\hat{S}}^{0}_{\alpha\nu ,q_j}$ onto the 
corresponding one-pseudofermion state of the many-pseudofermion virtual 
state shifts its discrete bare-momentum value $q_j$ to the bare-momentum 
value $q_j+Q_{\alpha\nu}^0/L$, where $Q_{\alpha\nu}^0$ is given in 
Eq. (\ref{pic0an}).

Finally, the third step consists of a set of two-pseudofermion
scattering events. It corresponds to the ``in"-state - ``out"-state transition,
where the latter state is the PS excited energy eigenstate under
consideration. The generator of that transition is the following operator,
\begin{equation}
{\hat{S}}^{\phi} = \prod_{\alpha\nu}\prod_{j=1}^{N^*_{\alpha\nu}}
{\hat{S}}^{\phi}_{\alpha\nu ,q_j} \, ,
\label{Sphi}
\end{equation}
where ${\hat{S}}^{\phi}_{\alpha\nu ,q_j}$ is a 
well-defined one-pseudofermion scattering 
unitary operator. The one-pseudofermion states belonging to the 
many-pseudofermion ``out" state are the ``out" asymptote pseudofermion
scattering states. Application of ${\hat{S}}^{\phi}_{\alpha\nu ,q_j}$ onto the 
corresponding one-pseudofermion state of the many-pseudofermion ``in"
state shifts its discrete bare-momentum value $q_j+Q_{\alpha\nu}^0/L$ to the
``out"-state discrete canonical-momentum value 
$q_j+Q_{\alpha\nu} (q_j)/L$. It follows that the generator of the 
virtual-state - ``out"-state transition is the unitary operator,
\begin{equation}
{\hat{S}}\equiv {\hat{S}}^{\phi}{\hat{S}}^{0} = \prod_{\alpha\nu}\prod_{j=1}^{N^*_{\alpha\nu}}
{\hat{S}}_{\alpha\nu ,q_j} \, ,
\label{Sope}
\end{equation}
where ${\hat{S}}_{\alpha\nu ,q_j}$ is the one-pseudofermion or hole unitary 
${\hat{S}}_{\alpha\nu ,q_j}={\hat{S}}^{\phi}_{\alpha\nu ,q_j}{\hat{S}}^{0}_{\alpha\nu ,q_j}$ 
operator. Application of the latter operator onto the corresponding 
one-pseudofermion state of the many-pseudofermion virtual 
state shifts its discrete bare-momentum value $q_j$ directly to the ``out"-state 
discrete canonical-momentum value $q_j+Q_{\alpha\nu} (q_j)/L$.

The virtual state, ``in" state, and ``out" state are PS excited energy eigenstates, as 
further discussed below. Thus, that the one-pseudofermion states of the 
many-pseudofermion ``in" state and ``out" state are the ``in" and ``out" asymptote 
pseudofermion scattering states, respectively, implies that the one-pseudofermion 
Hamiltonian $\hat{H}_{\alpha\nu,q_j}$ plays the role of the unperturbed 
Hamiltonian $\hat{H}_0$ of the 
spin-less one-particle nonrelativistic scattering theory \cite{Taylor}. Indeed, 
the unitary ${\hat{S}}_{\alpha\nu ,q_j}$ operator (and
the scattering unitary ${\hat{S}}^{\phi}_{\alpha\nu ,q_j}$ operator) commutes with the 
Hamiltonian $\hat{H}_{\alpha\nu,q_j}$ and thus the one-pseudofermion ``in"
and ``out" asymptote scattering states are energy eigenstates of $\hat{H}_{\alpha\nu,q_j}$
and eigenstates of ${\hat{S}}_{\alpha\nu ,q_j}$ (and ${\hat{S}}^{\phi}_{\alpha\nu ,q_j}$). 
It follows that the matrix elements between one-pseudofermion states
of  ${\hat{S}}_{\alpha\nu ,q_j}$ (and ${\hat{S}}^{\phi}_{\alpha\nu ,q_j}$) are
diagonal and thus these operators are fully defined by the set of their eigenvalues
belonging to these states. The same applies to the generator ${\hat{S}}$ 
(and ${\hat{S}}^{\phi}$) given in Eq. (\ref{Sope}) (and Eq. (\ref{Sphi})). 
The matrix elements of that generator between many-pseudofermion virtual states 
(and ``in" states) are also diagonal and thus it is fully defined by the set of 
its eigenvalues belonging to such states. Importantly, the virtual state and the ``in"
state of a given ground-state transition correspond to the same excited energy 
eigenstate of the 1D Hubbard model, which by construction is the ``out" state.
Indeed, the virtual state, ``in" state, and ``out" state only differ by mere 
overall phase factors whose general functional expression is given below. 
That the many-pseudofermion ``in" and ``out" states which are a
direct product of one-pseudofermion ``in" and ``out" asymptote pseudofermion 
scattering states, respectively, are PS excited energy eigenstates plays a major 
role in the pseudofermion scattering theory. 

Since ${\hat{S}}^{\phi}_{\alpha\nu ,q_j}$ and ${\hat{S}}_{\alpha\nu ,q_j}$ are unitary, 
each of their eigenvalues has modulus one and can be written as the exponent of a 
purely imaginary number given by,
\begin{eqnarray}
S^{\Phi}_{\alpha\nu} (q_j) & = & e^{iQ^{\Phi}_{\alpha\nu}(q_j)} =
\prod_{\alpha'\nu'}\,\prod_{j'=1}^{N^*_{\alpha'\nu'}}\,S_{\alpha\nu ,\,\alpha'\nu'} (q_j, q_{j'}) \, ;
\hspace{0.25cm} j=1,2,..., N^*_{\alpha\nu} \nonumber \\
S_{\alpha\nu} (q_j) & = & e^{iQ_{\alpha\nu}(q_j)} =
e^{i\,Q_{\alpha\nu}^0}  S^{\Phi}_{\alpha\nu} (q_j) \, ;
\hspace{0.25cm} j=1,2,..., N^*_{\alpha\nu}
\, . \label{San}
\end{eqnarray}
Here $Q^{\Phi}_{\alpha\nu}(q_j)$ and $Q_{\alpha\nu}(q_j)$ are the 
functionals defined by Eqs. (\ref{qcan1j}) and (\ref{Qcan1j}), respectively. By use of the
former functional we find that,
\begin{equation}
S_{\alpha\nu ,\,\alpha'\nu'} (q_j, q_{j'}) =
e^{i2\pi\,\Phi_{\alpha\nu,\,\alpha'\nu'}(q_j,q_{j'})\, \Delta N_{\alpha'\nu'}(q_{j'})}
\, , \label{Sanan}
\end{equation}
where the functions $\pi\,\Phi_{\alpha\nu,\,\alpha'\nu'}(q_j,q_{j'})$ are
given in Eq. (\ref{Phi-barPhi}). The main point is that except for the occupancy 
configuration changes produced by the ground-state - virtual-state transition,
the only effect of under a ground-state - excited-energy-eigenstate transition moving the 
$\alpha\nu$ pseudofermion or hole of initial virtual-state canonical-momentum 
${\bar{q}}_j=q_j$ once around the length $L$ lattice ring is that its wave function acquires 
the overall phase factor $S_{\alpha\nu} (q_j)$ given in Eq. (\ref{San}).
This property follows from the form of the energy spectrum of the pseudofermions,
which in contrast to that of the corresponding pseudoparticles of Ref. \cite{I} has no 
residual interaction terms \cite{IIIb}.

The phase factor $S_{\alpha\nu ,\,\alpha'\nu'} (q_j, q_{j'})$ of Eq.  (\ref{Sanan}) 
in the wave function of the $\alpha\nu$ pseudofermion or hole of bare-momentum
$q_j$ results from an elementary two-pseudofermion 
zero-momentum forward-scattering event whose scattering
center is a $\alpha'\nu'$ pseudofermion ($\Delta N_{\alpha'\nu'}(q_{j'})=1$)
or $\alpha'\nu'$ pseudofermion hole ($\Delta N_{\alpha'\nu'}(q_{j'})=-1$) created
under the ground-state - excited-state transition. Thus, the third step 
of that transition involves a well-defined set of elementary two-pseudofermion 
scattering events where all $\alpha\nu$ pseudofermions and $\alpha\nu$ 
pseudofermion holes of bare-momentum $q_j+Q_{\alpha\nu}^0/L$ of the ``in"
state are the scatterers, which leads to the overall scattering phase factor 
$S^{\Phi}_{\alpha\nu} (q_j)$ in their wave function given in Eq. (\ref{San}). 
That the scattering centers are the $\alpha'\nu'$ pseudofermions or 
pseudofermion holes of bare momentum $q_{j'}+Q_{\alpha\nu}^0/L$ created 
under the ground-state - ``in"-state transition is confirmed 
by noting that $S_{\alpha\nu ,\,\alpha'\nu'} (q_j,q_{j'})=1$ for 
$\Delta N_{\alpha'\nu'}(q_{j'}) =0$. Thus, out of the scatterers whose 
number equals that of the one-pseudofermion states
given in Eq. (\ref{DimSm}), the scattering centers are only those whose 
bare-momentum distribution-function deviation is finite. However, out of the 
above scatterers, only a subclass of scatterers contributes significantly to the 
spectral properties \cite{V,V-1}.

The following properties play an important role in the pseudofermion scattering
theory:
\begin{enumerate}

\item
The elementary two-pseudofermion scattering processes associated with the phase
factors (\ref{Sanan}) conserve the total energy and total momentum. Such an energy
conservation is further discussed in Appendix A.

\item
The elementary two-pseudofermion scattering processes are of zero-momentum 
forward-scattering type and thus conserve the individual ``in"
 asymptote $\alpha\nu$ 
pseudofermion momentum value $q_j+Q_{\alpha\nu}^0/L$ and energy.

\item
These processes also conserve the $\alpha\nu$ branch, usually called {\it channel} in
the scattering language \cite{Taylor}.

\item
The scattering amplitude does not connect quantum objects with different $\eta$ spin or
spin.

\item
For each $\alpha\nu$ pseudofermion or pseudofermion hole of virtual-state bare-momentum
$q_j$, the $S$ matrix associated with the ground-state - excited-energy-eigenstate transition 
is simply the phase factor $S_{\alpha\nu} (q_j)$ given in Eq. (\ref{San}).
\end{enumerate}

The one-particle phase factor $s_l (E)$ of Eq. (6.8) of Ref. \cite{Taylor} whose 
expression is given in Eq. (6.9) of the same reference corresponds 
to the one-$\alpha\nu$-pseudofermion or hole phase factor $S^{\Phi}_{\alpha\nu} (q_j)$ 
with the energy $E$ and the quantum numbers $l$ and $m$ replaced 
by the bare-momentum $q_j$. Indeed, while the $\alpha\nu$ pseudofermion 
or hole energy is uniquely defined by the absolute value $\vert q_j\vert$,
in 1D the sign of $q_j$ corresponds to the three-dimensional 
angular-momentum quantum numbers. Another difference is that $s_l (E)$ 
is associated with a single scattering event whereas $S^{\Phi}_{\alpha\nu} (q_j)$ 
results in general from several scattering events. Each of such events 
corresponds to a well defined factor $S_{\alpha\nu ,\,\alpha'\nu'} (q_j, q_{j'})$
of form (\ref{Sanan}) in the expression of $S^{\Phi}_{\alpha\nu} (q_j)$
given in Eq. (\ref{San}). There are as many of such factors as 
$\alpha'\nu'$ pseudofermion and hole scattering centers created 
under the transition to the virtual state and corresponding
excited energy eigenstate under consideration. 
The factor $2$ in the phase factor of Eq. (6.9) 
of Ref. \cite{Taylor} corresponds to the phase-shift definition of the standard 
nonrelativistic scattering theory for spin-less particles. 
As discussed below, we use here such 
a definition which introduces the overall scattering phase shift 
$\delta^{\Phi}_{\alpha\nu} (q_j)= Q^{\Phi}_{\alpha\nu} (q_j)/2$ and
overall phase shift $\delta_{\alpha\nu} (q_j)= Q_{\alpha\nu} (q_j)/2$. 
However, if instead we insert a factor $1$, we would introduce the 
overall scattering phase shift $Q^{\Phi}_{\alpha\nu} (q_j)$ and
overall phase shift $Q_{\alpha\nu} (q_j)$ whose values are defined
only to within addition of an arbitrary multiple of $2\pi$. That is the
choice for the phase shift definition used in Refs. \cite{S0,S,relation}.
For our definition the phase shifts are instead given only to within
addition of an arbitrary multiple of $\pi$, as in Ref. \cite{Taylor}.   

Application of the unitary ${\hat{S}}_{\alpha\nu ,q_j}$ operator onto
its one-pseudofermion state of the many-pseudofermion 
virtual state generates the corresponding one-pseudofermion 
state of the many-pseudofermion ``out" state. The latter one-pseudofermion 
state equals the former one multiplied by the phase factor 
$S_{\alpha\nu} (q_j)$ of Eq. (\ref{San}).
(Applying the scattering unitary ${\hat{S}}^{\Phi}_{\alpha\nu ,q_j}$ operator 
onto its one-pseudofermion state of the many-pseudofermion
``in" state also generates the corresponding one-pseudofermion 
state of the many-pseudofermion ``out" state; The latter one-pseudofermion 
state equals the former one multiplied by the phase factor 
$S^{\Phi}_{\alpha\nu} (q_j)$ of Eq. (\ref{San}).) It follows that the 
many-pseudofermion virtual states (and ``in" states) are eigenstates
of the generator ${\hat{S}}$ (and ${\hat{S}}^{\phi}$) given in Eq. 
(\ref{Sope}) (and Eq. (\ref{Sphi})). The eigenvalue of ${\hat{S}}$
belonging to a PS virtual state is given by,
\begin{equation}
S_T = e^{i2\delta_T} = \prod_{\alpha\nu}\prod_{j=1}^{N^*_{\alpha\nu}}
S_{\alpha\nu} (q_j) \, ; \hspace{0.5cm}
\delta_T = \sum_{\alpha\nu}\sum_{j=1}^{N^*_{\alpha\nu}} 
Q_{\alpha\nu}(q_j)/2 \, .
\label{ST}
\end{equation}
Thus, the corresponding ``out" state equals the virtual state multiplied by the
phase factor $S_T$. Furthermore, the eigenvalue of ${\hat{S}}^{\Phi}$
belonging to a PS ``in"
 state is given by,
\begin{equation}
S^{\Phi}_T = e^{i2\delta^{\Phi}_T} = \prod_{\alpha\nu}\prod_{j=1}^{N^*_{\alpha\nu}}
S^{\Phi}_{\alpha\nu} (q_j) \, ; \hspace{0.5cm}
\delta^{\Phi}_T = \sum_{\alpha\nu}\sum_{j=1}^{N^*_{\alpha\nu}} 
Q^{\Phi}_{\alpha\nu}(q_j)/2 \, .
\label{S-Phi}
\end{equation}
Again, the corresponding ``out" state equals the ``in" state multiplied by the
phase factor $S^{\Phi}_T$. Since the ``out" state is by construction an energy 
eigenstate of the 1D Hubbard model, this result confirms that the corresponding 
virtual and ``in" states are also energy eigenstates of the model: the ``out" state
only differs from the latter two states by the phase factors 
given in Eqs. (\ref{ST}) and (\ref{S-Phi}), respectively. The general expressions 
(\ref{qcan1j}) and (\ref{Qcan1j}) for the functionals $Q_{\alpha\nu}^{\Phi}(q_j)$ 
and $Q_{\alpha\nu} (q_j)$ define uniquely the eigenvalues $S^{\Phi}_T$
and $S_T$ of ${\hat{S}}^{\phi}$ and $\hat{S}$ for the whole set of ``in" states and virtual 
states, respectively, corresponding to the excited energy eigenstates that
span the PS.

The factorization of the BA bare $S$ matrix for the original spin $1/2$ electrons is
associated with the so called Yang-Baxter Equation (YBE) \cite{Natan,S}. On
the other hand, the factorization of the $\alpha\nu$ pseudofermion or hole $S$ matrix
$S_{\alpha\nu} (q_j)$ given in Eq. (\ref{San}), in terms of the elementary 
two-pseudofermion $S$ matrices $S_{\alpha\nu ,\,\alpha'\nu'} (q_j, q_{j'})$, 
Eq. (\ref{Sanan}), is commutative. Such commutativity is stronger than
the symmetry associated with the YBE. The pseudofermion $S$ matrix commutative
factorization is required by the form of the pseudofermion occupancy configurations that
describe the PS excited energy eigenstates. These states are direct products of
the one-pseudofermion scattering states of the theory and are described by well defined
occupancy configurations of rotated electrons. All such configurations have the same
number of rotated-electron occupied sites, unoccupied sites, spin-up singly occupied
sites, and spin-down singly occupied sites. However, the relative position of these
quantum objects is different in each occupancy configuration. There is a one-to-one
correspondence between these rotated-electron configurations and the local pseudofermion,
Yang holon, and HL spinon occupancy configurations that describe the same state. Again,
the number of local $\alpha\nu$ pseudofermions belonging to $\alpha\nu$ branches with
finite occupancy in the virtual state are the same for all occupancy configurations. As
for the rotated electrons, the relative position of these quantum objects is different in
each configuration. Thus, when under a specific ground-state - excited-energy-eigenstate
transition a $\alpha\nu$ pseudofermion or hole moves around the lattice ring, it
scatters the same $\alpha'\nu'$ pseudofermion or hole scattering centers, but in
different order for different occupancy configurations. However, it is required that the
phase factor $e^{iQ_{\alpha\nu}(q_j)}$ acquired by the $\alpha\nu$ pseudofermion 
or hole should be the same, independently of that order. This implies the commutativity 
of the $S$-matrix factors $S_{\alpha\nu ,\,\alpha'\nu'} (q_j, q_{j'})$ in the overall
$S$ matrix $S_{\alpha\nu} (q_j)$ whose expression is given in Eq. (\ref{San}). 
Such commutativity follows from the elementary $S$ matrices $S_{\alpha\nu
,\,\alpha'\nu'} (q_j, q_{j'})$ given in Eq. (\ref{Sanan}) being simple phase factors,
instead of matrices of dimension larger than one. This seems to be inconsistent with all
energy eigenstates being described by occupancy configurations which, besides $c0$
pseudofermions, involve finite-spin $1/2$ spinons and $\eta$-spin $1/2$ holons \cite{I}.
Indeed, the $S$ matrix of finite-$\eta$-spin or spin quantum objects is a matrix of
dimension larger than one \cite{relation,Taylor}. However, in spite of these finite-$\eta$-spin
and finite-spin objects, due to symmetry the system self organizes in such a way that, in
addition to the $\eta$-spin-less and spin-less $c0$ pseudofermions, the scatterers and 
scattering centers are the $\eta$-spin zero $2\nu$-holon composite $c\nu$ pseudofermions 
and spin zero $2\nu$-spinon composite $s\nu$ pseudofermions. Symmetry requirements 
also imply that the $\eta$-spin $1/2$ Yang holons and spin $1/2$ HL spinons \cite{I,II} are purely
scatter-less objects, as discussed in the following.

Thus, it is the $\eta$-spin-neutral (and spin-neutral) character of the $2\nu$-holon (and $2\nu$-spinon) 
composite pseudofermion scatterers and scattering centers and the $\eta$-spin-less and spin-less 
character of the $c0$ pseudofermion scatterers and scattering centers which is behind the
dimension of their $S$ matrix $S_{\alpha\nu} (q_j)$ given in Eq. (\ref{San}). Interestingly, in the 
pseudofermion scattering theory the relation of the composite $c\nu$ pseudofermion 
(and $s\nu$ pseudofermion) scatterers and scattering centers to the holons (and spinons) 
has similarities with that of the composite physical particles to the quarks in chromodynamics 
\cite{Martinus}. Within the latter theory the quarks have color but all quark-composite physical 
particles are color-neutral. Here the holons (and spinons) have $\eta$ spin $1/2$ (and spin $1/2$) 
but the $2\nu$-holon (and $2\nu$-spinon) composite pseudofermion scatterers and scattering 
centers are $\eta$-spin-neutral (and spin-neutral).

In turn, the holon-holon and spinon-spinon $S$ matrices of the conventional
spinon-holon scattering theory of Refs. \cite{Natan,S0,S} do not have the above 
commutative properties. Indeed, within that theory the scatterers and scattering 
centers have $\eta$-spin $1/2$ or spin $1/2$ and for initial ground states with 
densities in the ranges $0<n<1$ and $0<m<n$ many one-particle scattering states 
do not refer to energy eigenstates \cite{relation}. (The above mentioned requirement 
for commutative factorization of the $S$ matrix applies when the one-particle 
scattering states belong to many-particle energy eigenstates and the scatterers and 
scattering centers are $\eta$-spin-neutral and/or spin-neutral.) On the other hand, 
for the $n=1$ and $m=0$ initial ground state considered in Refs. \cite{Natan,S0,S} 
the one-particle scattering states of the conventional spinon-holon representation
of these references belong to many-particle energy eigenstates \cite{relation}. 
However, since the scatterers and scattering centers of that theory have $\eta$-spin $1/2$ 
or spin $1/2$, instead of the commutative factorization it is required that the $S$ matrix
has a YBE like factorization, as the BA bare $S$ matrix of the original spin $1/2$ 
electrons.

%%%%%%%%%%%%%%%%%%%%%%%%%%%%%%%%%%%%%%%%%%%%%%%%%%%%%%%%%%%%%%%%
\subsection{SCATTERERS, SCATTERING CENTERS, AND SYMMETRY}

The three generators of both the $\eta$-spin and spin $SU(2)$ algebras commute with
the electron - rotated-electron unitary operator \cite{I,II}. As a result, the symmetry
transformations of the $\alpha\nu$ pseudofermions, $\pm 1/2$ Yang holons, and $\pm 1/2$
HL spinons play an important role in their scattering properties.

For initial ground states with densities such that $0< n< 1$ and $0<m< n$ there is
only finite occupancy for $c0$ pseudofermions and holes, $s1$ pseudofermions and holes,
$+1/2$ Yang holons, and $+1/2$ HL spinons. The corresponding occupation numbers read
$N_{c0}=N$, $N^h_{c0}=[N_a-N]$, $N_{s1}=N_{\downarrow}$,
$N^h_{s1}=[N_{\uparrow}-N_{\downarrow}]$, $L_{c,\,+1/2}=[N_a -N]$, and
$L_{s,\,+1/2}=[N_{\uparrow}-N_{\downarrow}]$. However, that for such ground states
$N^h_{c0}=L_{c,\,+1/2}$ and $N^h_{s1}=L_{s,\,+1/2}$ does not imply that the $c0$
pseudofermion holes and the $s1$ pseudofermion holes are the same quantum objects as the
$+1/2$ Yang holons and $+1/2$ HL spinons, respectively. For instance, the $[N_a -N]$
$+1/2$ Yang holons have finite $\eta$ spin $1/2$ and 
the $[N_{\uparrow}-N_{\downarrow}]$ $+1/2$ HL spinons finite spin $1/2$ and 
both these objects are dispersion-less and have zero momentum and energy. In
contrast, the $c0$ pseudofermion holes are $\eta$-spin-less and spin-less and the 
$s1$ pseudofermion holes have spin zero and are $\eta$-spin-less and both
these types of pseudofermion holes have a momentum-dependent energy
dispersion, $\epsilon_{c0} ({\bar{q}})$ and $\epsilon_{s1} ({\bar{q}})$, respectively. 
The ground-state canonical-momentum distributions of the $[N_a -N]$ $c0$ 
pseudofermion holes and $[N_{\uparrow}-N_{\downarrow}]$ $s1$ pseudofermion 
holes correspond to compact domains such that $2k_F<\vert{\bar{q}}\vert<\pi$ and
$k_{F\downarrow}<\vert{\bar{q}'}\vert<k_{F\uparrow}$, respectively. Moreover, while under
a ground-state - excited-energy-eigenstate transition the $c0$ and $s1$ pseudofermion-hole 
discrete canonical-momentum values ${\bar{q}}_j$ and ${\bar{q}'}_j$ acquire a shift
given by $Q_{c0} (q_j)/L$ and $Q_{s1} ({q'}_j)/L$, respectively, the momentum zero of
the $+1/2$ Yang holons and $+1/2$ HL spinons remains unchanged. Also
the momentum values $\pi$ and zero of the $-1/2$ Yang holons and $-1/2$ HL spinons,
respectively, remain unchanged under such a transition.

The form of the scattering part of the overall phase shift (\ref{Qcan1j}), Eq.
(\ref{qcan1j}), reveals that the value of such a phase-shift functional is independent of
the changes in the occupation numbers of the $\pm 1/2$ Yang holons and $\pm 1/2$ HL
spinons. Thus, these objects are not scattering centers. Moreover, they are not
scatterers, once their momentum values remain unchanged under the transition
from the ``in" state to the ``out" state (excited energy eigenstate). 
Such a scatter-less character of the $\pm 1/2$ Yang holons and $\pm 1/2$ HL spinons 
is related to the above symmetry. Indeed, the $\pm1/2$ Yang holons (and $\pm 1/2$ HL 
spinons) are created and annihilated by the $\eta$-spin (and spin) off-diagonal generators. 
Since these generators commute with the electron - rotated-electron unitary operator, 
they have the same $U$-independent expressions both in terms of electronic and 
rotated-electronic operators \cite{I,II}. That the $\pm 1/2$ Yang holons and $\pm 1/2$ 
HL spinons are neither scatterers nor scattering centers is consistent with that property.

In contrast, since the pseudofermions and pseudofermion holes are not in general invariant
under the electron - rotated-electron unitary transformation \cite{I,IIIb}, their
creation and annihilation operators have $U$-dependent expressions in terms of electronic
operators. That these quantum objects are scatterers and scattering centers is consistent
with such a property. Moreover, these objects are not transformed by the $\eta$-spin and
spin generators. This latter symmetry is behind the $c0$ pseudofermion being a
$\eta$-spin-less and spin-less object and for $\nu >0$ the $\alpha\nu$ pseudofermions
being $\eta$-spin $(\alpha =c)$ and spin $(\alpha =s)$ singlet $2\nu$-holon and
$2\nu$-spinon composite objects, respectively. 

The particular case of the invariance under the electron - rotated-electron unitary
transformation of $\alpha\nu\neq c0,\,s1$ pseudofermions of limiting canonical-momentum
values $\bar{q}=\pm q^0_{\alpha\nu}$ and the corresponding separation of these objects
into $2\nu$ independent holons ($\alpha\nu=c\nu$) or spinons ($\alpha\nu=s\nu$) and
$\alpha\nu\neq c0,\,s1$ FP ({\it Fermi}-point) current scattering centers is discussed 
below in Sec. IV-C.

%%%%%%%%%%%%%%%%%%%%%%%%%%%%%%%%%%%%%%%%%%%%%%%%%%%%%%%%%%%%%%%%
\subsection{PSEUDOFERMION $S$ MATRIX AND THE FINITE-ENERGY SPECTRAL PROPERTIES}

The simple form of the pseudofermion and hole $S$ matrix renders the pseudofermion
description particularly suitable for the study of the unusual finite-energy spectral
properties of the model. Such a simplicity of the $S$ matrix form results in part from 
all ``in" and ``out" states being energy eigenstates. Indeed, the latter states can be 
expressed as direct products of the ``in" and ``out" one-pseudofermion scattering
states of the theory. Fortunately, this allows the use of Lehmann representations for 
the study of the spectral functions \cite{V,V-1,LE}. Moreover, the simple form of the 
pseudofermion and hole $S$ matrix simplifies the evaluation of the spectral-function 
matrix elements between the initial ground state and the exact excited energy eigenstates, 
which are the ``out" states. Indeed, the anticommutation relations (\ref{pfacrGS}) can be expressed
solely in terms of the difference ${\bar{q}}-{\bar{q}}'$ of the two pseudofermion momenta
and the $S$ matrix $S_{\alpha\nu} (q)$ given in Eq. (\ref{San}) of the
pseudofermion associated with the excited energy eigenstate as follows,
\begin{equation}
\{f^{\dag }_{{\bar{q}},\,\alpha\nu},\,f_{{\bar{q}}',\,\alpha'\nu'}\} =
{\delta_{\alpha\nu,\,\alpha'\nu'}\over N^*_{\alpha\nu}}\,\Bigl[S_{\alpha\nu}
(q)\Bigr]^{1/2}\,e^{-i({\bar{q}}-{\bar{q}}')/ 2}\,{{\Im}\Bigl[S_{\alpha\nu}
(q)\Bigr]^{1/2}\over\sin ([{\bar{q}}-{\bar{q}}']/2)} \, , \label{pfacrGS-S}
\end{equation}
and the anticommutators between two $\alpha\nu$ pseudofermion creation or annihilation
operators vanish. This reveals that the $S$ matrix $S_{\alpha\nu} (q_j)$ 
fully controls the pseudofermion anticommutators. Since within the PDT these
anticommutators determine the value of the matrix elements between the ground state and
the excited energy eigenstates \cite{V,V-1,LE}, it follows that the $S$ matrix 
$S_{\alpha\nu} (q_j)$ controls the finite-energy spectral properties. If it had
dimension larger than one, the pseudofermion algebra would be much more involved, for the
pseudofermion anticommutators would also be matrices of dimension larger than one. 
The problem of the evaluation of the spectral-function matrix elements 
between energy eigenstates simplifies for the pseudofermion representation because 
the PS subspaces associated with a given one- or two-electron
spectral function can be expressed in terms of direct products corresponding to each
of the $\alpha\nu$ branch quantum-number occupancy configurations
of branches with finite pseudofermion occupancy \cite{V,V-1,LE}. 
For these matrix elements the direct product is associated 
with the commutative factorization of the $S$ matrix (\ref{San}) 
in terms of the elementary $S$ matrices $S_{\alpha\nu ,\,\alpha'\nu'} (q_j,
q_{j'})$, Eq. (\ref{Sanan}). Therefore, the use of the 
pseudofermion description considerably simplifies the study of the exotic 
quantum-liquid finite-energy spectral properties \cite{V-1,LE,spectral,spectral0,super}.

The PDT of Refs. \cite{V,V-1,LE} confirms that the overall $\alpha\nu$ pseudofermion and
hole phase-shift functional (\ref{Qcan1j}) associated with the $S$ matrix $S_{\alpha\nu} (q_j)$
of Eq. (\ref{San}) fully controls the one- and two-electron spectral properties through 
the pseudofermion anticommutators. An one- or two-electron
excitation ${\cal{O}}^{\dag}\vert GS\rangle$ is contained in a well defined direct sum
(\ref{cphses}) of CPHS ensemble subspaces and is a superposition of excited energy
eigenstates. Each of these states is described by a set of deviations $\{\Delta
N_{\alpha\nu} (q_j)\}$, Eq. (\ref{DNq}), involving a finite number of scattering centers.
Such deviations obey the sum rules (18) and (19) of Ref. \cite{V} and the selection rules
given in Eq. (21) of the same reference. 

One can consider that the order of the direct sum (\ref{cphses}) of CPHS ensemble
subspaces associated with a given one- or two-electron excitation is that of the
increasing values of the minimum energy $\omega_0 (D_r,\,S_r)$, Eq. (\ref{omega0}), of
each CPHS ensemble subspace relative to the initial ground state. For fixed value of
$S_r$ and $D_r$, these energy values are such that,
\begin{equation}
\omega_0 (0,\,S_r)<\omega_0 (1,\,S_r)<\omega_0 (2,\,S_r)<... \, , \label{orderom-s1}
\end{equation}
and
\begin{equation}
\omega_0 (D_r,\,0)<\omega_0 (D_r,\,1)<\omega_0 (D_r,\,2)<... \, , \label{orderomDr}
\end{equation}
respectively. Note that for fixed values of electronic density $n$ and spin density $m$,
the ordering of the energies $\omega_0 (D_r,\,S_r)$ is well defined and such that,
\begin{equation}
\omega_0 (0,\,0)<{\rm min}\{\omega_0 (0,\,1),\,\omega_0 (1,\,0)\}<... \, .
\label{orderom}
\end{equation}

An important property is that for a given energy range $0<\Delta E <\omega_0$ the number
of CPHS ensemble subspaces of the direct sum (\ref{cphses}) is finite. Moreover, as each
energy $\omega_0 (D_r,\,S_r)$ is reached, an increasing number of channels open up that
correspond to CPHS ensemble subspaces of increasing energy. By new channels we mean here
$\alpha\nu$ pseudofermion branches of increasing $\nu$ value. Thus, for a given energy
range $0<\Delta E <\omega_0$ there is a well defined set of pseudofermion $S$ matrices,
whose ``in" and ``out" asymptote one-pseudofermion scattering states correspond to the 
set of excited energy eigenstates spanning the direct sum (\ref{cphses}) of CPHS ensemble 
subspaces of energy smaller than or equal to $\omega_0$. The pseudofermion $S$ matrices 
belonging to that finite set control the spectral properties for energies in the above range 
$0<\Delta E <\omega_0$.

The pseudofermion representation is valid for $L>>1$. Indeed, the expressions
of all quantities derived by use of that representation are physical up to first order in $1/L$. Thus, 
for the study of some properties one can replace the discrete bare-momentum $q_j$ and
canonical-momentum ${\bar{q}}_j$ by a continuum bare-momentum variable $q$ and
canonical-momentum variable $\bar{q}$, respectively. The ground-state is then described
by a $c0$ occupancy for $\vert q\vert <2k_F$ and unoccupancy for $2k_F <\vert q\vert<\pi$
and a $s1$ occupancy for $\vert q\vert < k_{F\downarrow}$ and unoccupancy for
$k_{F\downarrow} <\vert q\vert<k_{F\uparrow}$. (All remaining $\alpha\nu$ pseudofermion
bands are unoccupied for the ground state.) Thus, in the continuum momentum limit, the global
canonical-momentum shift $Q_{\alpha\nu}(q)/L = Q_{\alpha\nu}^0/L + Q^{\Phi}_{\alpha\nu}
(q)/L$ contributes to the spectral properties mainly through the $\alpha\nu =c0,\,s1$
branches for $q$ values in the vicinity of the {\it Fermi points} $\pm q^0_{\alpha\nu}$,
as confirmed by the studies of Refs. \cite{V,V-1,LE}. However, often such a limit
must be taken in the end of the calculations. Otherwise, one would loose the information
contained in the overall pseudofermion or hole phase shifts studied below, which correspond
to canonical-momentum shifts, $Q_{\alpha\nu}(q)/L$, of the order of $1/L$.

%%%%%%%%%%%%%%%%%%%%%%%%%%%%%%%%%%%%%%%%%%%%%%%%%%%%%%%%%%%%%%%%
\section{THE PSEUDOFERMION SCATTERING THEORY: PSEUDOFERMION PHASE SHIFTS}

In this section we study the $\alpha\nu$ pseudofermion phase shifts associated with the
$S$ matrix introduced above.

%%%%%%%%%%%%%%%%%%%%%%%%%%%%%%%%%%%%%%%%%%%%%%%%%%%%%%%%%%%%%%%%
\subsection{PHASE-SHIFT DEFINITION}

The effective $\alpha\nu$ lattices have the same length $L$ as the rotated-electron
and electronic lattice \cite{IIIb}. As above, our analysis refers to periodic boundary 
conditions and $L>>1$. Depending on the asymptote coordinate
choice, there are two possible definitions for the $\alpha\nu$ pseudofermion phase shifts
associated with the $S$ matrix $S_{\alpha\nu} (q_j)$ given in Eq. (\ref{San}). The choice of
either definition is a matter of taste and the uniquely defined quantity is the $S$
matrix. The two choices of asymptote coordinates for the $\alpha\nu$ pseudofermion or
$\alpha\nu$ pseudofermion hole correspond to $x\in (-L/2,\,+L/2)$ and $x\in (0,\,+L)$.

If when moving around the lattice ring the $\alpha\nu$ pseudofermion (or hole) departures
from the point $x=-L/2$ and arrives to $x=L/2$, one finds that,
\begin{equation}
\lim_{x\rightarrow L/2}\,\bar{q}\,x = q\,x +Q_{\alpha\nu}^0/2 + Q^{\Phi}_{\alpha\nu}
(q)/2 = q\,x + \delta_{\alpha\nu} (q) \, , \label{qr2}
\end{equation}
where
\begin{equation}
\delta_{\alpha\nu} (q) = Q_{\alpha\nu} (q)/2 \, . \label{danq}
\end{equation}
For this asymptote coordinate choice, $\delta_{\alpha\nu} (q)$ is the overall $\alpha\nu$
pseudofermion or hole phase shift whose value is given only to within addition of an arbitrary 
multiple of $\pi$. Moreover, from analysis of Eqs. (\ref{qcan1j}) and (\ref{Qcan1j}) it follows that
$\pi\,\Phi_{\alpha\nu,\,\alpha'\nu'}(q_j,q_{j'})$ is an elementary two-pseudofermion phase shift. 
This phase-shift definition corresponds to that used in standard quantum non-relativistic scattering 
theory \cite{Taylor}, such that the $S$ matrix $S_{\alpha\nu} (q_j)$ given in Eq. (\ref{San}) 
can be written as,
\begin{equation}
S_{\alpha\nu} (q_j) = e^{i2\delta_{\alpha\nu}(q_j)} \, , \hspace{0.5cm} j=1,2,...,
N^*_{\alpha\nu} \, . \label{San3}
\end{equation}
The factor $2$ appearing in the exponential argument of Eq. (\ref{San3}) corresponds to
the usual form of the $S$ matrix for that theory. In reference \cite{Impurity} it is
found that such a phase-shift definition is consistent with an exact theorem due to Fumi
\cite{Mahan}. Note that for the phase-shift definition (\ref{danq}),
$Q_{\alpha\nu}^0/2=0,\,\mp\pi/2$ corresponds to the scatter-less term $-l\pi/2$ of the
three-dimensional partial-wave problem of angular momentum $l$ \cite{Taylor,Mahan}.
Although the angular momentum vanishes in 1D, the scatter-less phase shift
(\ref{pic0an}) plays a similar role. Here we follow the definition of the standard
quantum non-relativistic scattering theory and choose the overall $\alpha\nu$
pseudofermion phase shift definition $Q_{\alpha\nu} (q)/2$ associated with Eq.
(\ref{qr2}).

In turn, if when moving around the lattice ring the $\alpha\nu$ pseudofermion (or hole)
departures from the point $x=0$ and arrives to $x=L$, one finds that,
\begin{equation}
\lim_{x\rightarrow L}\,\bar{q}\,x = q\,x +Q_{\alpha\nu}^0 + Q^{\Phi}_{\alpha\nu} (q) =
q\,x + Q_{\alpha\nu} (q) \, , \label{qr1}
\end{equation}
where $q$ refers to the initial ground state. For this asymptote coordinate choice,
$Q_{\alpha\nu} (q)$ is the overall $\alpha\nu$ pseudofermion (or hole) phase shift 
whose value is given only to within addition of an arbitrary multiple of $2\pi$ and 
$2\pi\,\Phi_{\alpha\nu,\,\alpha'\nu'}(q_j,q_{j'})$ is an elementary two-pseudofermion
phase shift.

The studies of Ref. \cite{relation} reveal that the overall pseudofermion phase-shift choice 
$Q_{\alpha\nu} (q)=Q_{\alpha\nu}^0 + Q^{\Phi}_{\alpha\nu} (q)$ associated
with the asymptote condition (\ref{qr1}) corresponds to a generalization of the conventional 
phase-shift definition previously used in the BA literature for
the particular case of the $n=1$ and $m=0$ initial ground state. (All
the discussions and analysis presented below in this paper also apply to the phase-shift definition
$Q_{\alpha\nu} (q) =Q_{\alpha\nu}^0 + Q^{\Phi}_{\alpha\nu} (q)$, provided that the
$\alpha\nu$ phase shifts $\delta_{\alpha\nu} (q) = Q_{\alpha\nu} (q)/2$ are multiplied by
two.)

%%%%%%%%%%%%%%%%%%%%%%%%%%%%%%%%%%%%%%%%%%%%%%%%%%%%%%%%%%%%%%%%
\subsection{THE TWO-PSEUDOFERMION PHASE SHIFTS: BARE-MOMENTUM
DEPENDENCE AND LEVINSON'S THEOREM}

The scattering $\alpha\nu$ pseudofermion or hole phase-shift $Q^{\Phi}_{\alpha\nu} (q)/2$
given in Eq. (\ref{qcan1j}) results from the set of elementary two-pseudofermion
scattering events associated with the transition from the ``in" state to the ``out"
state (excited energy eigenstate). In contrast, the transition from the
ground state to the ``in" state has no scattering character and leads to the
scatter-less phase-shift $Q_{\alpha\nu}^0/2=0,\,\mp\pi/2$ of Eq. (\ref{pic0an}).

The bare-momentum two-pseudofermion phase shifts $\Phi_{\alpha\nu,\,\alpha'\nu'}(q,q')$ 
in units of $\pi$ are related to the rapidity two-pseudofermion phase shifts 
$\bar{\Phi }_{\alpha\nu,\,\alpha'\nu'} (r ,\,r')$ by Eq. (\ref{Phi-barPhi}). The latter phase 
shifts are defined by the integral equations (A1)-(A13) of Ref. \cite{IIIb}.
In Appendix B we provide a set of simplified equations which define the rapidity
elementary two-pseudofermion phase shifts $\bar{\Phi }_{\alpha\nu,\,\alpha'\nu'} (r,\,r')$ 
in the limit $m\rightarrow 0$. Furthermore, in that Appendix we also provide
closed form expressions valid in the specific limit $m\rightarrow 0$ and 
$n\rightarrow 1$. 

In figures 1-6 we plot the two-pseudofermion phase shifts which 
contribute most to these properties as a function of the bare-momenta 
$q$ and $q'$ for electronic density $n=0.59$, spin density $m\rightarrow 0$, 
and $U/t\rightarrow 0$ , $U/t= 0.3,\,4.9$,\,$100$. (The electronic density value
$n=0.59$ and the value $U/t=4.9$ are those used in Ref. \cite{spectral} for the
description of the TCNQ photoemission dispersions observed in the quasi-1D organic
compound TTF-TCNQ.) Analytical expressions valid for $U/t\rightarrow 0$ as 
$m\rightarrow 0$ of the same two-pseudofermion phase shifts are provided in 
Appendix C.

Note that the function $\Phi_{c0,\,c1}(q,q')$ plotted in Fig. 5 can have values in the
domain $\Phi_{c0,\,c1}(q,q')\in (-1,1)$. Thus, within the standard quantum
non-relativistic scattering theory phase shift definition given in Eq. (\ref{qr2}), the
corresponding phase shift $\pi\,\Phi_{c0,\,c1}(q,q')$ has values in the domain
$\pi\,\Phi_{c0,\,c1}(q,q')\in (-\pi,\pi)$. Note that the width of this domain is $2\pi$,
whereas the definition (\ref{qr1}) would lead to a domain width of $4\pi$.

\begin{figure}
\subfigure{\includegraphics[width=7cm,height=7cm]{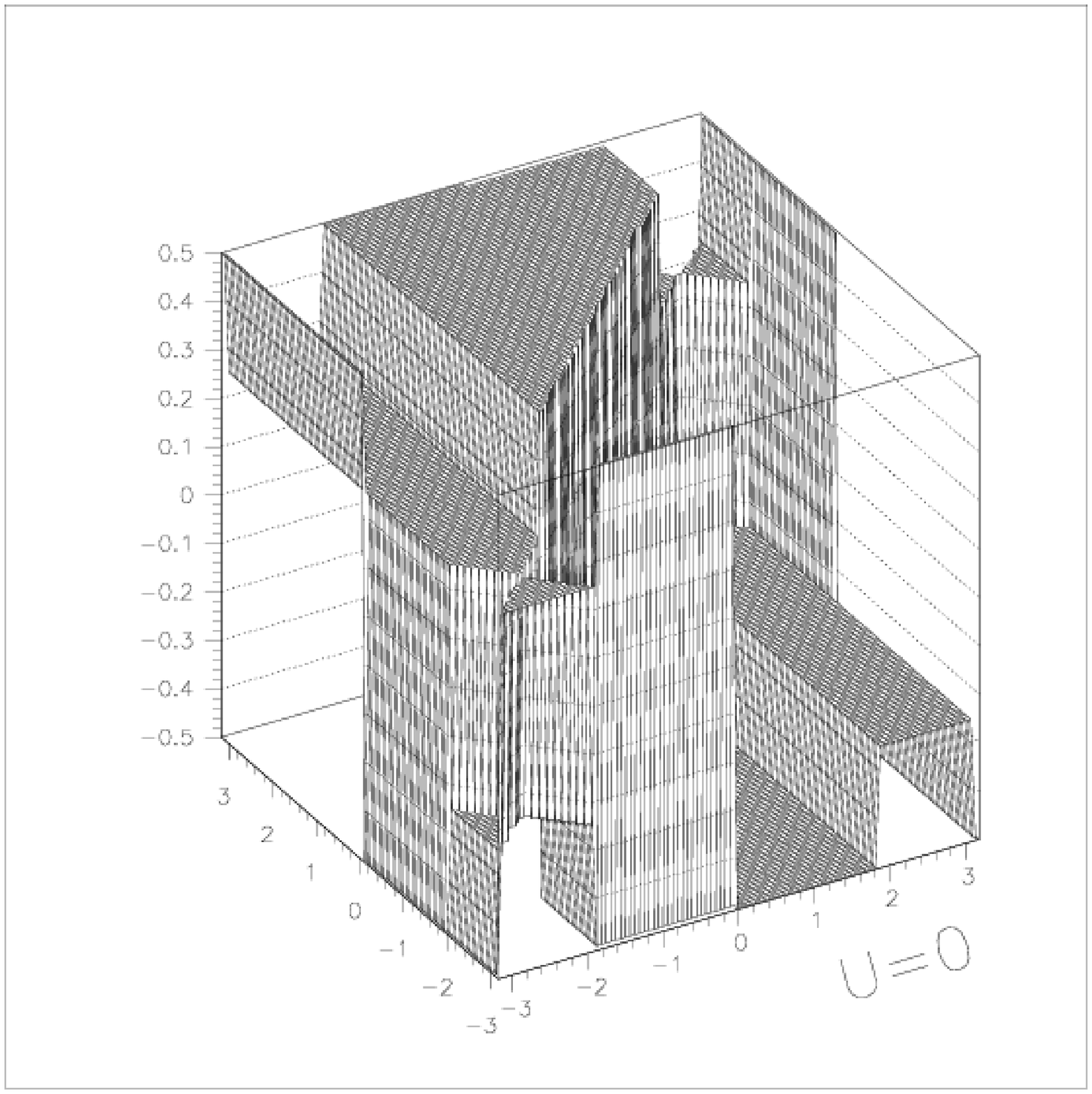}}
\subfigure{\includegraphics[width=7cm,height=7cm]{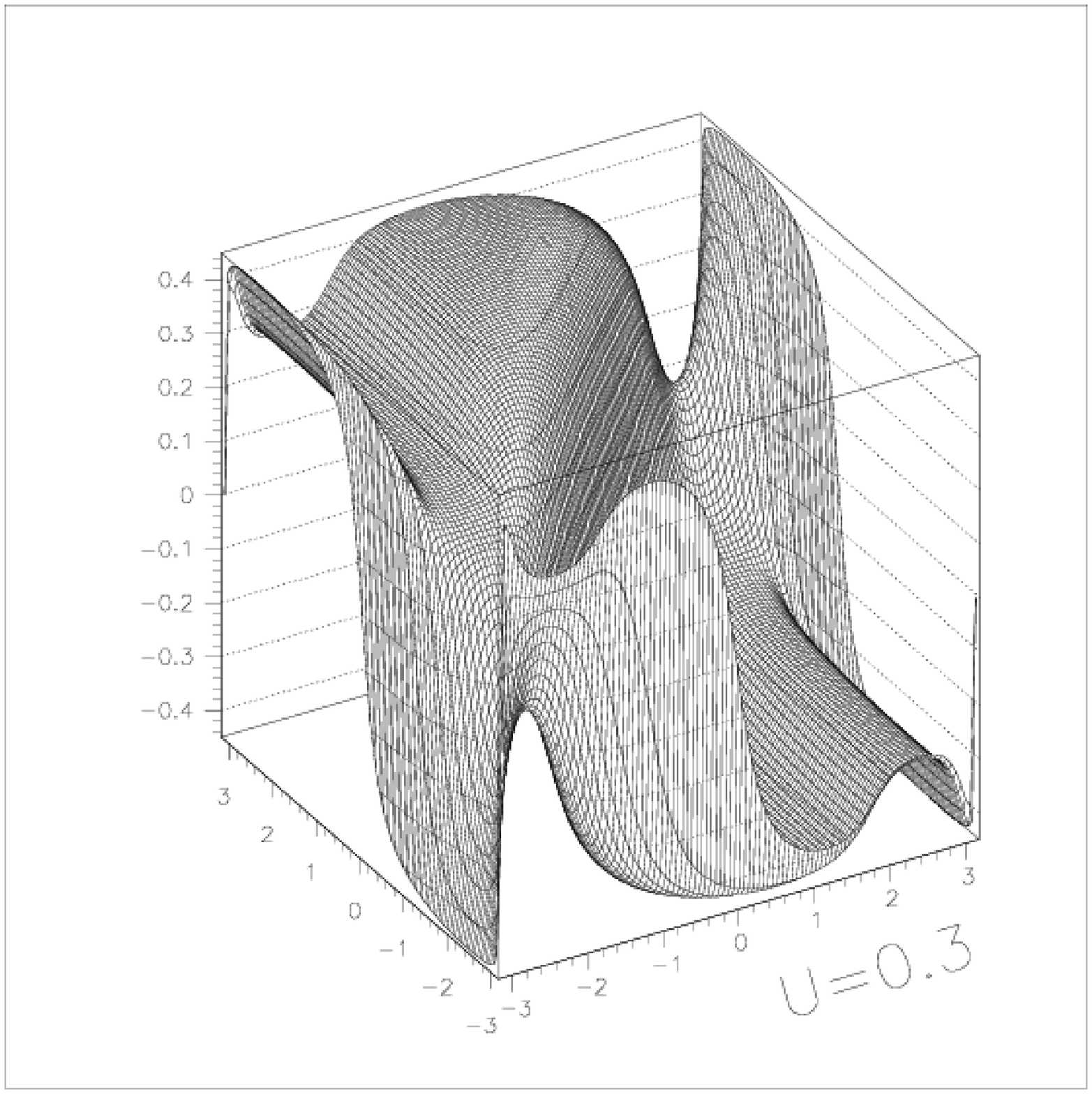}}
\subfigure{\includegraphics[width=7cm,height=7cm]{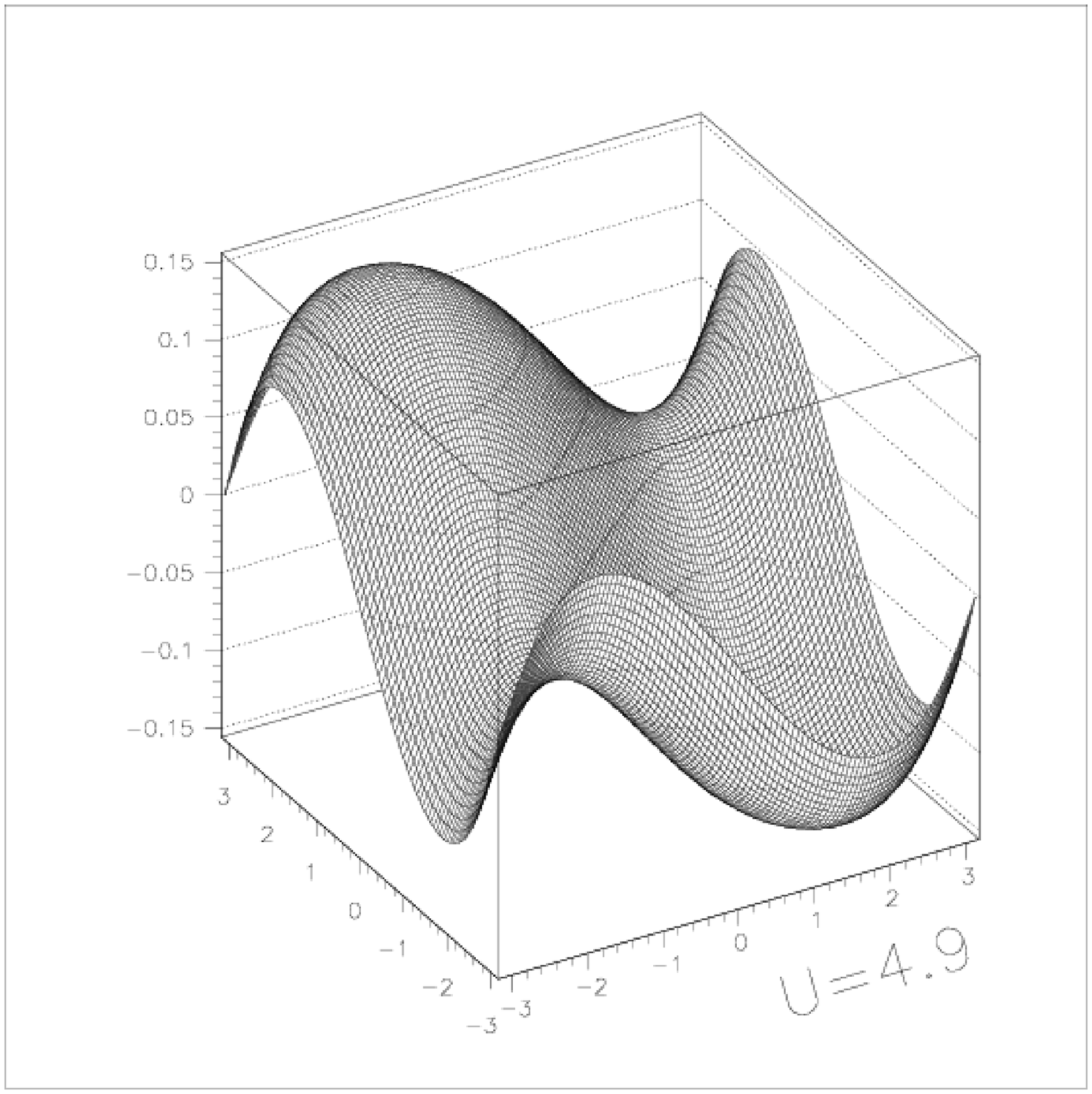}}
\subfigure{\includegraphics[width=7cm,height=7cm]{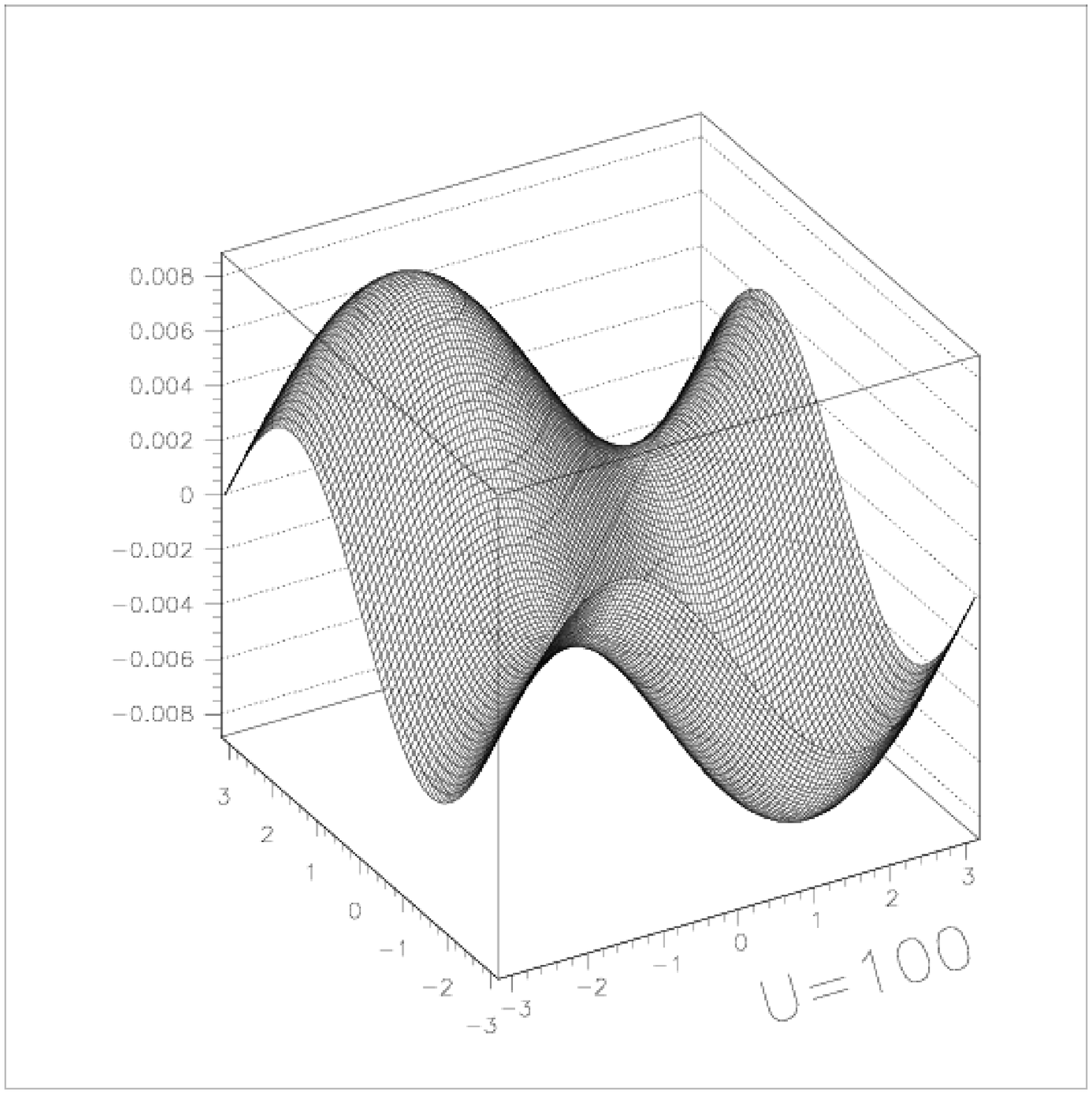}} \caption{The elementary
two-pseudofermion phase shift $\Phi_{c0,\,c0}(q,\,q')$ in units of $\pi$ as a function of
$q$ and $q'$ for $n=0.59$, $m=0$, and (a) $U/t\rightarrow 0$, (b) $U/t=0.3$, (c) $U/t=4.9$, and (d)
$U/t= 100$. The bare-momentum values $q$ and $q'$ correspond to the right and left 
axis of the figures, respectively.}
\end{figure}

\begin{figure}
\subfigure{\includegraphics[width=7cm,height=7cm]{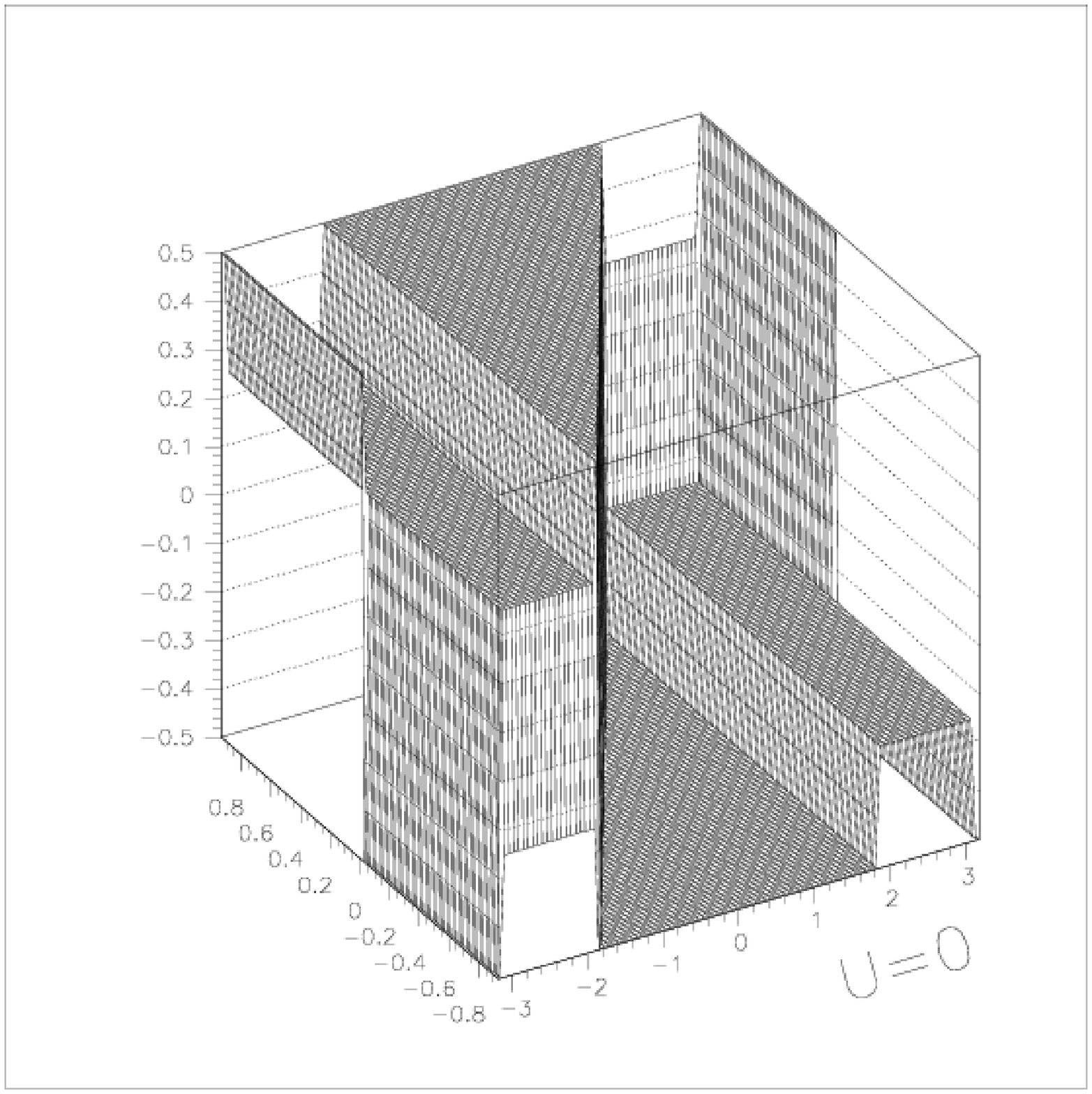}}
\subfigure{\includegraphics[width=7cm,height=7cm]{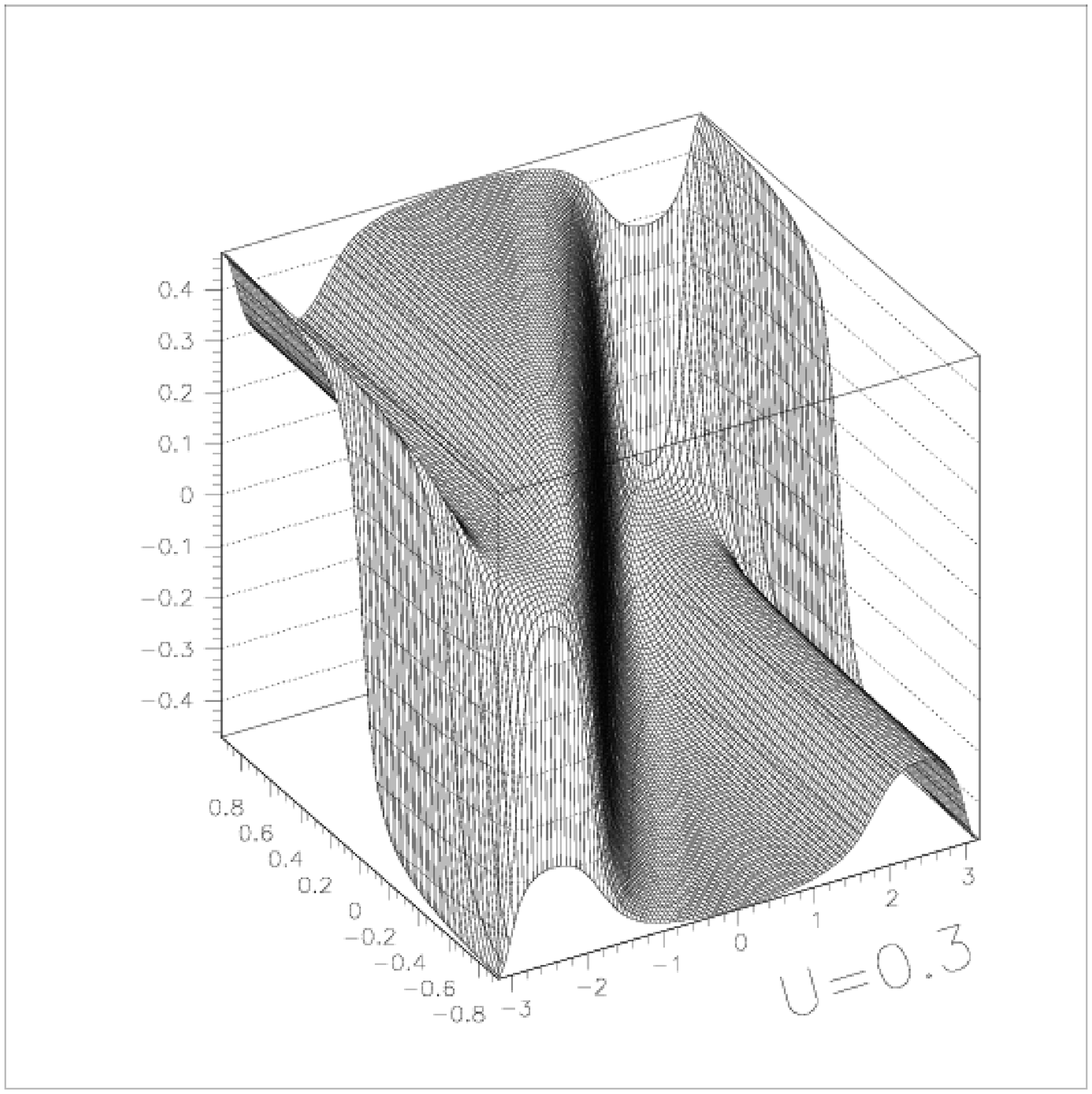}}
\subfigure{\includegraphics[width=7cm,height=7cm]{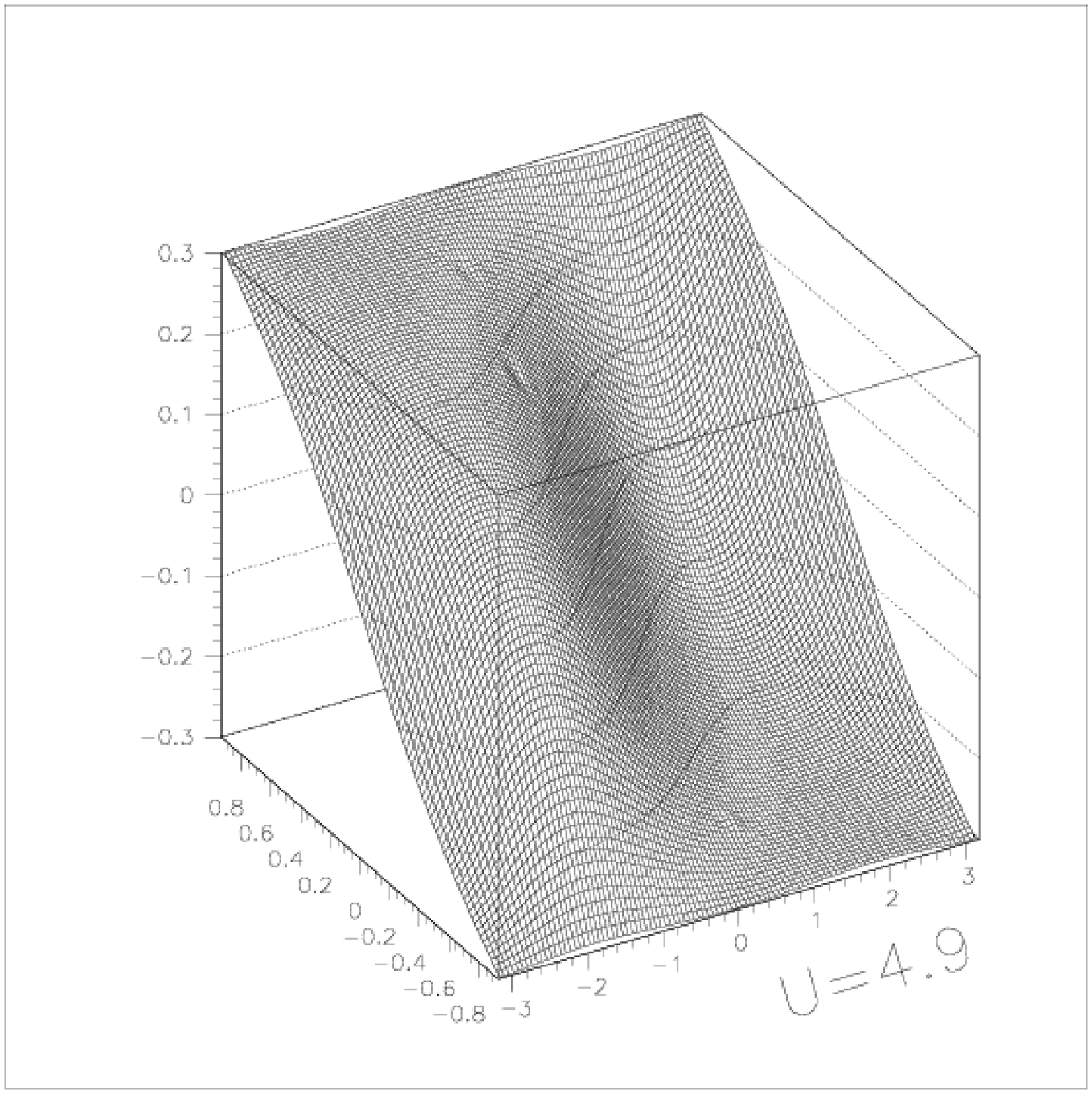}}
\subfigure{\includegraphics[width=7cm,height=7cm]{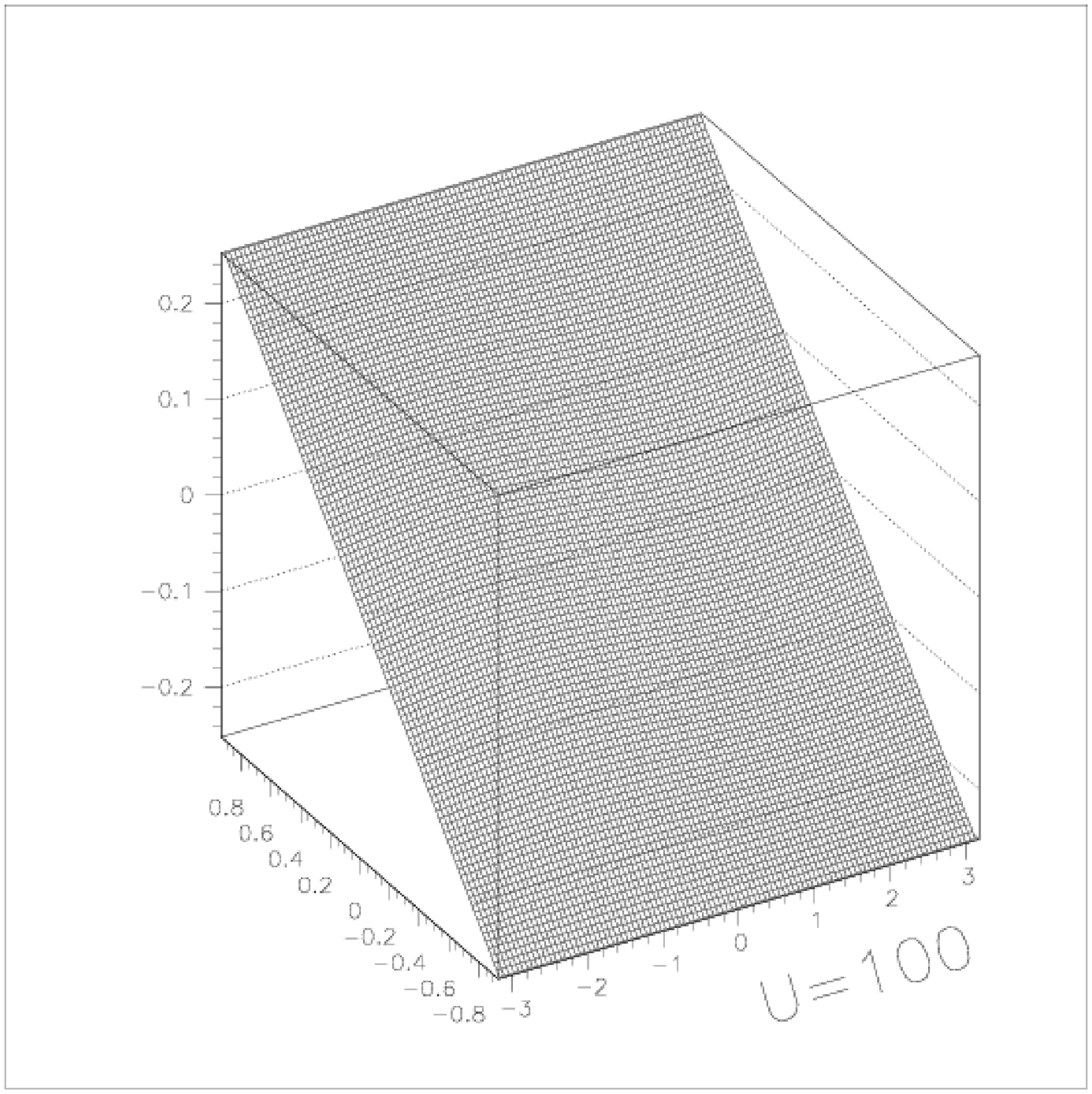}} \caption{The elementary
two-pseudofermion phase shift $\Phi_{c0,\,s1}(q,\,q')$ in units of $\pi$ as a function of
$q$ and $q'$ for $n=0.59$, $m=0$, and (a) $U/t\rightarrow 0$, (b) $U/t=0.3$, (c) $U/t=4.9$, and (d)
$U/t= 100$. The bare-momentum values $q$ and $q'$ correspond to the right and left 
axis of the figures, respectively.}
\end{figure}

\begin{figure}
\subfigure{\includegraphics[width=7cm,height=7cm]{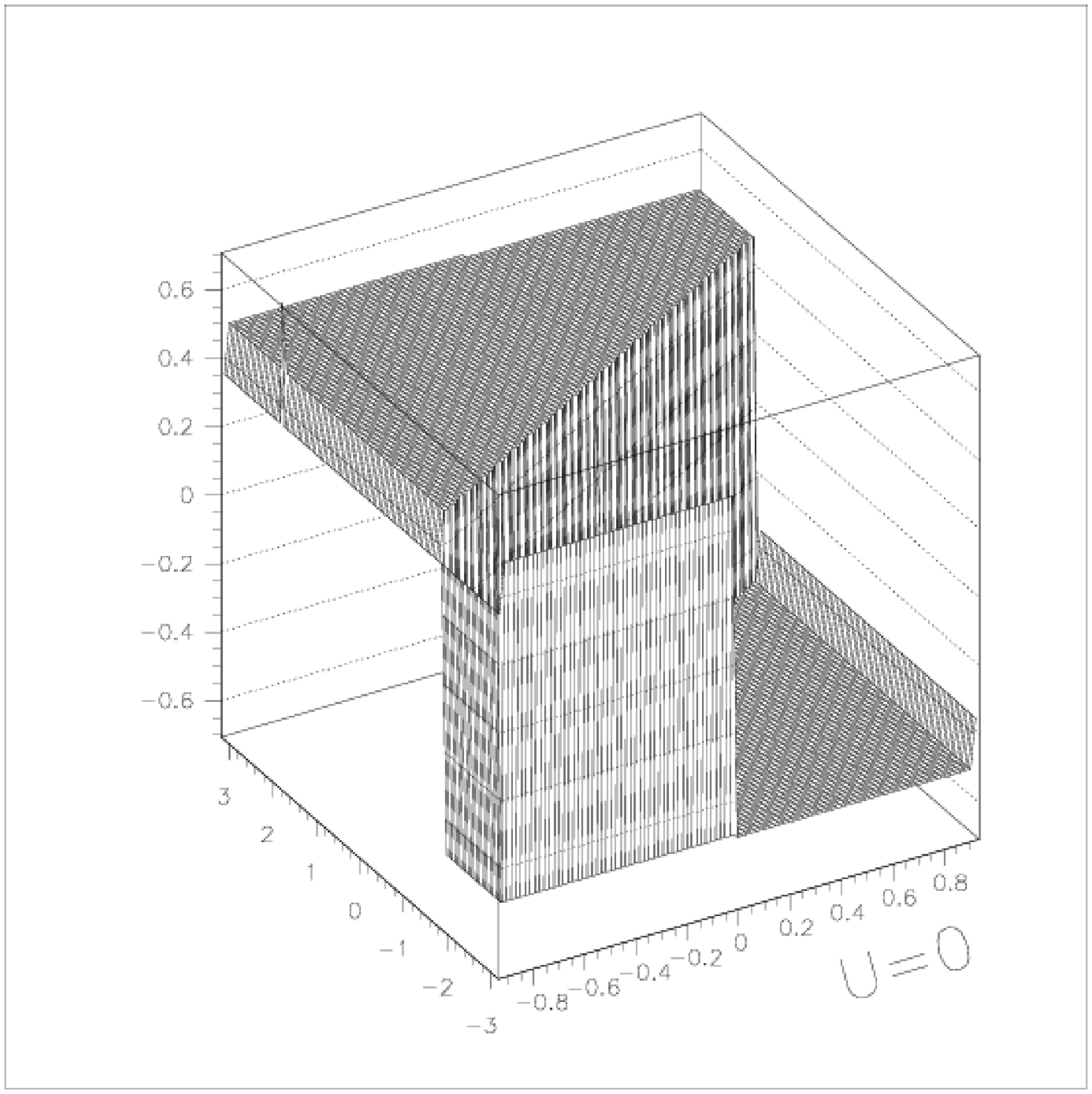}}
\subfigure{\includegraphics[width=7cm,height=7cm]{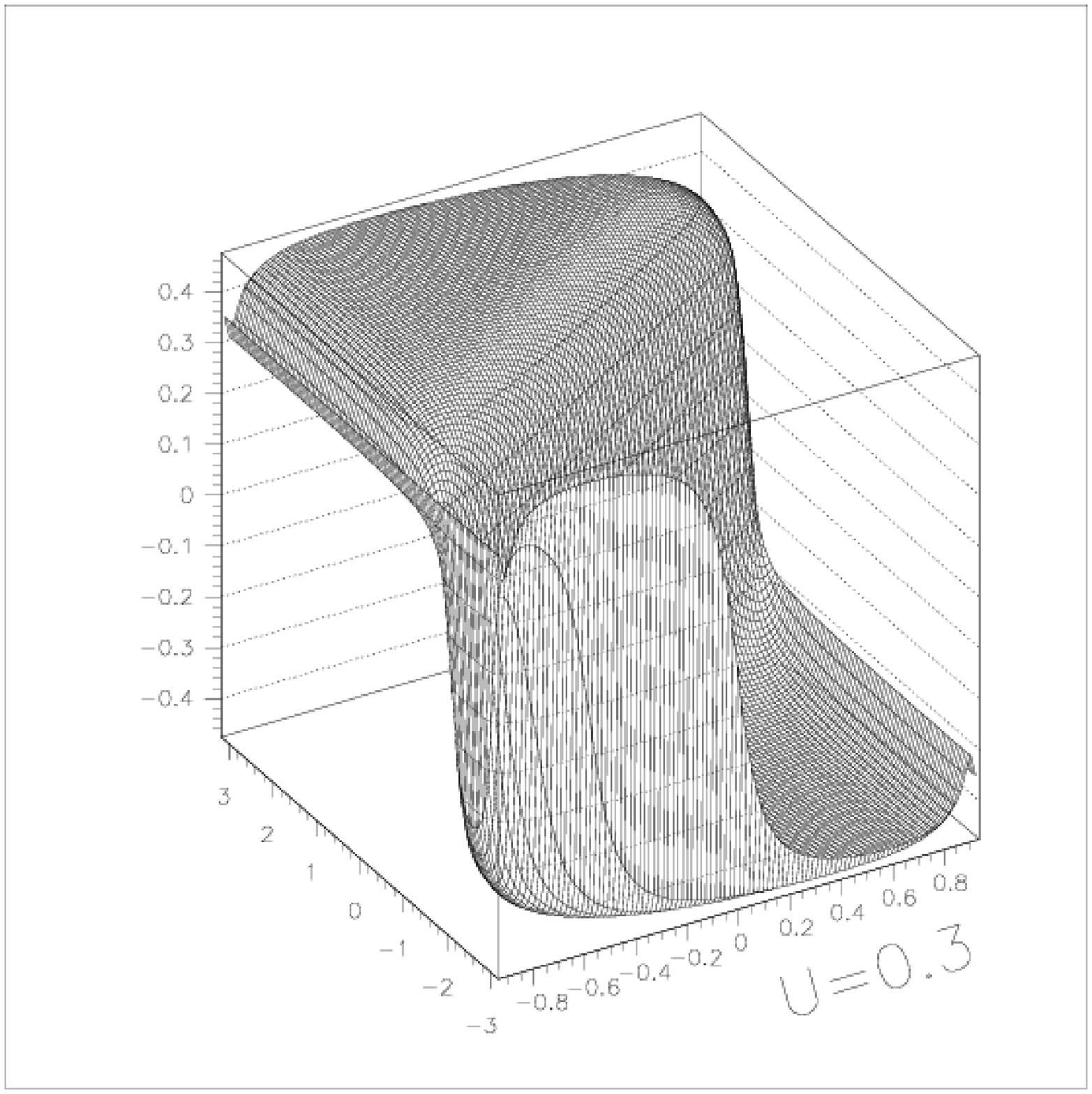}}
\subfigure{\includegraphics[width=7cm,height=7cm]{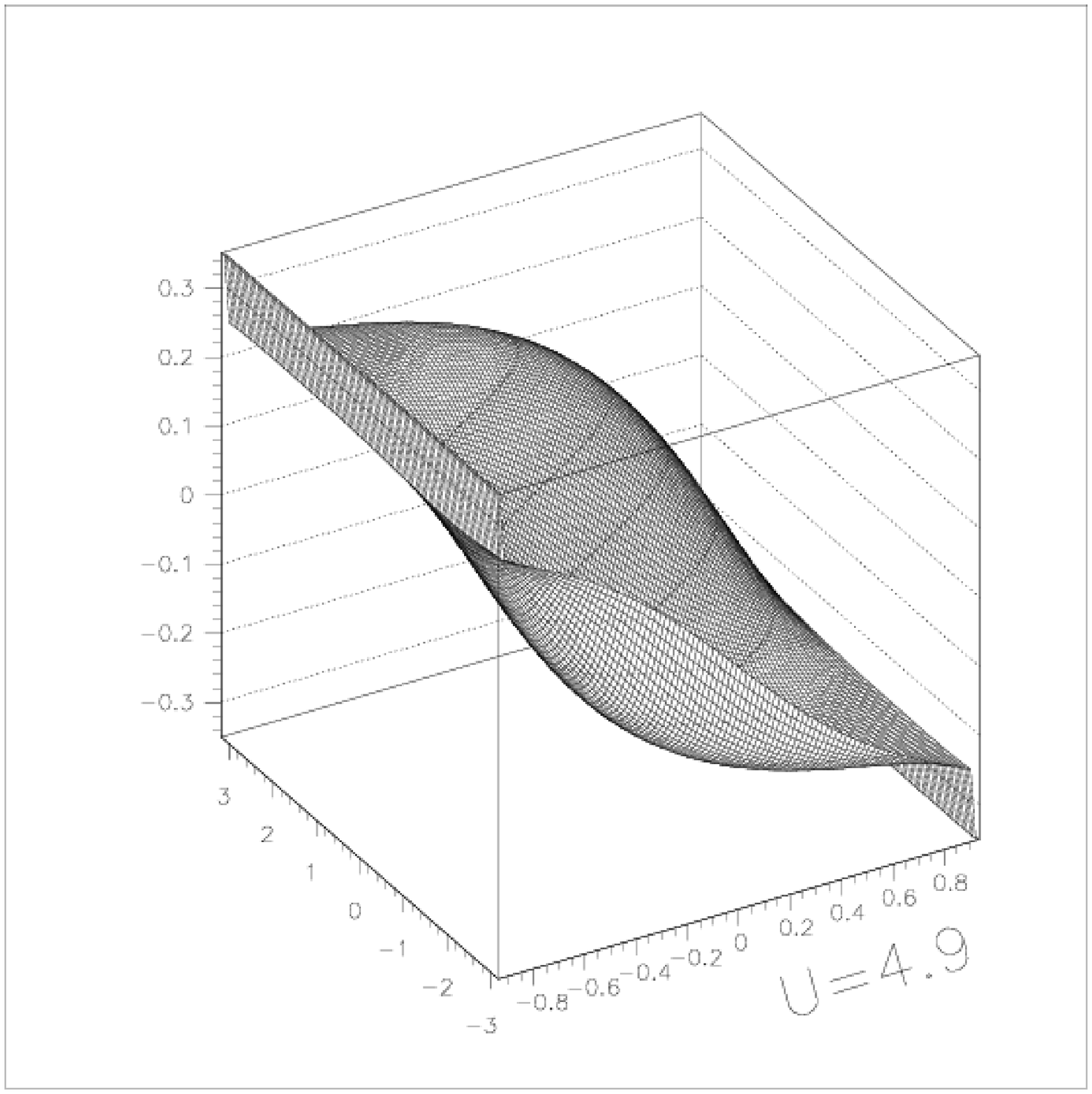}}
\subfigure{\includegraphics[width=7cm,height=7cm]{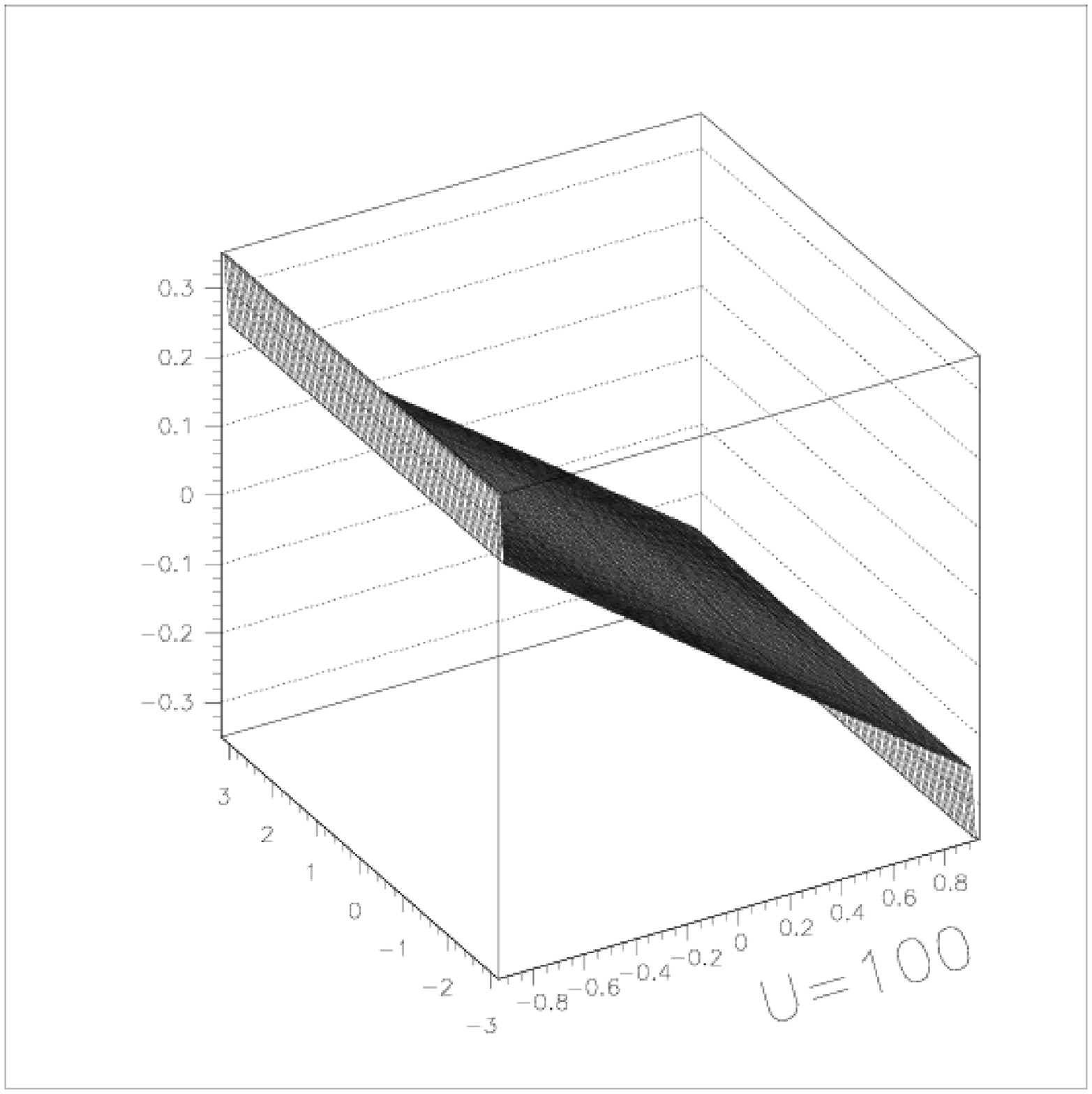}} \caption{The elementary
two-pseudofermion phase shift $\Phi_{s1,\,c0}(q,\,q')$ in units of $\pi$ as a function of
$q$ and $q'$ for $n=0.59$, $m=0$, and (a) $U/t\rightarrow 0$, (b) $U/t=0.3$, (c) $U/t=4.9$, and (d)
$U/t= 100$. The bare-momentum values $q$ and $q'$ correspond to the right and left 
axis of the figures, respectively.}
\end{figure}

\begin{figure}
\subfigure{\includegraphics[width=7cm,height=7cm]{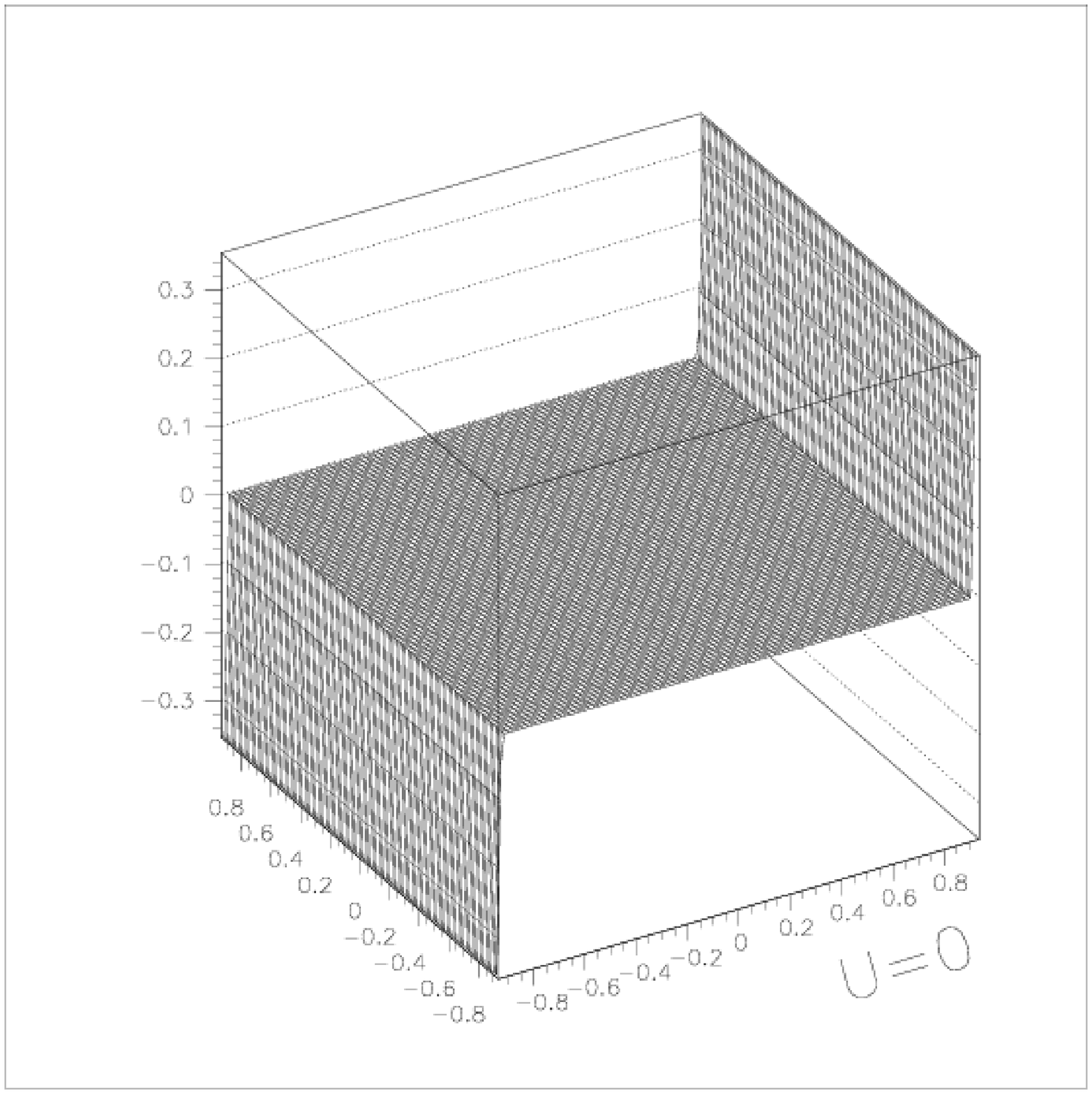}}
\subfigure{\includegraphics[width=7cm,height=7cm]{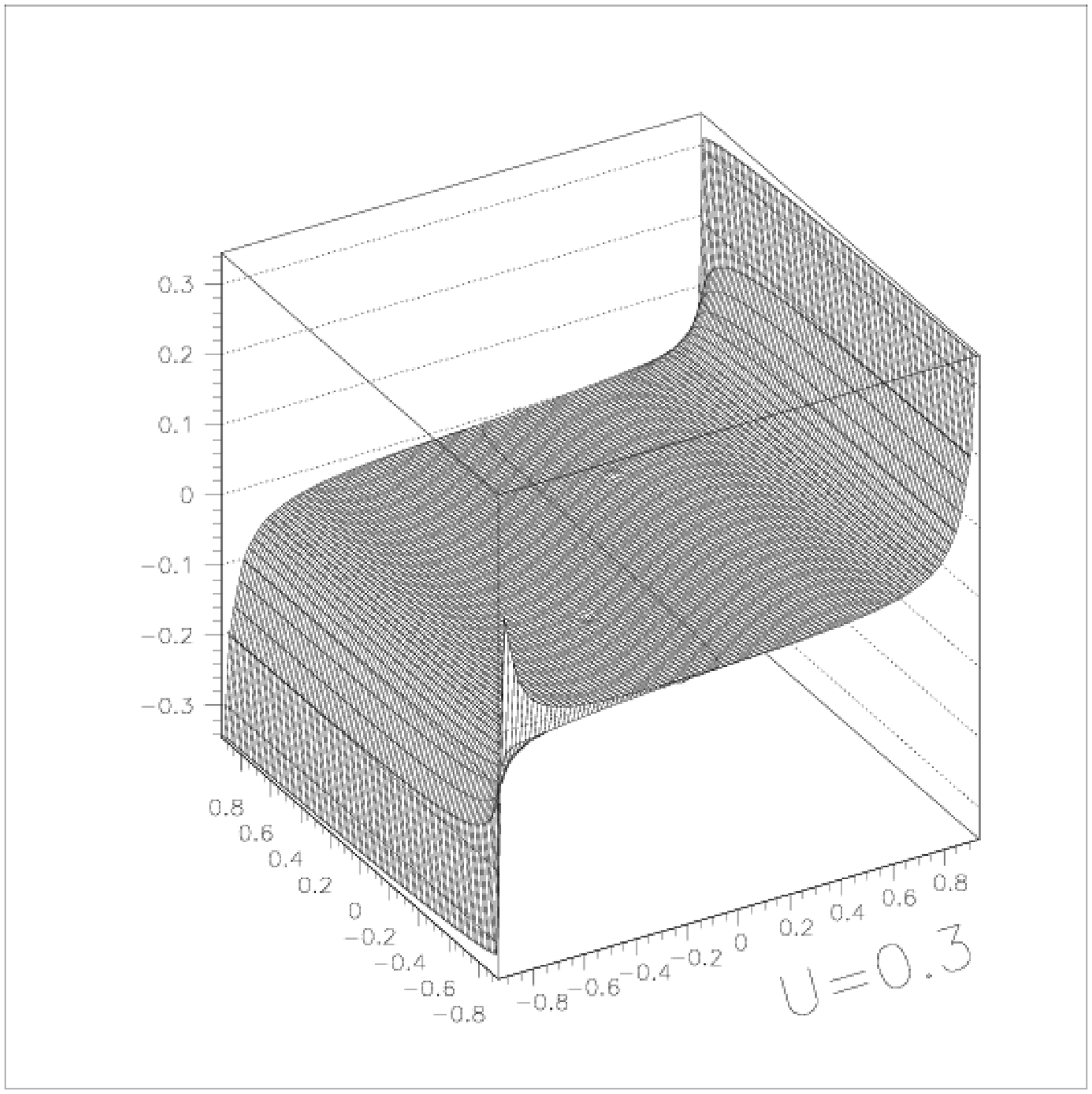}}
\subfigure{\includegraphics[width=7cm,height=7cm]{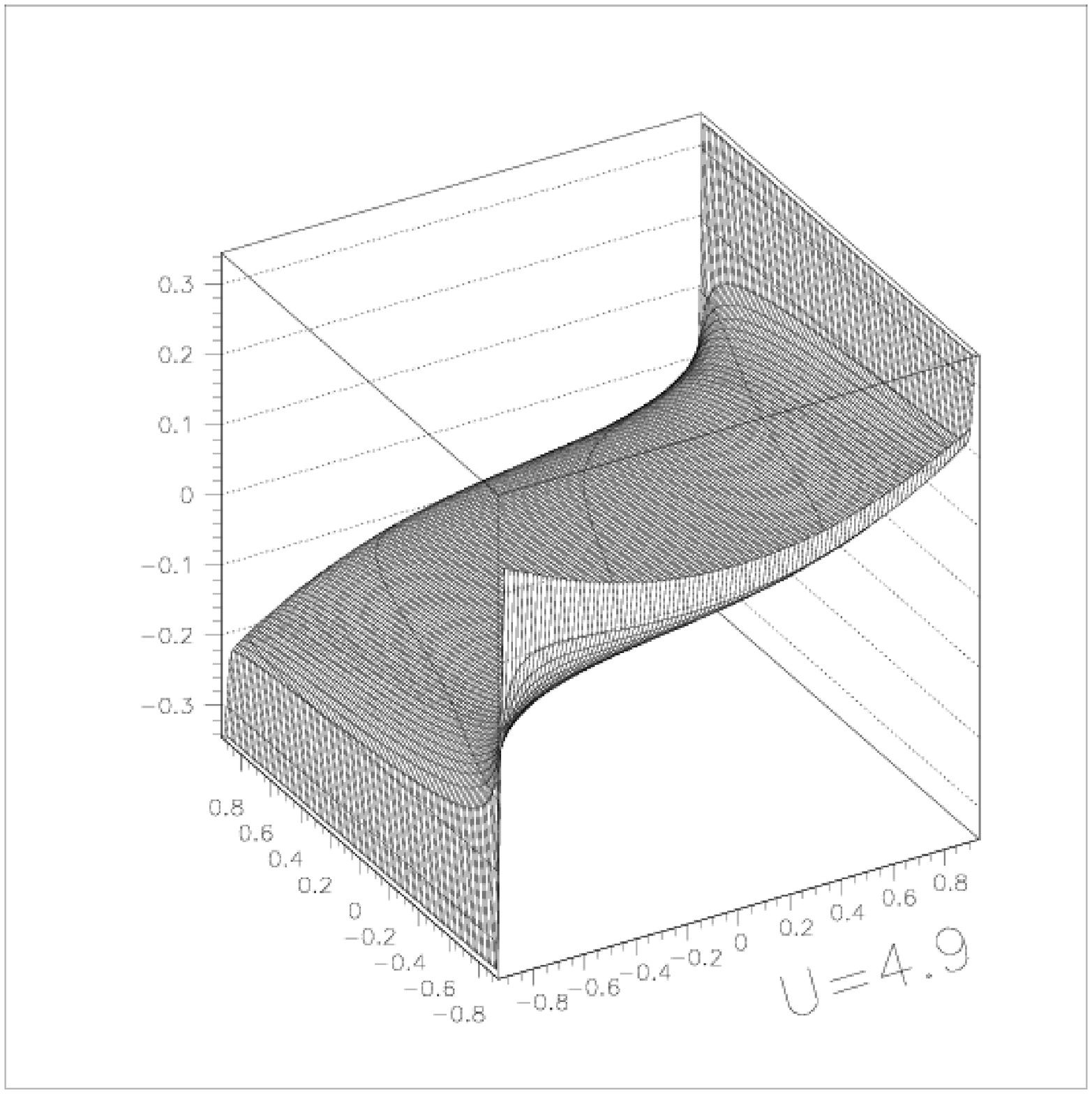}}
\subfigure{\includegraphics[width=7cm,height=7cm]{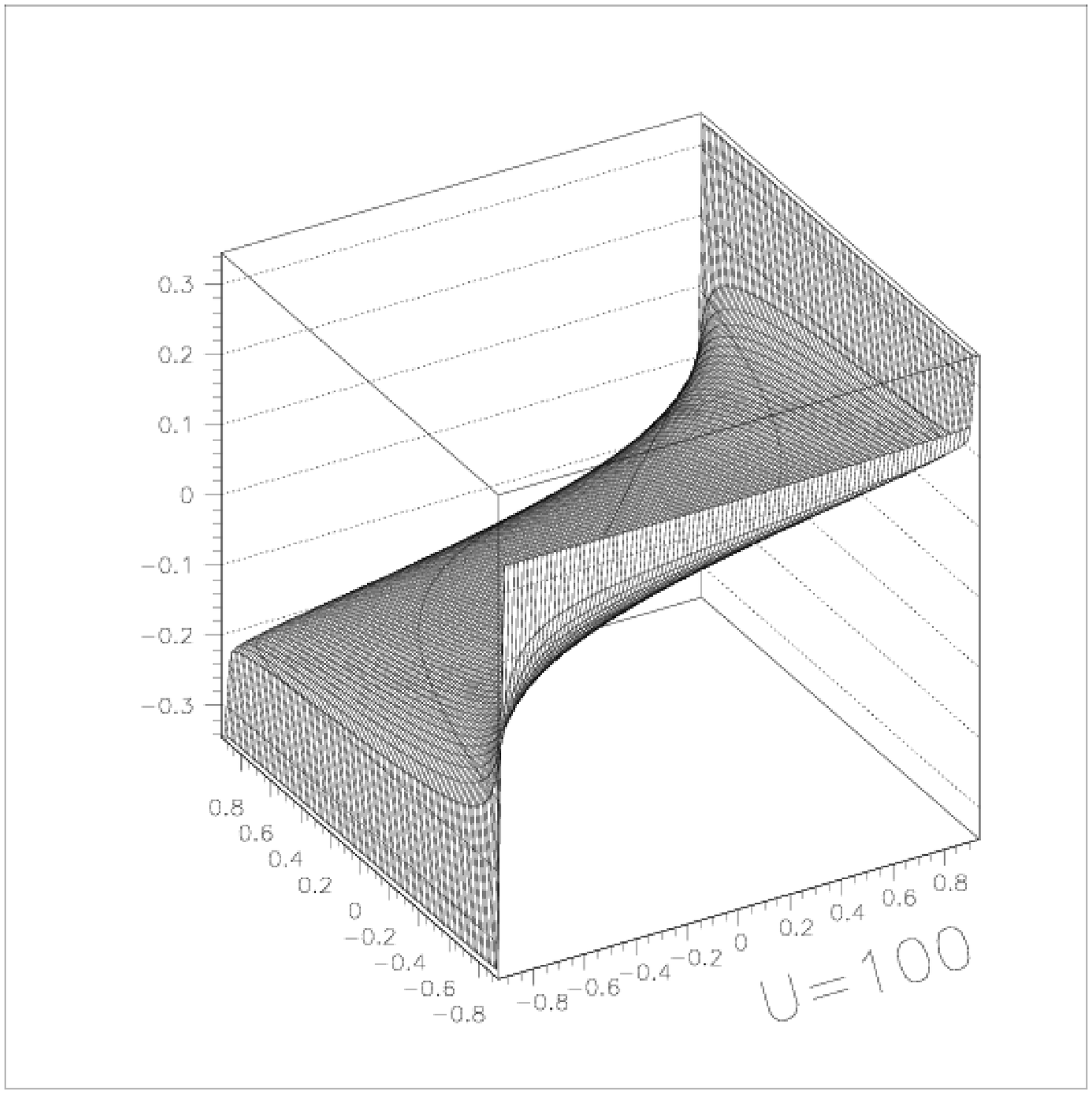}} \caption{The elementary
two-pseudofermion phase shift $\Phi_{s1,\,s1}(q,\,q')$ in units of $\pi$ as a function of
$q$ and $q'$ for $n=0.59$, $m=0$, and (a) $U/t\rightarrow 0$, (b) $U/t=0.3$, (c) $U/t=4.9$, and (d)
$U/t= 100$. The bare-momentum values $q$ and $q'$ correspond to the right and left 
axis of the figures, respectively.}
\end{figure}

\begin{figure}
\subfigure{\includegraphics[width=7cm,height=7cm]{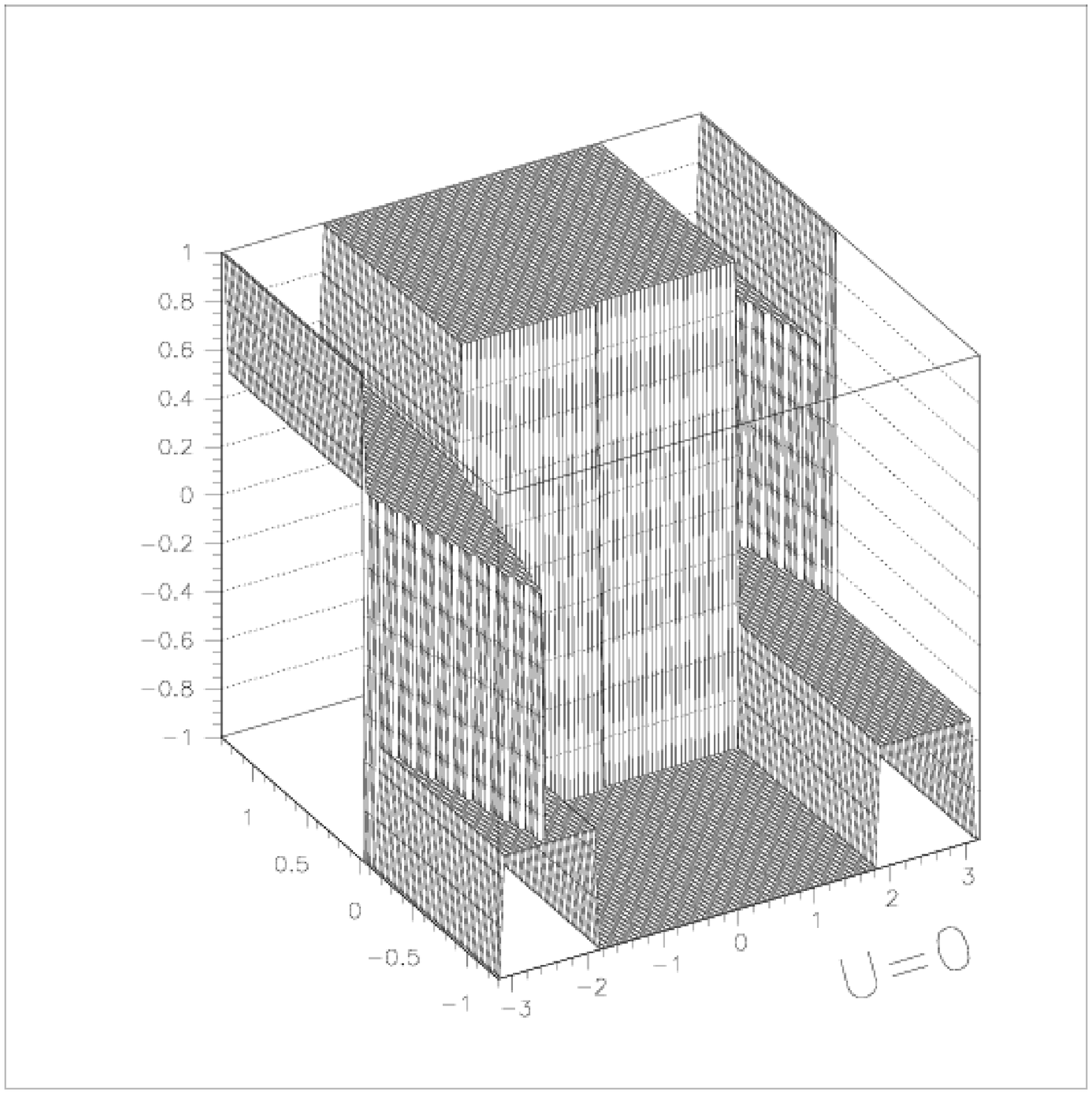}}
\subfigure{\includegraphics[width=7cm,height=7cm]{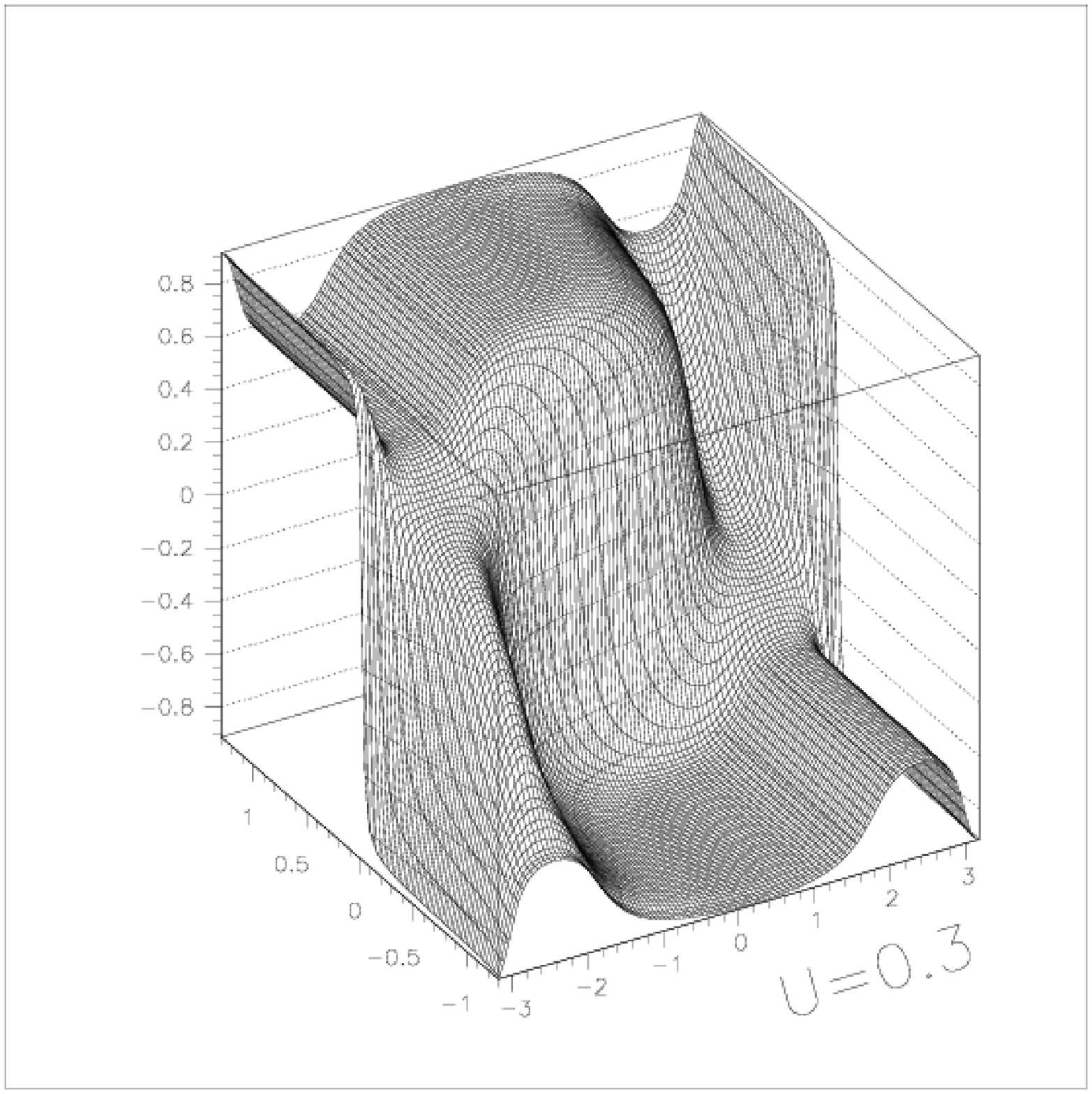}}
\subfigure{\includegraphics[width=7cm,height=7cm]{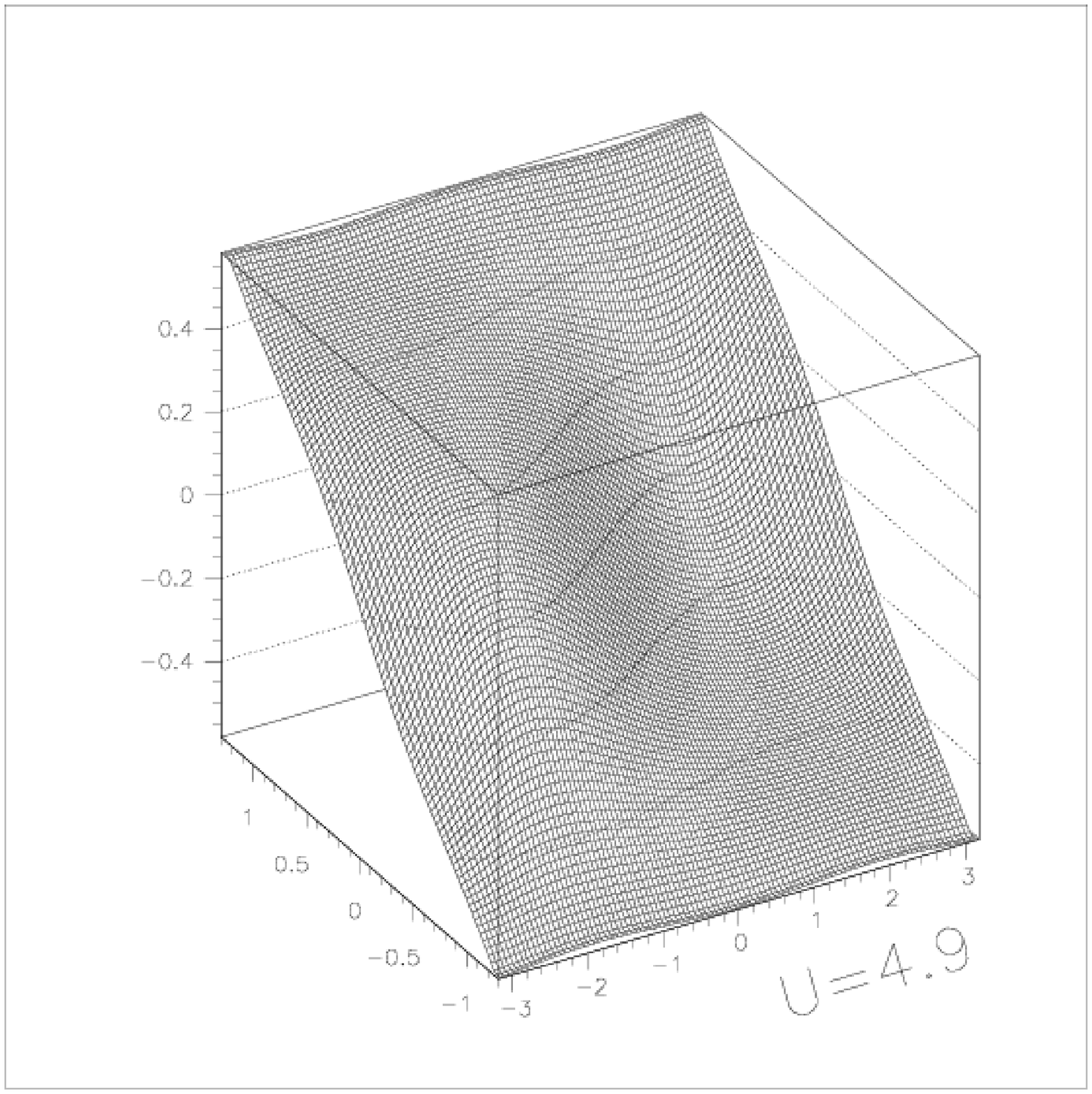}}
\subfigure{\includegraphics[width=7cm,height=7cm]{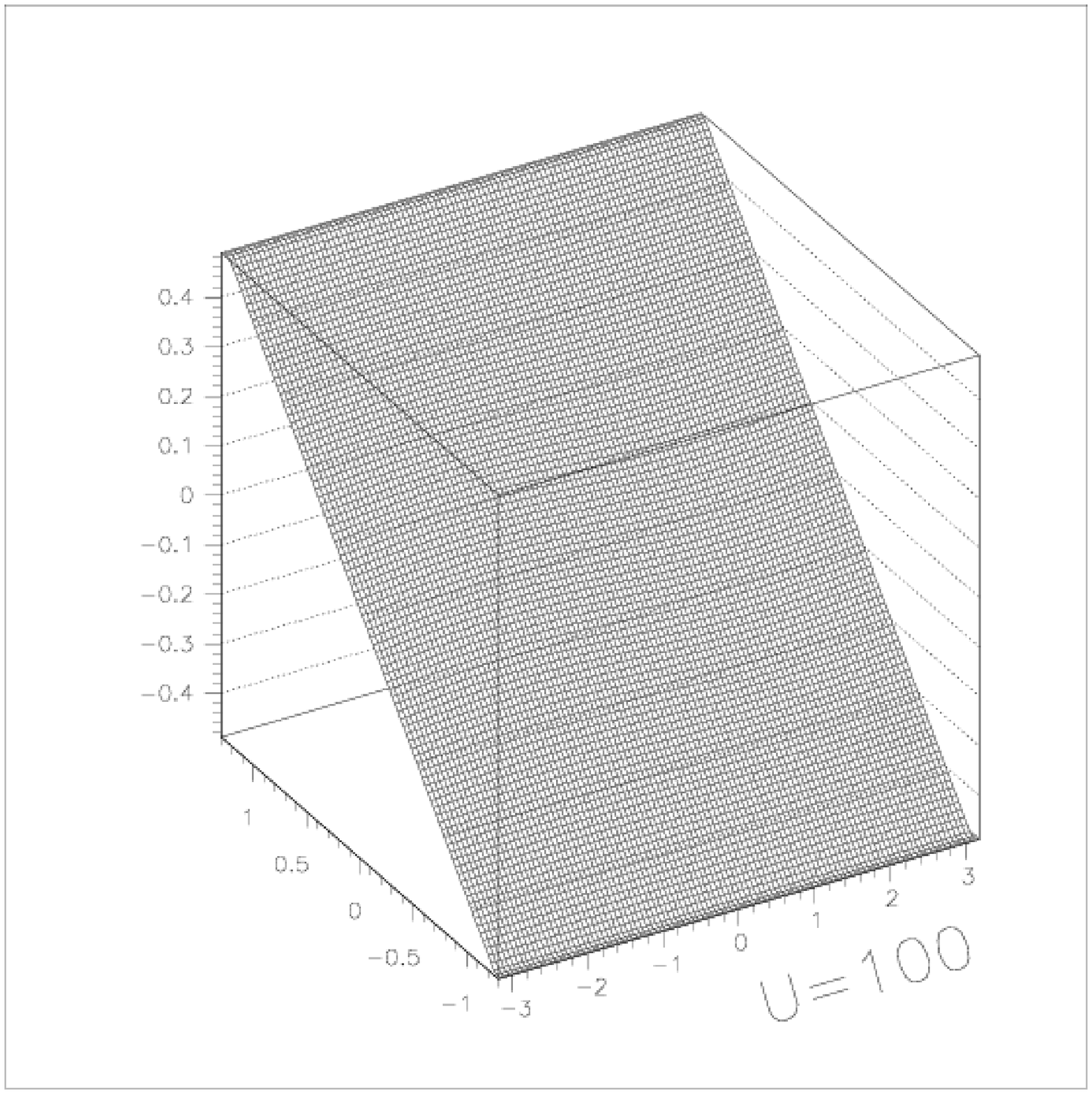}} \caption{The elementary
two-pseudofermion phase shift $\Phi_{c0,\,c1}(q,\,q')$ in units of $\pi$ as a function of
$q$ and $q'$ for $n=0.59$, $m=0$, and (a) $U/t\rightarrow 0$, (b) $U/t=0.3$, (c) $U/t=4.9$, and (d)
$U/t= 100$. The bare-momentum values $q$ and $q'$ correspond to the right and left 
axis of the figures, respectively.}
\end{figure}

\begin{figure}
\subfigure{\includegraphics[width=7cm,height=7cm]{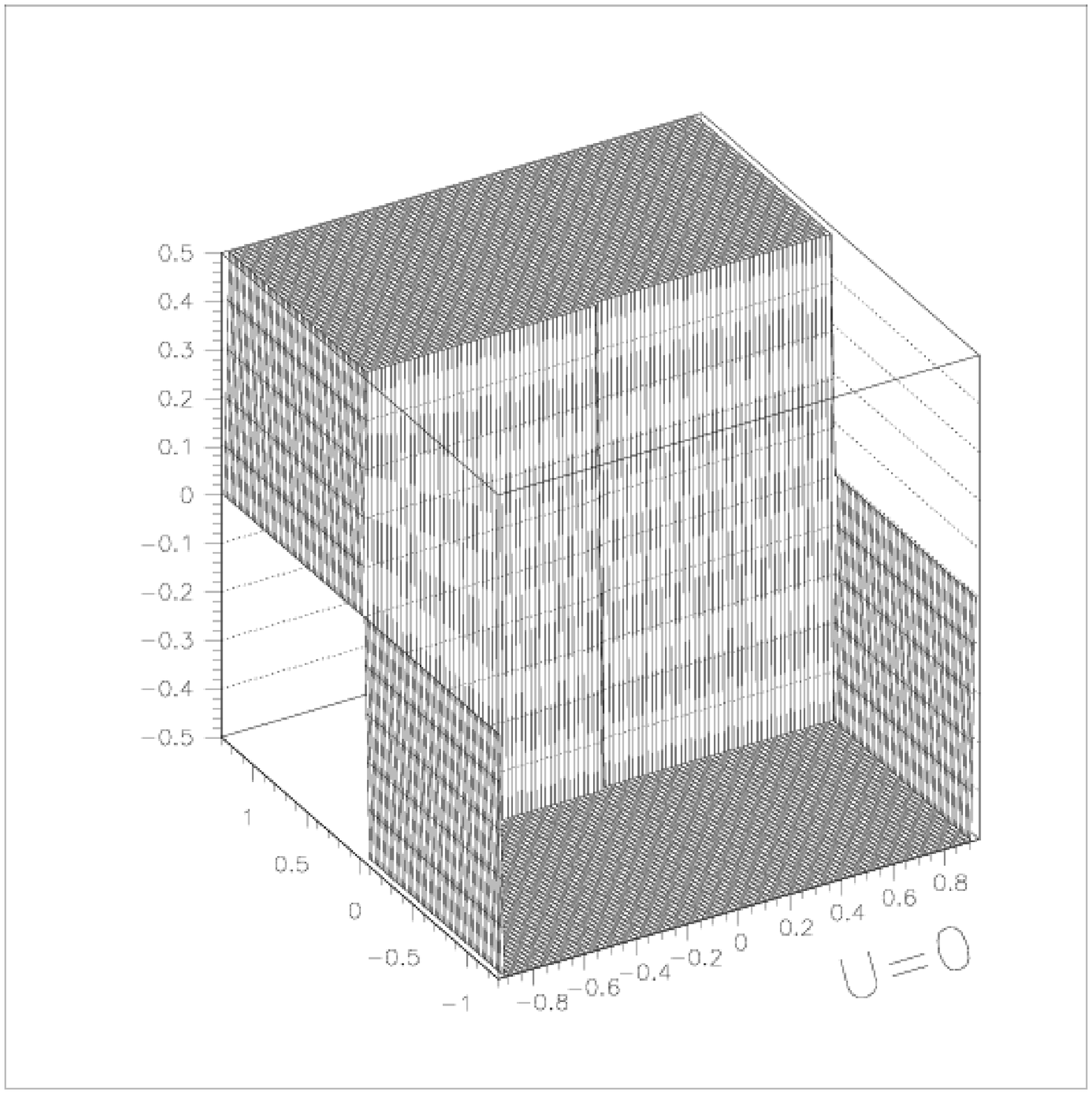}}
\subfigure{\includegraphics[width=7cm,height=7cm]{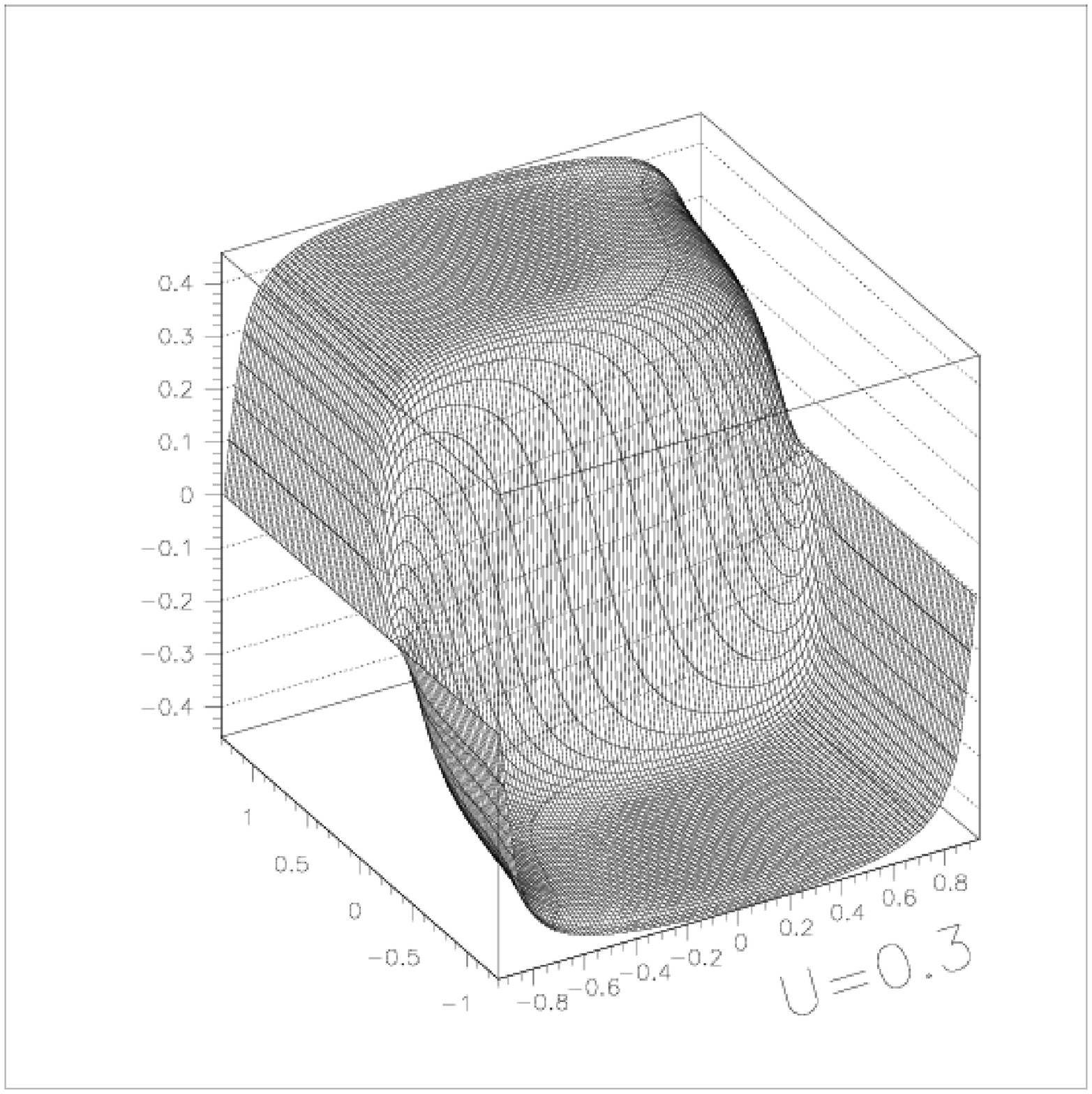}}
\subfigure{\includegraphics[width=7cm,height=7cm]{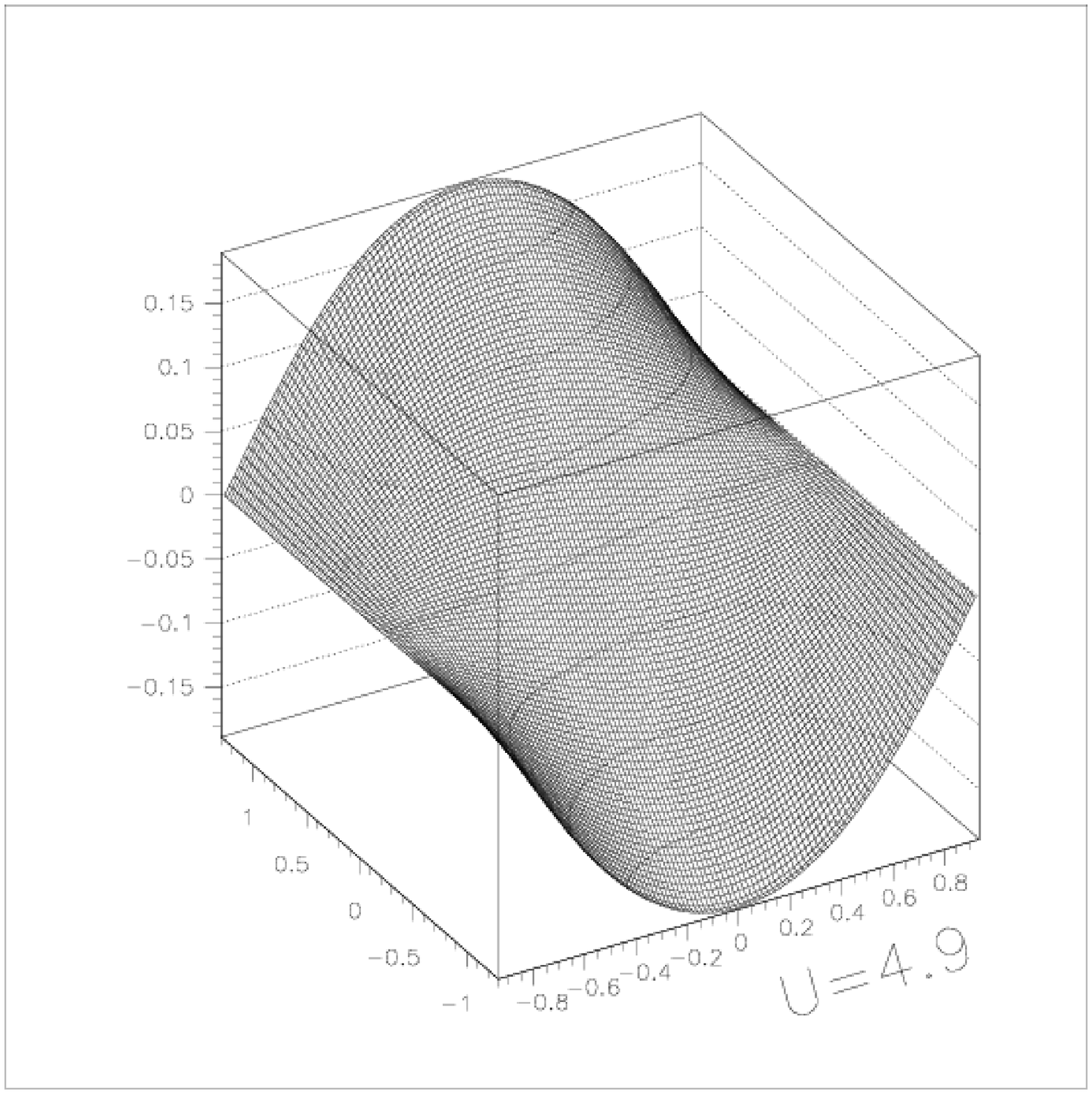}}
\subfigure{\includegraphics[width=7cm,height=7cm]{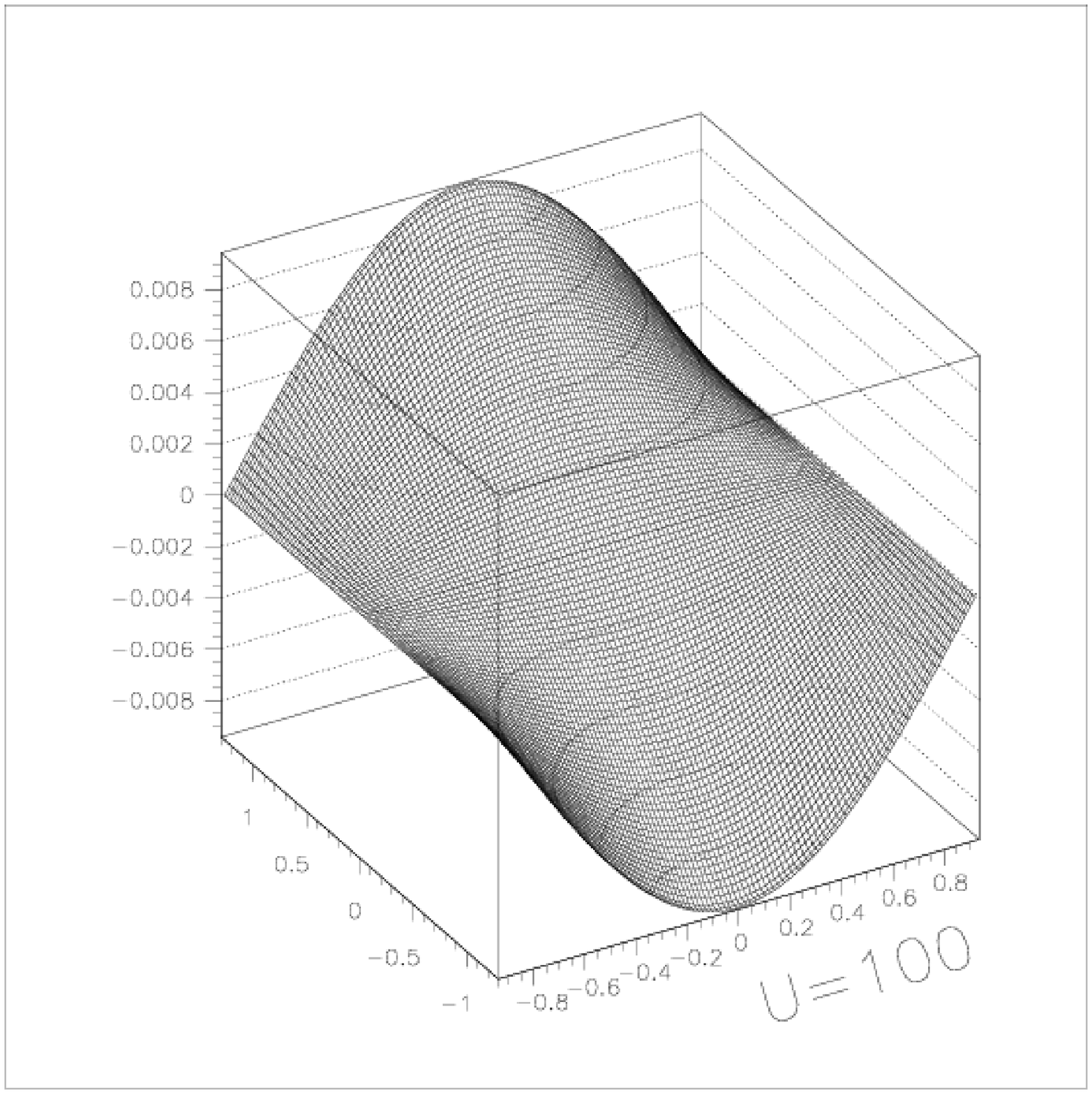}} \caption{The elementary
two-pseudofermion phase shift $\Phi_{s1,\,c1}(q,\,q')$ in units of $\pi$ as a function of
$q$ and $q'$ for $n=0.59$, $m=0$, and (a) $U/t\rightarrow 0$, (b) $U/t=0.3$, (c) $U/t=4.9$, and (d)
$U/t= 100$. The bare-momentum values $q$ and $q'$ correspond to the right and left 
axis of the figures, respectively.}
\end{figure}

For $m\rightarrow 0$ the two-pseudofermion phase shifts have a similar 
qualitative behavior for a large domain of $U/t$ values corresponding 
to $U/t>4$. Thus, in order to obtain further information about the $q$ 
and $q'$ dependence of such phase shifts, it is useful to derive an 
analytical expansion in $t/U$ for $m\rightarrow 0$, which provides the 
exact behavior for $U/t\gg 1$. By use of manipulations similar to those 
reported in Appendix C, we find that for 
$m\rightarrow 0$ and $U/t\gg 1$ the elementary two-pseudofermion 
phase shifts plotted in Figs. 1-6 can be written as follows,
\begin{equation}
\pi\,\Phi_{c0,\,c0}(q,\,q') = -{\pi\,(\xi_0 -1/\xi_0 )\over 2}\Bigl({\sin (q) -\sin (q')\over 2\sin
(\pi n)}\Bigr) + \pi\,\Bigl[{(\xi_0 +1/\xi_0 )\over 2}-1\Bigr]{\sin (q')\over\sin (\pi n)}\, ,
\label{PhiccUinf}
\end{equation}
\begin{equation}
\pi\,\Phi_{c0,\,s1}(q,\,q') = {1\over 2}\,{\rm arc}\tan\Bigl(\sinh \Bigl(-{2\pi t\over U}
\sin (q)+{\rm arc}\sinh\Bigl(\tan\Bigl({q'\over n}\Bigr)\Bigr)\Bigr)\Bigr) + {(\xi_0
-1)\over 4}{2q'\over n}\, , \label{PhicsUinf}
\end{equation}
\begin{eqnarray}
\pi\,\Phi_{s1,\,c0}(q,\,q') & = & -{1\over 2}\,{\rm arc}\tan\Bigl(\sinh \Bigl({\rm
arc}\sinh\Bigl(\tan\Bigl({q\over n}\Bigr)\Bigr) -{2\pi t\over U} \sin (q')\Bigr)\Bigr) \,
; \hspace{0.5cm} q\neq \pm
k_F \nonumber \\
& = & -{{\rm sgn} (q)\pi\over 2\sqrt{2}} \, ; \hspace{0.5cm} q = \pm k_F \, ,
\label{PhiscUinf}
\end{eqnarray}
\begin{eqnarray}
\pi\,\Phi_{s1,\,s1}(q,\,q') & = & \int_{0}^{\infty}
d\omega{\sin\Bigl(\omega\,{2\over\pi}[{\rm arc}\sinh\Bigl(\tan\Bigl({q\over
n}\Bigr)\Bigr)- {\rm arc}\sinh\Bigl(\tan\Bigl({q'\over n}\Bigr)\Bigr)]\Bigr)\over
\omega \Bigl(1+e^{2\omega}\Bigr)} \nonumber \\
& + & {t\over U}{\sin (\pi n)\over 2}{2q'\over n}\cos\Bigl({q\over n}\Bigr) \, ;
\hspace{0.3cm} q\neq\pm k_F \nonumber
\\
& = & {{\rm sgn} (q)\pi\over 2\sqrt{2}} \, ; \hspace{0.5cm} q = \pm k_F \, , \hspace{0.5cm}
q' \neq \pm k_F \nonumber
\\
& = & [{\rm sgn} (q)]\pi\Bigl({3\over 2\sqrt{2}}-1\Bigr) \, ; \hspace{0.5cm} q = q' = \pm
k_F \, , \, , \label{PhissUinf}
\end{eqnarray}
\begin{equation}
\pi\,\Phi_{c0,\,c1}(q,\,q') = {\rm arc}\tan\Bigl(-{4t\over U} \sin
(q)+\tan\Bigl({q'\over 2\delta}\Bigr)\Bigr) + {(\xi_0 -1)\over 2}{q'\over\delta} \, ,
\label{Phicc1Uinf}
\end{equation}
and
\begin{equation}
\pi\,\Phi_{s1,\,c1}(q,\,q') = {t\over U}\sin (\pi\delta){2q'\over\delta}\cos\Bigl({q\over
n}\Bigr) \, , \label{Phisc1Uinf}
\end{equation}
respectively. In these equations $\delta =1-n$ and $\xi_0$ is the 
interaction-dependent parameter which for zero spin density appears 
in the expressions of the low-energy quantities 
\cite{Woy,Ogata,Kawakami,Frahm,Brech,93-94,Karlo,CFT,Bozo,Carmelo91,Carmelo92}.
For instance, it is defined in Eq. (74) of 
Ref. \cite{Carmelo92}. In the above $U/t\gg 1$ two-pseudofermion expansions 
(\ref{PhiccUinf}), (\ref{PhicsUinf}), and (\ref{Phicc1Uinf}), $\xi_0$ should be 
replaced by its first-order expansion in $t/U$,
\begin{equation}
\xi_0 = 1 + {4t\over\pi U}\ln 2 \sin (\pi n) \, ; \hspace{0.5cm} U/t \gg 1 \, .
\label{xi0Uinf}
\end{equation}
Note that the $t/U$ leading-order term of the quantity $(\xi_0 +1/\xi_0)$ appearing in
Eq. (\ref{PhiccUinf}) is of second order. However, since the $t/U$ second order term of
the parameter $\xi_0$, which is not given in Eq. (\ref{xi0Uinf}), does not contribute to
$(\xi_0 +1/\xi_0)$, that leading-order term should be considered.

For $U/t=100$ and $n=0.59$ the bare-momentum dependence of the two-pseudofermion
phase-shift analytical expansions (\ref{PhiccUinf})-(\ref{Phisc1Uinf}) is similar to the exact
dependence plotted in Figs. 1-6. In figure 7 such phase-shift expansions are plotted for 
$U/t=4.9$ and $n=0.59$. In spite that they reproduce the exact behavior for $U/t>>1$ only, 
note that their use for $U/t=4.9$ is indeed a reasonably good approximation. This can be
confirmed by comparison of the two-pseudofermion phase-shift expansions plotted in Fig. 7
with the corresponding exact phase-shift values plotted in Figs. 1-6. This result is consistent
with the $U/t>>1$ physics being dominant from $U/t\approx 4$ until $U/t\rightarrow\infty$. It follows 
that the phase-shift analytical expressions (\ref{PhiccUinf})-(\ref{Phisc1Uinf}) can be used as an 
approximation for the two-pseudofermion phase-shift expressions for such a range of $U/t$ 
values, becoming an increasingly better approximation as $U/t$ increases.

\begin{figure}
\subfigure{\includegraphics[width=7cm,height=7cm]{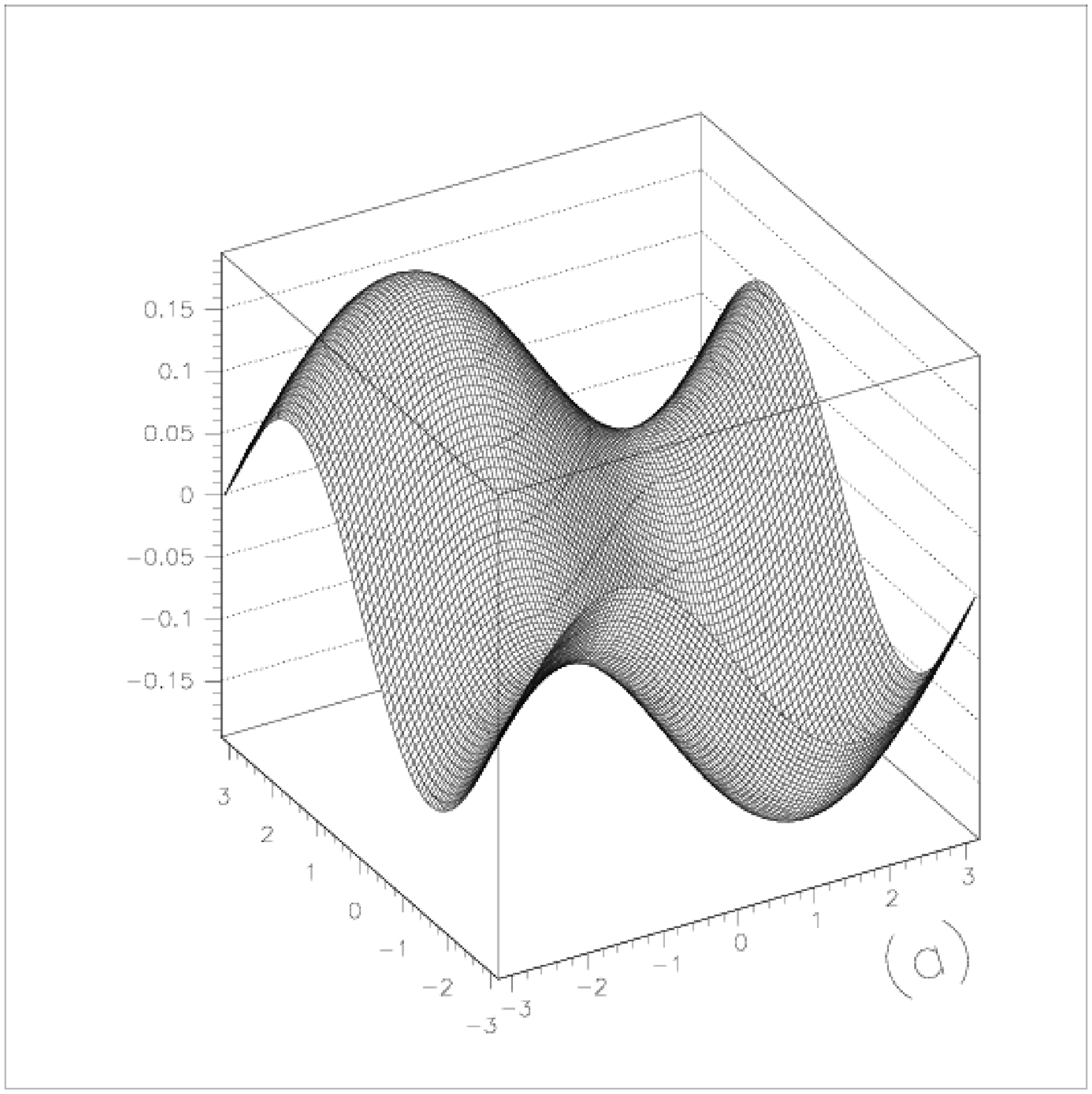}}
\subfigure{\includegraphics[width=7cm,height=7cm]{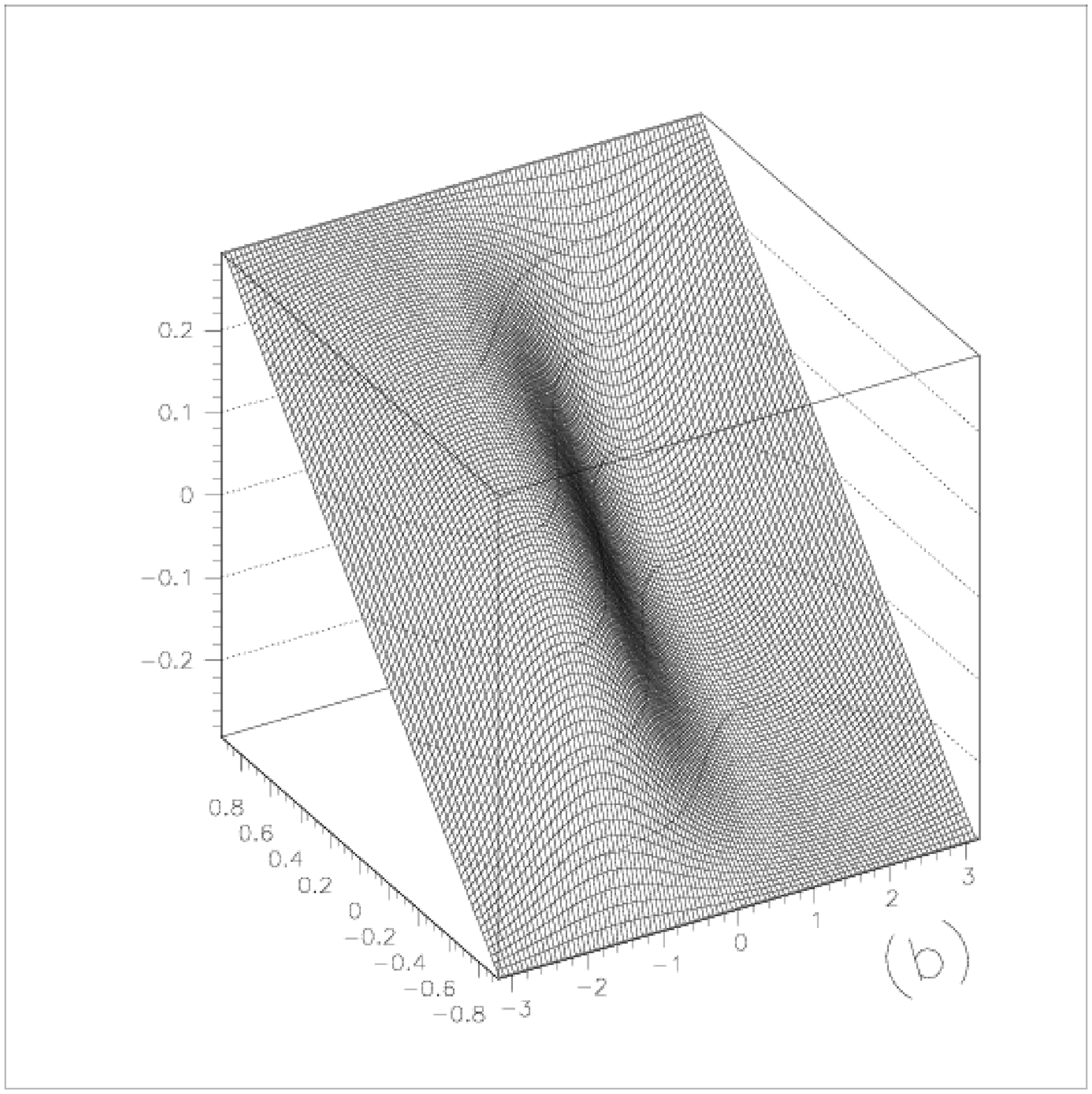}}
\subfigure{\includegraphics[width=7cm,height=7cm]{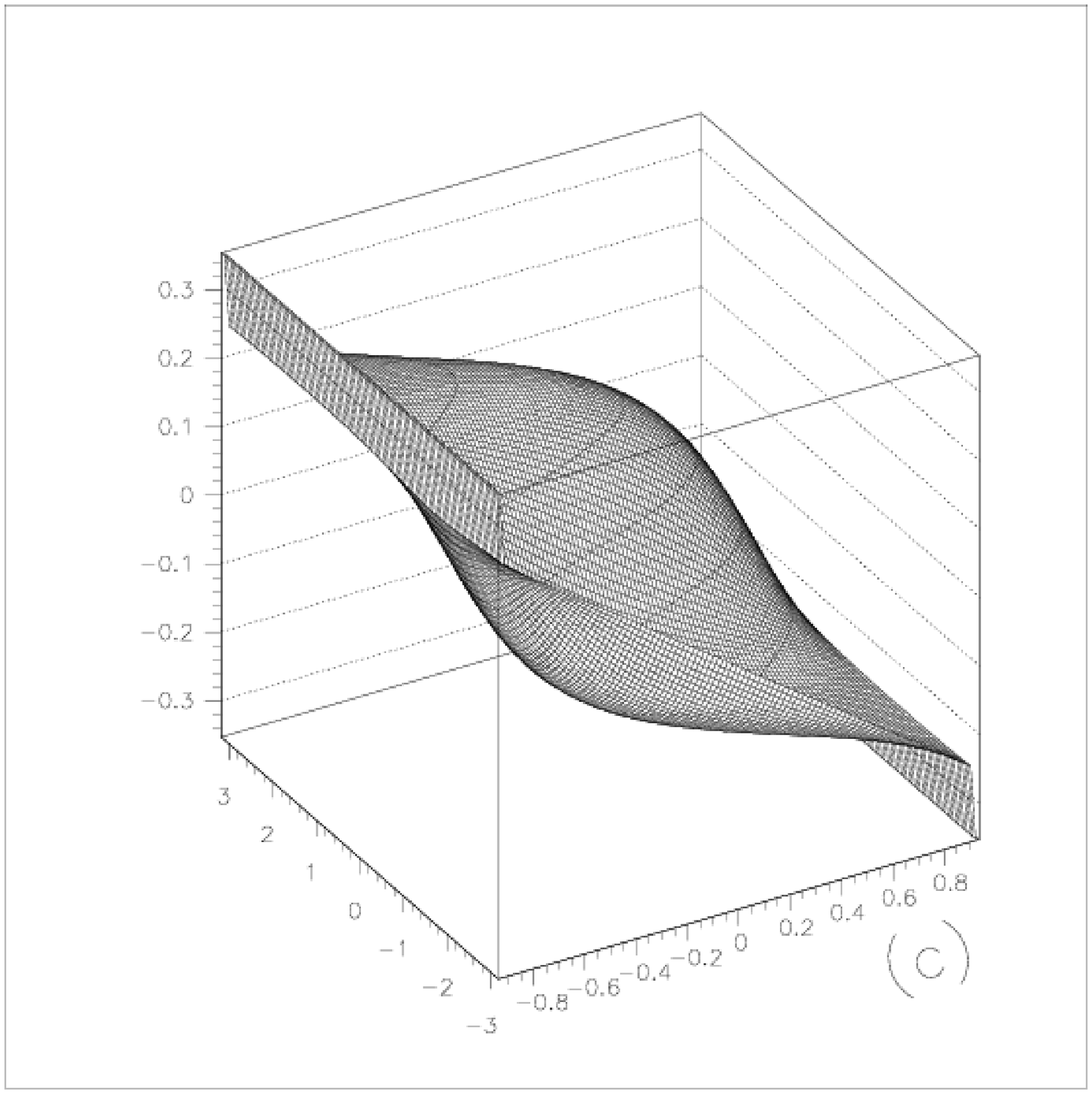}}
\subfigure{\includegraphics[width=7cm,height=7cm]{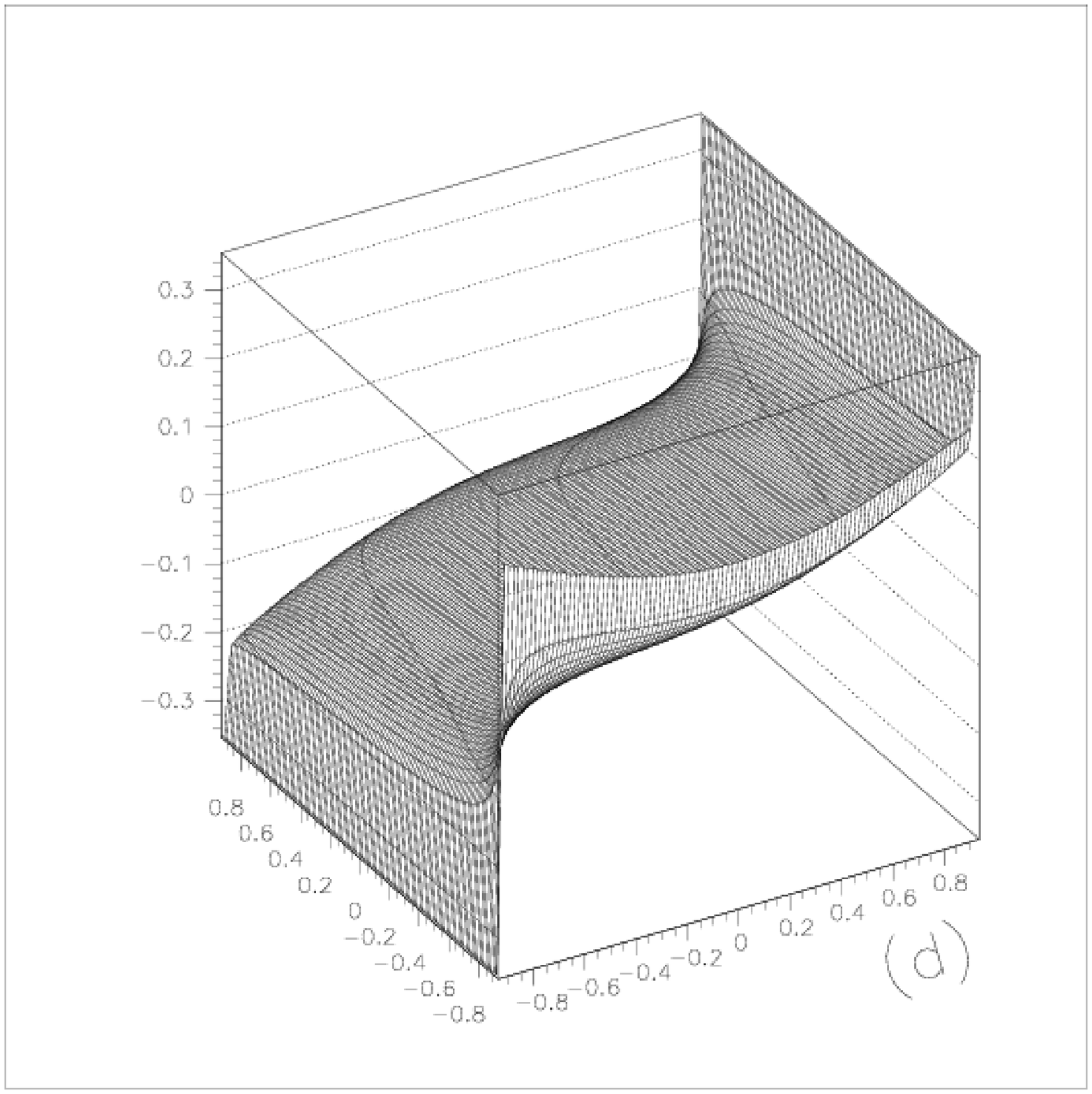}}
\subfigure{\includegraphics[width=7cm,height=7cm]{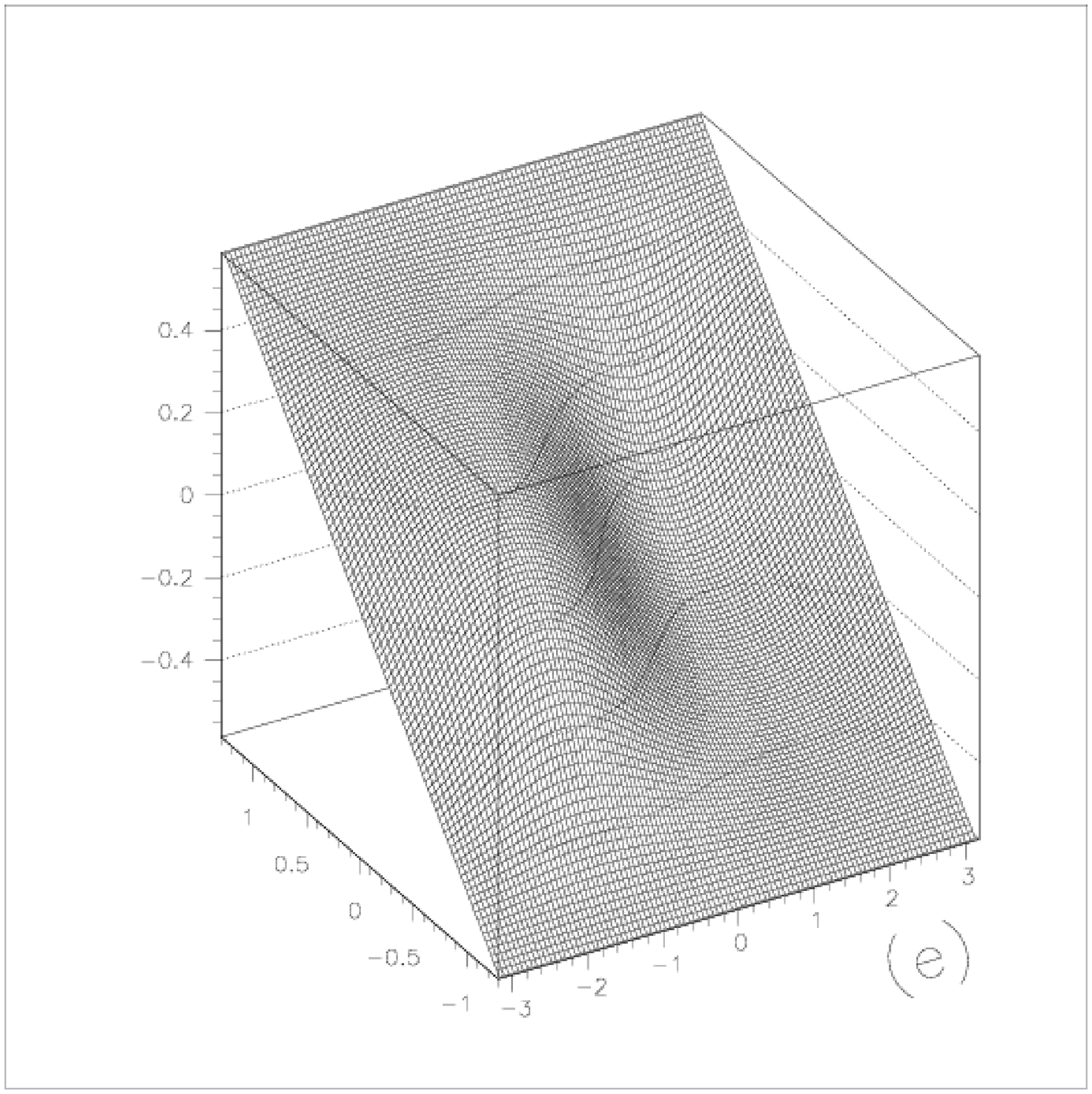}}
\subfigure{\includegraphics[width=7cm,height=7cm]{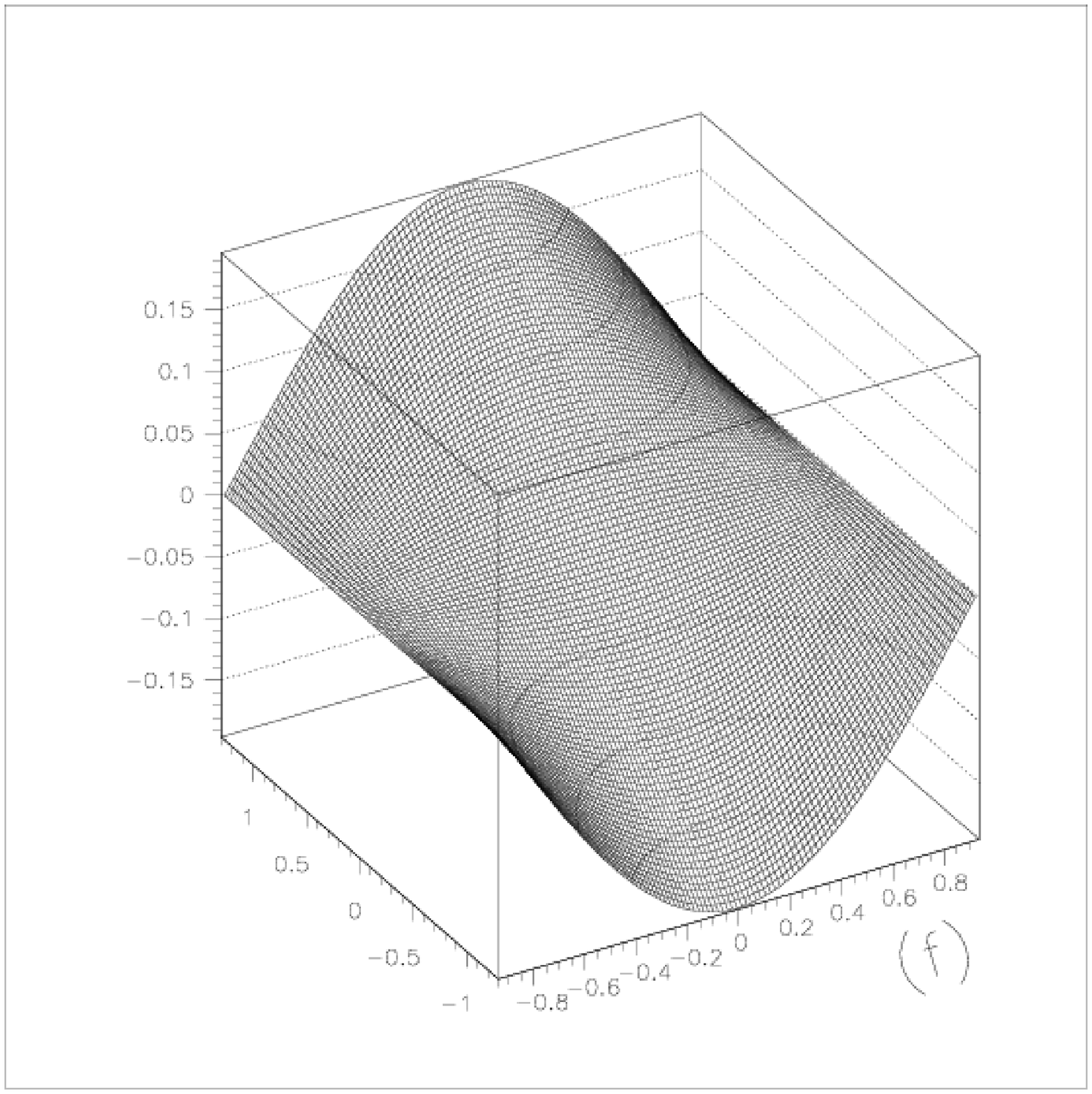}} \caption{The 
$U/t>>1$ two-pseudofermion phase-shift expansions (\ref{PhiccUinf})-(\ref{Phisc1Uinf}) 
in units of $\pi$ for (a) $\Phi_{c0,\,c0}(q,\,q')$, (b) $\Phi_{c0,\,s1}(q,\,q')$, (c) 
$\Phi_{s1,\,c0}(q,\,q')$, (d) $\Phi_{s1,\,s1}(q,\,q')$, (e) $\Phi_{c0,\,c1}(q,\,q')$, 
and (f) $\Phi_{s1,\,c0}(q,\,q')$, respectively,  as a function of $q$ and $q'$ for 
$n=0.59$ and $U/t=4.9$. The bare-momentum values $q$ and $q'$ correspond 
to the right and left axis of the figures, respectively. For $U/t=4.9$ the phase-shift 
large $U/t$ expansions (\ref{PhiccUinf})-(\ref{Phisc1Uinf}) are a reasonably 
good approximation for the corresponding exact phase shifts plotted in Figs. 1-6.}
\end{figure}

In order to confirm that our theory is consistent with the expected general properties of
the standard scattering theory, let us check whether the phase shifts
$\pi\,\Phi_{\alpha\nu,\,\alpha'\nu'}(q,q')$ associated with the elementary
two-pseudofermion scattering events obey Levinson's Theorem \cite{Ohanian}. Such a
theorem just states that when in the reference frame of the scattering center the 
momentum of the scatterer tends to zero the phase shift is given by $\pi N_b$, where 
$N_b$ is the number of bound states. In that frame the phase shift
$\pi\,\Phi_{\alpha\nu,\,\alpha'\nu'}(q,q')$ reads
$\pi\,\Phi_{\alpha\nu,\,\alpha'\nu'}(q-q',0)$. Moreover, in our case there are no bound
states and thus $N_b=0$. Therefore, for the pseudofermion scattering theory Levinson's
Theorem requires that,
\begin{equation}
\lim_{q-q'\rightarrow 0}\pi\,\Phi_{\alpha\nu,\,\alpha'\nu'}(q-q',0) = 0 \, . \label{LT}
\end{equation}
The validity of this result is confirmed by suitable analysis of the integral equations
(A1)-(A13) of Ref. \cite{IIIb}, which reveals that
$\pi\,\bar{\Phi}_{\alpha\nu,\,\alpha'\nu'}\left(r,r'\right)=-\pi\,\bar{\Phi
}_{\alpha\nu,\,\alpha'\nu'}\left(-r,-r'\right)$. This result combined with the use of Eq.
(\ref{Phi-barPhi}) and the odd character of the ground-state rapidity functions, such
that $\Lambda_{\alpha\nu}^0 (q)=-\Lambda_{\alpha\nu}^0 (-q)$ \cite{V-1}, leads then to,
\begin{equation}
\pi\,\Phi_{\alpha\nu,\,\alpha'\nu'}(q,q ') = - \pi\,\Phi_{\alpha\nu,\,\alpha'\nu'}(-q,-q') \, . \label{antiPP}
\end{equation}
This latter symmetry implies that $\pi\,\Phi_{\alpha\nu,\,\alpha'\nu'}(q-q',0)$ is a odd
function of $q-q'$, what confirms the validity of the Levinson's Theorem (\ref{LT}).

Finally, let us address the relation of the pseudofermion phase shifts to the phase
shifts considered in Refs. \cite{Carmelo91,Carmelo92}. It is straightforward to show that
the elementary phase shifts defined by Eqs. (32)-(38) of Ref. \cite{Carmelo92} correspond
to a particular case of the rapidity elementary two-pseudofermion phase shifts defined by
the integral equations (A1)-(A13) of Ref. \cite{IIIb}. If one considers the PS subspace
spanned by energy eigenstates such that $N_{\alpha\nu}=0$ for the $\alpha\nu\neq c0,\,s1$
branches and $L_{\alpha,\,-1/2}=0$ for $\alpha =c,\,s$, the general integral equations
(A1)-(A13) of Ref. \cite{IIIb} reduce to the integral equations (32)-(38) of
Ref. \cite{Carmelo92}. Thus, the phase shifts previously considered in Refs.
\cite{Carmelo91,Carmelo92} correspond to a particular case of the general elementary
two-pseudofermion phase shifts. In contrast to the interpretation of Refs.
\cite{Carmelo91,Carmelo92}, the scatterers and scattering centers associated with the
phase shifts defined by Eqs. (32)-(38) of Ref. \cite{Carmelo92} are the pseudofermions
and pseudofermion holes, rather than the corresponding pseudoparticles and pseudoparticle
holes considered in that reference. Indeed, under the ground-state - excited-state transitions 
the discrete bare-momentum values of the pseudoparticles do not acquire the scattering 
momentum shift $Q^{\Phi}_{\alpha\nu} (q_j)/L$ given in Eq. (\ref{qcan1j}). Instead, within
the pseudoparticle representation the functional $Q^{\Phi}_{\alpha\nu} (q_j)$ appears 
in the energy spectrum, where it leads to $f$-function energy terms associated with the 
two-pseudoparticle residual interactions \cite{IIIb}. In turn, such interactions do not exist
for the pseudofermion representation.

%%%%%%%%%%%%%%%%%%%%%%%%%%%%%%%%%%%%%%%%%%%%%%%%%%%%%%%%%%%%%%%%
\subsection{INVARIANCE UNDER THE ELECTRON - ROTATED-ELECTRON UNITARY TRANSFORMATION}

In the previous section it was mentioned that the pseudofermions are not in general
invariant under the electron - rotated-electron unitary transformation. However, the
exception is for the $\alpha\nu\neq c0,\,s1$ branches as the canonical momentum 
$\bar{q}$ approaches the limiting values, $\bar{q}\rightarrow \pm q_{\alpha\nu}^0$. 
This is consistent with the result of Ref. \cite{II} that creation of one $c\nu\neq c0$
pseudoparticle at $ \pm q_{c\nu}^0=\pm [\pi -2k_F]$ (and one $s\nu\neq s1$
pseudoparticle at $ \pm q_{s\nu}^0=\pm [k_{F\uparrow}-k_{F\downarrow}]$) 
leads to a change $\nu$ in the number of lattice sites doubly occupied by
both electrons and rotated electrons (and singly occupied by both spin-down 
electrons and spin-down rotated electrons). The same result applies to the 
corresponding $c\nu\neq c0$ pseudofermion (and $s\nu\neq s1$
pseudofermion). Here we study the role of such a symmetry in the 
scattering properties of the $\alpha\nu\neq c0,\,s1$ pseudofermions for
initial ground states with densities in the ranges $0<n<1$ and $0<m<n$. (As mentioned above, 
the problem is addressed for initial ground states with densities $n=1$ and/or $m=0$ 
in Sec. IV-D.) First we use the phase-shift expressions involving $\alpha\nu\neq c0,\,s1$
pseudofermions of canonical momentum $\bar{q}=\pm q_{\alpha\nu}^0$ to find important
information about their properties as scatterers and scattering centers. Next we use
such properties to clarify the effects of their invariance under the electron - rotated-electron 
unitary transformation. Below in this subsection we use in general the notations $\alpha\nu$
and $\alpha'\nu'$ to denote the branch of the pseudofermion or pseudofermion-hole scatterers 
and scattering centers, respectively.

Our first goal is to show that the $\alpha\nu\neq c0,\,s1$ pseudofermions of limiting canonical
momentum $\bar{q}=\pm q_{\alpha\nu}^0$ are not active scatterers. By active 
scatterers we mean scatterers whose overall phase shifts generated by the ground-state - 
excited-energy-eigenstate transitions lead to a shift of the corresponding canonical-momentum 
values. Let us confirm that the ground-state limiting canonical-momentum values 
$\bar{q}=\pm q_{\alpha\nu}^0$ of the pseudofermions belonging to $\alpha\nu\neq c0,\,s1$ 
branches are not shifted by the ground-state - excited-energy-eigenstate transitions. 
In contrast to the $c0$ pseudoparticles and the usual band particles and
Fermi-liquid quasi-particles, the band bare-momentum limiting values 
$\pm q_{\alpha\nu}^0=\pm \pi\,[N^*_{\alpha\nu}-1]/L$ of the 
$\alpha\nu\neq c0,\,s1$ pseudoparticles can be changed by the above transitions. Since
$\pm \Delta q_{\alpha\nu}^0=\pm\pi\,\Delta N^*_{\alpha\nu}/L$, such an exotic 
behavior occurs when the deviation $\Delta N^*_{\alpha\nu}$, Eq. (\ref{DN*s1an}), generated by 
the ground-state - excited-energy-eigenstate transition is finite. From use of Eqs. (A8)-(A13)
of Ref. \cite{IIIb} we find that the two-pseudofermion shifts 
$\pi\,\Phi_{\alpha\nu,\,\alpha'\nu'}(\iota\,q^0_{\alpha\nu},q)$ which can be written as 
$\pi\,\Phi_{\alpha\nu,\,c\nu'}(\iota [\pi -2k_F],q)$ and 
$\pi\,\Phi_{\alpha\nu,\,s\nu'}(\iota [k_{F\uparrow}-k_{F\downarrow}],q)$ for 
$\alpha\nu =c\nu\neq c0$ and $\alpha\nu =s\nu\neq s1$, respectively, 
have the following expression,
\begin{equation}
\pi\,\Phi_{\alpha\nu,\,\alpha'\nu'}(\iota\,q^0_{\alpha\nu},q) =
{\iota\,\pi\over 2}\Bigl[\delta_{\alpha'\nu' ,\,c0}(\delta_{\alpha ,\,c}-\delta_{\alpha ,\,s}) 
+\delta_{\alpha ,\,\alpha'} (-\delta_{\nu ,\,\nu'} + \nu +\nu' -\vert\nu -\nu'\vert )\Bigr]  
\, ; \hspace{0.25cm} \iota =\pm 1 \, ; \hspace{0.25cm}
\alpha\nu\neq c0,\,s1 \, , \label{Phi-nas}
\end{equation}
where $q\neq\iota\,q^0_{\alpha\nu}$ for $\alpha'\nu'=\alpha\nu'\neq c0,\,s1$. Use of this
two-pseudofermion phase-shift expression in the general overall scattering
phase-shift expression (\ref{qcan1j}) leads to,
\begin{equation}
{Q^{\Phi}_{\alpha\nu} (\iota q_{\alpha\nu}^0)\over 2} = 
-{\iota\,\pi\over 2}\Bigl[\Delta N_{\alpha\nu} - [\delta_{\alpha ,\,c}
- \delta_{\alpha ,\,s}]\Delta N_{c0} -\sum_{\nu'=1}^{\infty}
(\nu +\nu' -\vert\nu -\nu'\vert )\Delta N_{\alpha\nu'}\Bigr] \, ; \hspace{0.25cm} \iota =\pm 1
\, ; \hspace{0.25cm} \alpha\nu\neq c0,\,s1 \, . 
\label{QPhi-limiting}
\end{equation}
Comparison of this expression with that of 
$\iota\Delta q_{\alpha\nu}=\iota\pi\,\Delta N^*_{\alpha\nu}/L$ where
$N^*_{\alpha\nu}=N_{\alpha\nu}+N^h_{\alpha\nu}$ and $N^h_{\alpha\nu}$ 
is provided in Eq. (\ref{Nhag}) confirms that $Q^{\Phi}_{\alpha\nu} (\iota q_{\alpha\nu}^0)/L
=\iota\Delta q_{\alpha\nu}$. The value of the two-pseudofermion phase 
shift (\ref{Phi-nas}) is not well defined for $q\neq\iota\,q^0_{\alpha\nu}$ and 
$\alpha'\nu'=\alpha\nu'\neq c0,\,s1$. However, from the rapidity expression 
$\Lambda_{\alpha\nu}(q) = \Lambda_{\alpha\nu}^0({\bar{q}}
(q))$ for the PS excited energy-egenstates, where $\Lambda_{\alpha\nu}^0(q)$
is the expression for the initial ground state \cite{IIIb}, one confirms that the relation
$Q^{\Phi}_{\alpha\nu} (\iota q_{\alpha\nu}^0)/L=\iota\Delta q_{\alpha\nu}$ 
is valid for all PS excited states. Indeed, that the rapidity 
functions of the excited energy states of a given initial ground state
equal those of the latter state provided that in the argument of such 
functions the ground-state bare momentum is replaced by the excited-state 
canonical momentum implies that the corresponding bare-momentum 
and canonical-momentum bands have precisely the same width.

This result has a deep physical meaning: the scattering phase shift leads to a 
canonical-momentum shift $Q^{\Phi}_{\alpha\nu} (\pm q_{\alpha\nu}^0)/L$ that 
exactly cancels the bare-momentum shift $\pm \Delta q_{\alpha\nu}$. This 
implies that the overall canonical-momentum shift $\pm\Delta {\bar{q}}_{\alpha\nu}
=\pm \Delta q_{\alpha\nu} +Q^{\Phi}_{\alpha\nu} (\pm q_{\alpha\nu}^0)/L$ 
indeed vanishes. It follows that for the $\alpha\nu\neq c0,\,s1$
pseudofermions the limiting canonical momenta have the same values, 
$\pm q_{\alpha\nu}^0$, both for the ground state and excited energy
eigenstates and thus for $\bar{q}=\pm q_{\alpha\nu}^0$ such objects are not 
active scatterers. 

Let us next investigate the properties of such pseudofermions as scattering 
centers. By use of Eqs. (A1)-(A13) of Ref. \cite{IIIb} and Eqs. (A.11)-(A.14) of 
Ref. \cite{LE} we find after some algebra that the two-pseudofermion shifts 
$\pi\,\Phi_{\alpha\nu,\,\alpha'\nu'}(q,\iota\,q^0_{\alpha'\nu'})$
which can be written as $\pi\,\Phi_{\alpha\nu,\,c\nu'}(q,\iota [\pi -2k_F])$ and 
$\pi\,\Phi_{\alpha\nu,\,s\nu'}(q,\iota [k_{F\uparrow}-k_{F\downarrow}])$ for 
$\alpha'\nu' =c\nu'\neq c0$ and $\alpha'\nu' =s\nu'\neq s1$, respectively,
have the following general form,
\begin{eqnarray}
\pi\,\Phi_{\alpha\nu,\,\alpha'\nu'}(q,\iota\,q^0_{\alpha'\nu'}) & = &
{\iota\,\pi\over 2}\Bigl[\delta_{\alpha\nu ,\,c0} -\delta_{\alpha ,\,\alpha'}
(\nu +\nu' -\vert\nu -\nu'\vert )\Bigr] \nonumber \\
& + & {\iota\over 2}\sum_{\iota'=\pm 1}\iota'\Bigl[\pi\,\Phi_{\alpha\nu\,c0} (q,\iota'\,2k_F) 
-\delta_{\alpha',\,s}\,2\pi\,\Phi_{\alpha\nu\,s1} (q,\iota'\,k_{F\downarrow})
\Bigr] \, ; \hspace{0.25cm} \iota =\pm 1 \, ; \hspace{0.25cm}
\alpha'\nu'\neq c0,\,s1 \, . \label{inde}
\end{eqnarray}
Here $\alpha\nu$ is any of the branches with finite pseudofermion occupancy 
for the excited state under consideration and the values of $q$ are such that 
$\vert q\vert <q^0_{\alpha'\nu'}$ for $\alpha\nu =\alpha'\nu\neq c0,\,s1$ and 
otherwise can have any value and thus correspond to {\it all} active $\alpha\nu$ 
scatterers of that state.
The form of these exact two-pseudofermion phase-shift expressions reveals that, 
except for the constant phase-shift terms, creation of one $c\nu'\neq c0$ 
pseudofermion (and one $s\nu'\neq s1$ pseudofermion) at canonical momentum 
$\iota [\pi -2k_F]$ (and $\iota [k_{F\uparrow}-k_{F\downarrow}]$)
is felt by the $\alpha\nu$ pseudofermion or hole
active scatterers as a shift  $\iota\pi/L$ of both $c0$ bare-momentum {\it Fermi} 
points (and a shift $\iota\pi/L$ of both $c0$ bare-momentum {\it Fermi} points 
and a shift $-\iota 2\pi/L$ of both $s1$ bare-momentum {\it Fermi} points). 
Thus, such scatterers effectively feel that they are scattered by
$c0$ {\it Fermi}-point current shifts (and $c0$ and $s1$ {\it Fermi}-point 
current shifts), rather than by the $c\nu'\neq c0$ (and  $s\nu'\neq s1$)
pseudofermion created at canonical momentum $\iota [\pi -2k_F]$ 
(and $\iota [k_{F\uparrow}-k_{F\downarrow}]$).

Active scattering centers are those which contribute to the scattering phase 
shift (\ref{qcan1j}). For instance, small-momentum and low-energy
$c0$ and $s1$ pseudofermion particle-hole processes in the vicinity of
the {\it Fermi} points, called elementary processes (C) in Ref. \cite{V-1},
do not generate active scattering centers. Indeed, within such processes
the phase shifts generated by the pseudofermion particle 
excitations exactly cancel those produced by the pseudofermion 
hole excitations. The part of the bare-momentum distribution-function 
deviation generated by $\alpha'\nu'$ active scattering centers can be written 
as $\Delta N^{NF}_{\alpha'\nu'} (q') + \Delta N^{F}_{\alpha'\nu'} (q')$. 
Here $\Delta N^{NF}_{\alpha'\nu'} (q')$ is generated by the processes
called elementary processes (A) in Ref. \cite{V-1}, which create and annihilate 
(and create) $\alpha'\nu' = c0,\,s1$ pseudofermions (and $\alpha'\nu'\neq 
c0,\,s1$ pseudofermions) away from the {\it Fermi points} (and away 
from the limiting values $\pm q_{\alpha'\nu'}^0$). In turn, 
$\Delta N^{F}_{\alpha'\nu'} (q')$ is generated by the processes 
called elementary processes (B) in the same reference, which create and 
annihilate (and create) $\alpha'\nu' = c0,\,s1$ pseudofermions (and $\alpha'\nu'\neq c0,\,s1$ 
pseudofermions) at the {\it Fermi points} (and at the limiting 
values $\pm q_{\alpha'\nu'}^0$). In this subsection we are
mostly interested in the scattering centers associated
with the deviation $\Delta N^{F}_{\alpha'\nu'} (q')$, whose
general expression reads,
\begin{eqnarray}
\Delta N^{F}_{\alpha'\nu'} (q') & = & \sum_{\iota =\pm 1}
\Bigl[{\Delta N^{F}_{\alpha'\nu'}\over 2}+
\iota\,\Delta J^{F}_{\alpha'\nu'}\Bigr]\,
\delta_{q',\,\iota\,q^0_{F\alpha'\nu'}} \, ; \hspace{0.25cm}
\alpha'\nu' = c0,\,s1 \, ; \nonumber \\
& = & \sum_{\iota =\pm 1}
\Bigl[{N^{F}_{\alpha'\nu'}\over 2}+
\iota\,J^{F}_{\alpha'\nu'}\Bigr]\,
\delta_{q',\,\iota\,q^0_{\alpha'\nu'}} \, ; \hspace{0.25cm}
\alpha'\nu'\neq c0,\,s1 \, . \label{DNqF-all}
\end{eqnarray}
Here the deviation numbers (and numbers) $\Delta N^F_{\alpha'\nu'}$ (and
$N_{\alpha'\nu'}^{F}$) are such that $\Delta N_{\alpha'\nu'} = 
\Delta N^F_{\alpha'\nu'}+\Delta N^{NF}_{\alpha'\nu'}$ (and $N_{\alpha'\nu'} 
= N_{\alpha'\nu'}^{F} + N_{\alpha'\nu'}^{NF}$). They can be expressed
as $\Delta N^F_{\alpha'\nu'}=\Delta N^F_{\alpha'\nu' ,\,+1}+\Delta N^F_{\alpha'\nu' ,\,-1}$ 
(and $N^{F}_{\alpha'\nu'} = N^{F}_{\alpha'\nu' ,\,+1}+N^{F}_{\alpha'\nu' ,\,-1}$), 
where $\Delta N^F_{\alpha'\nu' ,\,\pm 1}$ is the deviation in the number of 
$\alpha'\nu'$ pseudofermions at the right $(+1)$ and left right $(-1)$ {\it Fermi 
point} (and $N_{\alpha'\nu',\,\iota}^{F}$ is the number of $\alpha'\nu'$ 
pseudofermions created at $\iota\,q_{\alpha'\nu'}^0$ with $\iota =\pm 1$).
The associated deviation current numbers (and current numbers) read 
$2\Delta J^F_{\alpha'\nu'}=\Delta N^F_{\alpha'\nu' ,\,+1}-\Delta 
N^F_{\alpha'\nu' ,\,-1}$ (and $2J^F_{\alpha'\nu'}=N^F_{\alpha'\nu' ,\,+1}-
N^F_{\alpha'\nu',\,-1}$). 

According to the PDT of Refs. \cite{V,V-1}, the elementary
processes (A), (B), and (C) mentioned above lead to 
qualitatively different contributions to the spectral-weight distributions. 
The PDT studies of these references considered that creation of
$c\nu'\neq c0$ and $s\nu'\neq s1$ pseudofermions at the limiting 
bare-momentum values is felt by both the $c0$ and $s1$
scatterers as effective $c0$ scattering centers and effective $c0$ 
and $s1$ scattering centers, respectively. However, in these studies
that was only considered for $c0$ and $s1$ scatterers of 
bare-momentum value $q=\pm q^0_{F\alpha\nu}$. We emphasize 
that the general two-pseudofermion expression (\ref{inde}) generalizes 
that result to {\it all} active $\alpha\nu$ scatterers, including both
$\alpha\nu =c0,\,s1$ and $\alpha\nu\neq c0,\,s1$ active
scatterers of arbitrary bare momentum $q$. 

>From the linearity in the deviations of the overall scattering phase shift 
(\ref{qcan1j}) one can write $Q^{\Phi}_{\alpha\nu} (q)/2= 
Q^{\Phi (NF)}_{\alpha\nu} (q)/2+Q^{\Phi (F)}_{\alpha\nu} (q)/2$ where 
$Q^{\Phi (NF)}_{\alpha\nu} (q)/2$ and $Q^{\Phi (F)}_{\alpha\nu} (q)/2$ 
result from the  $\alpha'\nu'$ scattering centers associated with the 
deviations $\Delta N^{NF}_{\alpha'\nu'} (q')$ and 
$\Delta N^{F}_{\alpha'\nu'} (q')$, respectively. Also the part of the total 
momentum deviation (\ref{DP}) associated with the elementary processes 
(A) and (B) can be written as $\Delta P^{NF}+\Delta P^{F}$. After some 
algebra involving the use of Eqs. (\ref{qcan1j}), (\ref{DP}), (\ref{inde}), 
and (\ref{DNqF-all}) we reach the following expressions for these
quantities,
\begin{equation}
{Q^{\Phi (NF)}_{\alpha\nu} (q)\over 2} = \sum_{\alpha'\nu'}\,
\sum_{q'}\,\pi\,\Phi_{\alpha\nu,\,\alpha'\nu'}(q,q')\, \Delta
N^{NF}_{\alpha'\nu'}(q') \, , \label{qcan1j-NF}
\end{equation}
\begin{eqnarray}
{Q^{\Phi (F)}_{\alpha\nu} (q)\over 2} & = & \sum_{\alpha'\nu'=c0,\,s1}\,
\sum_{\iota'=\pm 1}\pi\,\Phi_{\alpha\nu,\,\alpha'\nu'}(q,\iota'\,q^0_{F\alpha'\nu'})\, 
{\Delta N^{F}_{\alpha'\nu'}\over 2} \nonumber \\
& + & \sum_{\iota'=\pm 1}\iota'\,\pi\,\Phi_{\alpha\nu,\,c0} (q,\iota'\,2k_F)
\Bigl[\Delta J^F_{c0} + \sum_{\nu' =1}^{\infty}\,J^F_{c\nu'} +
\sum_{\nu' =2}^{\infty}\,J^F_{s\nu'}\Bigr] \nonumber \\
& + & \sum_{\iota'=\pm 1}\iota'\,\pi\,\Phi_{\alpha\nu,\,s1} (q,\iota'\,k_{F\downarrow})
\Bigl[\Delta J^F_{s1} -
2\sum_{\nu' =2}^{\infty}\,J^F_{s\nu'}\Bigr] \nonumber \\
& + &\sum_{\alpha'\nu'\neq c0,\,s1} \pi\Bigl[\delta_{\alpha\nu,\,c0}
-\delta_{\alpha,\,\alpha'}\,(\nu +\nu' -\vert\nu -\nu'\vert )\Bigr] J^F_{\alpha'\nu'}
\, , \label{qcan1j-J-F-q}
\end{eqnarray}
and
\begin{eqnarray}
\Delta P^{NF} & = & \sum_{\alpha'\nu'=c0,\,s\nu'}\,
\sum_{q'}\, q'\, \Delta N^{NF}_{\alpha'\nu'}(q') + 
\sum_{c\nu'\neq c0}\,\sum_{q'}\, [\pi -q']\, 
\Delta N^{NF}_{c\nu'}(q') \, ; \nonumber \\
\Delta P^{F} & = & \pi\,[L_{c,\,-1/2}+\sum_{\nu' =1}^{\infty}\nu' N^F_{c\nu'}]+
4k_F\Bigl[\,\Delta J^F_{c0} + \sum_{\nu' =1}^{\infty}\,J^F_{c\nu'} + \sum_{\nu'
=2}^{\infty}\,J^F_{s\nu'}\Bigr] + 2k_{F\downarrow}[\,\Delta J^F_{s1} - 2\sum_{\nu'
=2}^{\infty}\,J^F_{s\nu'}\Bigr] \, . \label{PFI}
\end{eqnarray}
The general expression of the phase shift $Q^{\Phi (F)}_{\alpha\nu} (q)/2$
given in Eq. (\ref{qcan1j-J-F-q}) is valid for all active $\alpha\nu$ scatterers
which for the $\alpha\nu\neq c0,\,s1$ branches refer to bare-momentum
values in the range $\vert q\vert <q^0_{\alpha\nu}$. In the $\Delta P^{F}$
expression of Eq. (\ref{PFI}) we have included the contribution from the 
$-1/2$ Yang holons. (The momentum contributions from the $+1/2$ Yang 
holons and $\pm 1/2$ HL spinons vanish.) Note that the current contributions 
to the momentum spectrum $\Delta P^{F}$ given in Eq. (\ref{PFI}), which multiply 
twice the value of the $c0$ and $s1$ {\it Fermi momenta} $2k_F$ and 
$k_{F\downarrow}$ are identical to the current contributions to
the scattering phase shift (\ref{qcan1j-J-F-q}) which multiply the 
phase shifts $\pi\,\Phi_{\alpha\nu,\,c0} (q,\iota'\,2k_F)$ and
$\pi\,\Phi_{\alpha\nu,\,s1} (q,\iota'\,k_{F\downarrow})$, respectively.

That the $c\nu'\neq c0$ pseudofermions and $s\nu'\neq s1$ pseudofermions 
created at limiting canonical-momentum values are not active scatterers and 
are felt by the active scatterers as effective $c0$ scattering centers and
$c0$ and $s1$ scattering centers, respectively, follows from their invariance
under the electron - rotated-electron unitary transformation. Such properties
reflect the following important decoupling: as $\bar{q}\rightarrow \pm q_{\alpha'\nu'}^0$, 
the $c\nu'\neq c0$ pseudofermion (and $s\nu'\neq s1$ pseudofermion) separates 
into $2\nu'$ independent holons (and $2\nu'$ independent spinons) and a $c\nu'$ 
(and $s\nu'$) $FP$ current scattering center. By independent holons and spinons 
we mean those which remain invariant under the electron - rotated-electron 
unitary transformation. It follows that the Yang holons and HL spinons are also 
independent holons and spinons, respectively. By a $c\nu'$ (and $s\nu'$) $FP$ 
current scattering center we mean the elementary current $J^F_{c\nu'}=\iota/2$
(and $J^F_{s\nu'}=\iota/2$) generated by creation of one $c\nu'\neq c0$ 
pseudofermion (and one $s\nu'\neq s1$ pseudofermion) at ${\bar{q}}=\iota [\pi -2k_F]$ 
(and ${\bar{q}}=\iota [k_{F\uparrow}-k_{F\downarrow}]$). As confirmed by
the form of the phase shifts given in Eqs. (\ref{inde}) and (\ref{qcan1j-J-F-q}),
such elementary currents are felt by the $\alpha\nu$ active scatterers as
elementary shifts of both $c0$ {\it Fermi}-points (and both $c0$ and both $s1$ 
{\it Fermi}-points). This justifies the designation $FP$ (from {\it Fermi}-points)
current scattering center. 

The $c\nu'\neq c0$ and $s\nu'\neq s1$ pseudofermions of limiting 
canonical-momentum value $\pm q_{\alpha'\nu'}^0$ have energy $2\nu'\mu$ 
and $2\nu'\mu_0\,H$, respectively, relative to the ground state. The energy of 
one $-1/2$ independent holon and one $+1/2$ independent holon (and one
$-1/2$ independent spinon and one $+1/2$ independent spinon) relative to that 
state is $2\mu$ and zero (and $2\mu_0\,H$ and zero), respectively. Thus, for 
$\bar{q}=\pm q_{\alpha'\nu'}^0$ the $c\nu'$ pseudofermion (and $s\nu'$
pseudofermion) energy is additive in those of the corresponding $\nu'$ $-1/2$ holons and
$\nu'$ $+1/2$ holons (and $\nu'$ $-1/2$ spinons and $\nu'$ $+1/2$ spinons). This is
consistent with the $\alpha'\nu'\neq c0,\,s1$ pseudofermions loosing their composite 
character as $\bar{q}\rightarrow\pm q_{\alpha'\nu'}^0$. While the whole pseudofermion 
energy goes over to the $\nu'$ $-1/2$ holons or $\nu'$ $-1/2$ spinons, part or the whole 
pseudofermion momentum, respectively, is transferred over to the $FP$ current scattering 
center. In contrast, for $\vert\bar{q}\vert<q_{\alpha'\nu'}^0$ the $c\nu'$ 
pseudofermion (and $s\nu'$ pseudofermion) energy relative to the ground state reads 
$\epsilon_{c\nu'} (q)=2\nu'\mu + \epsilon^0_{c\nu'} (q)$ (and $\epsilon_{s\nu'} (q)=2\nu'\mu_0\,H 
+ \epsilon^0_{s\nu'} (q)$), where the energy $\epsilon^0_{\alpha'\nu'} (q)\neq 0$ 
is defined in Eqs. (C.17), (C.18), (C.20), and (C.21) of 
Ref. \cite{I}. That the energy $\epsilon^0_{\alpha'\nu'} (q)$ is finite 
is consistent with the composite character of the $2\nu'$ holons
and $2\nu'$ spinons of the $c\nu'$ and $s\nu'$ pseudofermion, respectively. In this general 
case the pseudofermions are not invariant under the electron - rotated-electron unitary
transformation. The holon or spinon degrees of freedom are then combined with the
pseudofermion scattering part and, therefore, the $c\nu'\neq c0$ and $s\nu'\neq s1$
pseudofermions created under a ground-state - excited-state transition are felt by 
the $\alpha\nu$ pseudofermion and $\alpha\nu$ pseudofermion-hole active scatterers 
as independent scattering centers, unrelated to the $c0$ and $s1$ {\it Fermi} points.

In what the independent holons associated with a $c\nu'$ pseudofermion of canonical 
momentum $\bar{q}=\pm q_{c\nu'}^0=\pm [\pi -2k_F]$ is concerned, the main difference 
between having (i) $L_{c,\,-1/2}=\nu'$ $-1/2$ Yang holons plus $L_{c,\,+1/2}=\nu'$ 
$+1/2$ Yang holons and (ii) one $c\nu'$ pseudofermion is that the $2\nu'$ Yang holons 
have a total $\eta$-spin value $S_c=\nu'$ whereas for the $2\nu'$ holons associated 
with the pseudofermion the total $\eta$-spin value is $S_c=0$ and thus 
corresponds to a $\eta$-spin singlet configuration. (The same holds for
the $s\nu'$ pseudofermions provided that one replaces holons by spinons and Yang
holons by HL spinons.) As the Yang holons and HL spinons, the $2\nu'$ 
independent holons ($\alpha' =c$) or $2\nu'$ independent spinons ($\alpha' =s$) 
of a $\alpha'\nu'$ pseudofermion of canonical momentum $\bar{q}=\pm q_{\alpha'\nu'}^0$ 
are neither scatterers nor scattering centers.

Within the PDT of Refs. \cite{V,V-1}, the $\alpha\nu = c0,\, s1$ canonical-momentum 
{\it Fermi}-point deviations $\iota\,\Delta {\bar{q}}_{F\alpha\nu}$ play a central role in 
the spectral-function expressions through the related quantity
$2\Delta^{\iota}_{\alpha\nu}=[\iota\,\Delta {\bar{q}}_{F\alpha\nu}/(2\pi/L)]^2$
where $\iota\,\Delta {\bar{q}}_{F\alpha\nu}/(2\pi/L) =\iota\,\Delta N^F_{\alpha\nu,\,\iota}+
Q^{\Phi}_{\alpha\nu} (\iota q^0_{F\alpha\nu})/2\pi$. We thus emphasize that the validity of 
the corresponding expressions (37) and (40) of Ref. \cite{V-1} follows from the general 
expression for the phase shift $Q^{\Phi(F)}_{\alpha\nu} (q)/2$ given in 
Eq. (\ref{qcan1j-J-F-q}).

%%%%%%%%%%%%%%%%%%%%%%%%%%%%%%%%%%%%%%%%%%%%%%%%%%%%%%%%%%%%%%%%
\subsection{INVARIANCE UNDER BOTH THE ELECTRON - ROTATED-ELECTRON
AND PSEUDOPARTICLE - PSEUDOFERMION TRANSFORMATIONS FOR
$n=1$ AND $m=0$ DENSITIES}

The general pseudofermion scattering theory also applies to initial ground states with
densities $n=1$ and/or $m=0$, provided that the specific features reported here 
are taken into account. For an initial ground state with electronic density  $n=1$ 
(and spin density $m=0$) one has that $N^*_{c\nu}=0$ (and $N^*_{s\nu}=0$) 
for the $c\nu$ (and $s\nu$) band and thus the corresponding pseudofermion 
branch does not exist. For simplicity, we focus our attention onto excited energy 
eigenstates of such an initial ground state with a single $c\nu\neq c0$ 
pseudofermion (and a  single $s\nu\neq s1$ pseudofermion). For these excited states, 
$N^*_{c\nu}=1$ (and $N^*_{s\nu}=1$) and the corresponding $c\nu$ (and $s\nu$) 
bare-momentum band reduces to the bare momentum zero. 

A property specific to $n=1$ (and $m=0$) initial ground states is that a transition 
to an excited state involving creation of one $c\nu\neq c0$ pseudofermion (and 
one $s\nu\neq s1$ pseudofermion) always also involves creation of $2\nu$ $c0$ 
pseudofermion holes (and $2(\nu -1)$ $s1$ pseudofermion holes).
Furthermore, if one considers the $n=1$ and $m=0$ initial ground state it 
follows from Eqs. (\ref{Phisnc}), (\ref{Phiccn-csn1}), (\ref{Phicn-ssn1}), and 
(\ref{Phicnc-cncn-cnsn1}) of Appendix B that $\Phi_{c0,\,s\nu'}(q,\,0) = 
\Phi_{c\nu,\,s\nu'}(q,\,0) = \Phi_{s\nu,\,c\nu'}(q,\,0) = 0$
for $\nu\geq 1$ and $\nu'\geq 1$. In turn, we find below that the value of 
the phase shifts $\pi\,\Phi_{c0,\,c\nu}(q,0)$ (and $\pi\,\Phi_{s1,\,s\nu}(q,0)$) is 
fully determined by the $2\nu$ (and $2(\nu -1)$) bare-momentum values
of the $c0$ pseudofermion-hole (and $2(\nu -1)$ $s1$ pseudofermion-hole)
scattering centers. Thus, for excited states of the $n=1$ and $m=0$ initial 
ground state the $c0$ and $s1$ scatterers feel the created $c\nu\neq c0$ 
pseudofermion (and $s\nu\neq s1$ pseudofermion) as $c0$ effective scattering 
centers (and $s1$ effective scattering centers). 

These effective scattering centers are different from those considered above for
the excited states of ground states with electronic density (and spin density) 
in the range $0<n<1$ (and $0<m<n$). Indeed, for the excited states of 
a $n=1$ (and $m=0$) initial ground state considered here the current number 
$J^F_{c\nu}$ (and $J^F_{s\nu}$) vanishes and thus there are no $c\nu\neq c0$ (and 
$s\nu\neq s1$) $FP$ current scattering centers. The occurrence of the 
type of $c0$ (and $s1$) effective scattering centers considered in this section
follows from the non-scatterer character of the corresponding $c\nu\neq c0$
pseudofermions (and $s\nu\neq s1$ pseudofermion), as discussed below.

In spite of the lack of ground-state $c\nu\neq c0$ (and $s\nu\neq s1$) pseudofermion bands, 
the scattering theory can be generalized to an initial ground state with electronic 
density $n=1$ (and $m=0$): we recall that the ``in" asymptote one-pseudofermion 
scattering states do not contribute to the direct-product expression of the 
initial ground state but rather to that of the ``in" state defined 
in Sec. III-C. For the excited states of $n=1$ (and $m=0$) ground states considered
here, the $c\nu\neq c0$ (and $s\nu\neq s1$) bare-momentum band 
corresponds to a single value at $q=0$. For initial ground states with densities $0<n<1$ 
and $0<m<n$ the scattering canonical-momentum shift $Q^{\Phi}_{\alpha\nu} (q)/L$, 
Eq. (\ref{qcan1j}), has the same value whether one uses the 
ground-state rapidity functions $\Lambda^{0}_{\alpha\nu}(q)$
and $\Lambda^{0}_{\alpha'\nu'}(q')$ or the corresponding ``out"-state
(excited-energy-eigenstate) rapidity functions $\Lambda_{\alpha\nu}(q)$ and
$\Lambda_{\alpha'\nu'}(q')$ in the general expression (\ref{Phi-barPhi}) for the
two-pseudofermion phase shifts. Indeed, these two alternative definitions of the
two-pseudofermion phase shifts lead to the same value for the functional
$Q^{\Phi}_{\alpha\nu} (q)/L$ up to contributions of order $1/L$. In turn, 
the general pseudofermion scattering theory introduced above also applies to initial 
ground states with densities $n=1$ and/or $m=0$ provided that the following
procedure is performed: 
\vspace{0.25cm}

-- Since for a $n=1$ and/or $m=0$ initial ground state there are no $c\nu\neq
c0$ and/or $s\nu\neq s1$ pseudofermion bands, the two-pseudofermion expression 
(\ref{Phi-barPhi}) should be replaced by
$\Phi_{\alpha\nu,\,\alpha'\nu'}(q,q') = \bar{\Phi }_{\alpha\nu,\,\alpha'\nu'}
(4t\,\Lambda_{\alpha\nu}(q)/U, 4t\,\Lambda_{\alpha'\nu'}(q')/U)$, where the 
rapidity function $\Lambda_{\alpha\nu}(q)$ (and $\Lambda_{\alpha'\nu'}(q')$)
is that of the excited state and $\alpha\nu =c\nu$ with $\nu >0$ and/or
$\alpha\nu =s\nu$ with $\nu >1$ (and $\alpha'\nu' =c\nu'$ with $\nu' >0$ and/or
$\alpha'\nu' =s\nu'$ with $\nu' >1$), and otherwise is that of the initial
ground state. Since the former rapidity functions are those of the
excited state under consideration, it follows that for the particular case of
such an initial ground state the quantity (\ref{Phi-barPhi}) is a functional 
rather than a function. 
\vspace{0.25cm}

We found in the previous subsection that for initial ground states with
densities in the ranges $0<n<1$ and $0<m<n$, the 
$\alpha\nu\neq c0,\,s1$ pseudofermions 
with limiting canonical momentum given by $\pm q_{\alpha\nu}^0$ are not 
active scatterers.  As discussed below, for the excited energy eigenstates 
considered in this subsection the $c\nu\neq c0$ (and $s\nu\neq s1$) 
band reduces to a single discrete canonical momentum value 
at $\bar{q}=0$. Such excited states have electronic density $n\rightarrow 1$ 
(and spin density $m\rightarrow 0$) and thus the single $\bar{q}=0$ value 
corresponds to the limiting canonical momentum values 
$\pm q_{c\nu}^0=\pm [\pi -2k_F]$ (and $\pm q_{s\nu}^0=
\pm [k_{F\uparrow}-k_{F\downarrow}]$) such that
$\pm q_{c\nu}^0\rightarrow 0$ as $n\rightarrow 1$ (and $\pm 
q_{s\nu}^0\rightarrow 0$ as $m\rightarrow 0$). It follows that
similarly to the pseudofermions considered in the previous
subsection, the $c\nu\neq c0$ pseudofermion (and $s\nu\neq s1$ 
pseudofermion) of the excited states considered 
here are invariant under the electron - rotated-electron unitary
transformation. Indeed, it follows from the results of Ref. \cite{II} 
that creation of such a $c\nu\neq c0$ pseudofermion (and $s\nu\neq s1$
pseudofermion) leads to a change $\nu$ in the number of lattice sites 
doubly occupied by both electrons and rotated electrons (and singly 
occupied by both spin-down electrons and spin-down rotated electrons). 
As a result, they separate into $2\nu$ independent holons (and 
spinons). Such $2\nu$ independent holons (and $2\nu$ independent spinons) 
are fully decoupled from the above mentioned $c0$ effective scattering centers 
(and $s1$ effective scattering centers) also associated with the $c\nu\neq c0$ 
pseudofermion (and $s\nu\neq s1$ pseudofermion). It is then required that 
they are not active scatterers. However, for the class of excited states 
considered here a necessary condition for such 
objects not being active scatterers is that they are not scatterers at all. Indeed, 
once the $c\nu\neq c0$ (and $s\nu\neq s1$) bare-momentum band of the ``in"
 
state has a single value at $q=0$, it is required that the corresponding 
canonical-momentum band of the ``out" state has also a single value at 
$\bar{q}=0$. This implies both that $Q^{\Phi}_{\alpha\nu}(0)/L=0$ and 
$Q_{\alpha\nu} (0)/L=0$, and thus that the $c\nu\neq c0$ (and $s\nu\neq s1$) 
pseudoparticle remains invariant under the pseudoparticle - pseudofermion 
unitary transformation. Indeed, since the $c\nu\neq c0$ (and $s\nu\neq s1$) 
band does not exist for the initial ground state, one has that the scatter-less 
overall phase shift vanishes, $Q_{\alpha\nu}^0/2=0$, and thus for the above 
reasoning the overall phase shift is such that $Q_{\alpha\nu} (0)/2=
Q^{\Phi}_{\alpha\nu}(0)/2=0$. It follows that the corresponding $c\nu\neq c0$ 
(and $s\nu\neq s1$) pseudofermion is not a scatterer.

However, the requirement that such a $c\nu\neq c0$ (and $s\nu\neq s1$) pseudofermion is
not a scatterer and thus that $Q^{\Phi}_{c\nu} (0)/L=0$ (and $Q^{\Phi}_{s\nu} (0)/L=0$)
imposes a specific form to the corresponding two-peudofermion phase shifts
$\Phi_{c\nu,\,\alpha'\nu'}(0,q')$ (and $\Phi_{s\nu,\,\alpha'\nu'}(0,q')$). Since such
objects are neither scatterers nor scattering centers, the quantities
$\pi\,\Phi_{c\nu,\,\alpha'\nu'}(0,q')$ (and $\pi\,\Phi_{s\nu,\,\alpha'\nu'}(0,q')$) are
not real two-peudofermion phase shifts: they are just effective two-peudofermion phase
shifts whose values are such that the overall scattering phase shift $Q^{\Phi}_{c\nu} (0)/2$ 
(and $Q^{\Phi}_{s\nu} (0)/2$) vanishes. Thus, there is no requirement that they do obey 
Levinson's Theorem. Also the phase shifts
$\pi\,\Phi_{\alpha'\nu',\,c\nu}(q',0)$ and $\pi\,\Phi_{\alpha'\nu',\,s\nu}(q',0)$ are not
required to obey that theorem for $\alpha'\nu' =c0,\,s1$, once they refer to effective
$c0$ and $s1$ scattering centers, as confirmed below. 

Let us consider three types of excited energy eigenstates of the above class. Those are
states with finite pseudofermion occupancy for the $c0$ and $s1$ bands plus (a) one
$c\nu\neq c0$ pseudofermion and one $s\nu'\neq s1$ pseudofermion, 
(b) one $c\nu\neq c0$ pseudofermion, and (c) one $s\nu'\neq s1$ pseudofermion. 
For simplicity, we consider excited energy eigenstates of the $n=1$ and $m=0$
initial ground state. Such a ground state is described by full $c0$ and $s1$ 
pseudofermion bands whose {\it Fermi} bare momentum reads $2k_F =\pi$ and 
$k_{F\downarrow}=k_{F\uparrow}=k_F=\pi/2$, respectively. Thus, from
the use of Eq. (B.11) of Ref. \cite{I} we have that $\Delta N_{c0}=-\Delta N_{c0}^h
=-2\nu$, $\Delta N_{s1} = -\nu - \nu'$, $\Delta N^h_{s1} = 2(\nu' - 1)$ for the excited
states (a), $\Delta N_{c0}=-\Delta N_{c0}^h =-2\nu$, $\Delta N_{s1} = -\nu$, and $\Delta
N^h_{s1} = 0$ for the excited states (b), and $\Delta N_{c0}=-\Delta N_{c0}^h = 0$,
$\Delta N_{s1} = -\nu'$, and $\Delta N^h_{s1} = 2(\nu' - 1)$ for the excited states (c).
On the other hand, according to Eqs. (\ref{Phisnc}), (\ref{Phisns1}), and
(\ref{Phicnc-cncn-cnsn1}) of Appendix B, for $m\rightarrow 0$ and $n\rightarrow 1$ the
two-pseudofermion phase shifts that contribute to $Q^{\Phi}_{c\nu} (0)/2$ and
$Q^{\Phi}_{s\nu'} (0)/2$ simplify to $\bar{\Phi }_{c\nu,\,s\nu'}\left(r,r'\right)
={\bar{\Phi }}_{s\nu',\,c0}\left(r',r\right) = \bar{\Phi
}_{s\nu',\,c\nu}\left(r',r\right) = 0$, ${\bar{\Phi }}_{c\nu,\,c0}\left(r,r'\right) =
{1\over{\pi}}\,{\rm arc}{\rm tan}\Bigl({r-r'\over {\nu}}\Bigr)$, and ${\bar{\Phi
}}_{s'\nu,\,s1}\left(r',r\right) = {1\over{\pi}}\,{\rm arc}{\rm tan}\Bigl({r'-r\over
{\nu' -1}}\Bigr)$ for $\nu' > 1$ and $\nu>0$. It follows that for the excited energy
eigenstates (a)-(c) the equation $Q^{\Phi}_{c\nu} (0)/2=0$ and/or 
$Q^{\Phi}_{s\nu'} (0)/2=0$ leads to,
\begin{eqnarray}
& & \sum_{l=1}^{2\nu}\,{\rm arc}{\rm tan}\left({4t\over \nu U}\,\Bigl[\Lambda_{c\nu}(0)-
\Lambda^{0}_{c0}(q_l)\Bigr]\right) =
0 \, , \hspace{0.5cm} \nu > 0 \, , \nonumber \\
& & \sum_{l=1}^{2(\nu'-1)}\,{\rm arc}{\rm tan}\left({4t\over (\nu'
-1)U}\,\Bigl[\Lambda_{s\nu'}(0)- \Lambda^{0}_{s1}({q'}_l)\Bigr]\right) = 0 \, ,
\hspace{0.5cm} \nu'
> 1 \, . \label{QIsnNV}
\end{eqnarray}
Here the first and second equations refer to the $c0$ branch and both the states (a) and
(b) and to the $s1$ branch and both the states (a) and (c), respectively. In these
equations the set of $2\nu$ values $\{q_l\}$ and of $2(\nu'-1)$ values $\{{q'}_l\}$
correspond to the excited-energy-eigenstate $c0$ pseudofermion holes and $s1$
pseudofermion holes, respectively. Moreover, the ground-state rapidity functions
$\Lambda^{0}_{c0}(q)\equiv\sin k_0 (q)$ and $\Lambda^{0}_{s1}(q)$ can be
defined in terms of their inverse functions given in Eq. (A.1) of Ref. \cite{V-1}.
For $n=1$ and $m=0$ the expressions provided in the latter equation lead to,
\begin{equation}
q = k_0 (q) + 2\int_0^{\infty}d\omega {\sin(\omega\,\sin k_0
(q))\over\omega\,(1+e^{2\omega\,U/4t})}\,J_0 (\omega) \, ; \hspace{0.5cm} q =
\int_0^{\infty}d\omega {\sin(\omega\,\Lambda^{0}_{s1}(q))\over\omega\,\cosh
(\omega\,U/4t)}\,J_0 (\omega) \, , \label{Bessel}
\end{equation}
where $J_0 (\omega)$ is a Bessel function. These equations define the inverse 
of the functions $k_0 (q)$
and $\Lambda^{0}_{s1}(q)$, respectively. (To arrive to the second equality of Eq.
(\ref{QIsnNV}) we used that $\Lambda^{0}_{s1}(\pm
k_F)=\Lambda^{0}_{s1}(\pm\pi/2)=\pm\infty$.) The form of the equalities given in Eq.
(\ref{QIsnNV}) reveals that the corresponding solutions
$\Lambda_{c\nu}(0)=\Lambda_{c\nu}(0,\{q_l\})$ and/or
$\Lambda_{s\nu'}(0)=\Lambda_{s\nu'}(0,\{{q'}_l\})$ are functions of the above sets of
bare-momentum values $\{q_l\}$ and $\{{q'}_l\}$, respectively. 

We emphasize that the solution of the BA equations (13)-(16) of Ref. \cite{I} for
the above excited states leads to functions $\Lambda_{c\nu}(0)=
\Lambda_{c\nu}(0,\{q_l\})$ and $\Lambda_{s\nu'}(0)=\Lambda_{s\nu'}(0,\{{q'}_l\})$
for the rapidities $\Lambda_{c\nu}(0)$ and $\Lambda_{s\nu'}(0)$, respectively, 
which also obey Eq. (\ref{QIsnNV}). This
confirms that for such excited states the exact solution of the BA equations is 
indeed equivalent to imposing the symmetry requirement $Q^{\Phi}_{c\nu} (0)/2=0$ 
and $Q^{\Phi}_{s\nu'} (0)/2=0$ associated with the non-scatterer character
of the corresponding $c\nu$ and $s\nu'$ pseudofermion, respectively.

The above functions $\Lambda_{c\nu}(0)=\Lambda_{c\nu}(0,\{q_l\})$ and
$\Lambda_{s\nu'}(0)=\Lambda_{s\nu'}(0,\{{q'}_l\})$ are to be
used in the following expressions,
\begin{eqnarray}
\pi\,\Phi_{c\nu,\,c0}(0,q') & = & \pi\,\bar{\Phi }_{c\nu,\,c0}
\Bigl({4t\,\Lambda_{c\nu}(0,\{q_l\})\over U}, {4t\,\Lambda^0_{c0}(q')\over U}\Bigr) =
{\rm arc}{\rm tan}\Bigl({4t\,[\Lambda_{c\nu}(0,\{q_l\})-\Lambda^0_{c0}(q')]\over
{\nu\,U}}\Bigr) \, ; \nonumber \\
\pi\,\Phi_{s\nu',\,s1}(0,q') & = & \pi\,\bar{\Phi }_{s\nu',\,s1}
\Bigl({4t\,\Lambda_{s\nu'}(0,\{{q'}_l\})\over U}, {4t\,\Lambda^0_{s1}(q')\over U}\Bigr) =
{\rm arc}{\rm tan}\Bigl({4t\,[\Lambda_{s\nu'}(0,\{{q'}_l\})-\Lambda^0_{s1}(q')]\over
{(\nu'-1)\,U}}\Bigr)\, , \label{Phis}
\end{eqnarray}
so that $Q^{\Phi}_{c\nu} (0)/2=0$ and/or $Q^{\Phi}_{s\nu'} (0)/2=0$. The simplest case
corresponds to $\nu =1$ and/or $\nu'=2$ such that solution of Eq. (\ref{QIsnNV}) leads to
$\Lambda_{c1}(0,q_1,q_2)=[\Lambda^{0}_{c0}(q_1)+\Lambda^{0}_{c0}(q_2)]/2$ and/or
$\Lambda_{s2}(0,{q'}_1,{q'}_2)=[\Lambda^{0}_{s1}({q'}_1)+\Lambda^{0}_{s1}({q'}_2)]/2$,
respectively.

The requirement that the $c\nu\neq c0$ pseudofermion (and $s\nu\neq s1$ pseudofermion)
considered here is not a scatterer implies that the corresponding rapidity function
$\Lambda_{c\nu}(0)=\Lambda_{c\nu}(0,\{q_l\})$ (and
$\Lambda_{s\nu'}(0)=\Lambda_{s\nu'}(0,\{{q'}_l\})$) does not in general vanishes and is
the unique solution of the first (and second) equality of Eq. (\ref{QIsnNV}). Combination
of this result with the two-pseudofermion phase shift of Eqs. (\ref{Phiccn-csn1}) and
(\ref{Phicn-ssn1}) of Appendix B reveals that the $c0$ scatterer two-pseudofermion phase shift
$\pi\,\Phi_{c0,\,c\nu}(q,0)$ and the $s1$ scatterer two-pseudofermion phase shift
$\pi\,\Phi_{s1,\,s\nu'}(q,0)$ are such that,
\begin{eqnarray}
\pi\,\Phi_{c0,\,c\nu}(q,0) & = & \pi\,\bar{\Phi }_{c0,\,c\nu}
\Bigl({4t\,\Lambda^0_{c0}(q)\over U}, {4t\,\Lambda_{c\nu}(0,\{q_l\})\over U}\Bigr) =
-{\rm arc}{\rm tan}\Bigl({4t\,[\Lambda^0_{c0}(q)-\Lambda_{c\nu}(0,\{q_l\})]\over
{\nu\,U}}\Bigr)\, ; \nonumber \\
\pi\,\Phi_{s1,\,s\nu'}(q,0) & = & \pi\,\bar{\Phi }_{s1,\,s\nu'}
\Bigl({4t\,\Lambda^0_{s1}(q)\over U}, {4t\,\Lambda_{s\nu'}(0,\{{q'}_l\})\over U}\Bigr) =
{\rm arc}{\rm tan}\Bigl({4t\,[\Lambda^0_{s1}(q)-\Lambda_{s\nu'}(0,\{{q'}_l\})]\over
{(\nu'-1)\,U}}\Bigr) \, ; \hspace{0.25cm} q\neq\pm k_F \nonumber \\
& = & \pm {\pi\over\sqrt{2}} \, ; \hspace{0.5cm} q= \pm k_F \, . \label{Phis-2}
\end{eqnarray}
We note that in addition to the $c0$ or $s1$ scatterer bare-momentum $q$, the
two-pseudofermion phase shifts provided in Eq. (\ref{Phis-2}) 
are functions of the set of $2\nu$ bare-momentum values
$\{q_l\}$ or $2(\nu'-1)$ bare-momentum values $\{{q'}_l\}$ of the $2\nu$ $c0$
pseudofermion-hole scattering centers or $2(\nu'-1)$ $s1$ pseudofermion-hole 
scattering centers, respectively, also created under the ground-state - 
excited-energy-eigenstate transition. As a result of the creation of the 
$c\nu\neq c0$ (and $s\nu'\neq s1$) pseudofermion, the $c0$ (and $s1$) scatterers 
acquire the phase shift $\pi\,\Phi_{c0,\,c\nu}(q,0)$ (and $\pi\,\Phi_{s1,\,s\nu'}(q,0)$)
whose value is fully controlled by the $2\nu$ (and $2(\nu'-1)$) bare-momentum values
of the $2\nu$  (and $2(\nu'-1)$) $c0$ (and $s1$) pseudofermion-hole scattering centers. 
Thus, through the $\{q_l\}$ (and $\{{q'}_l\}$) momentum dependence of the 
phase shift $\pi\,\Phi_{c0,\,c\nu}(q,0)$  (and $\pi\,\Phi_{s1,\,s\nu'}(q,0)$), the $c0$ 
(and $s1$) scatterers feel the created $c\nu\neq c0$ (and $s\nu'\neq s1$) pseudofermion 
as $2\nu$ $c0$ effective scattering centers (and $2(\nu'-1)$ $s1$ effective 
scattering centers).

Similar results are obtained for excited states of initial ground states of density 
$n=1$ and/or $m=0$ with finite occupancy for a larger finite number of $\alpha\nu$ 
pseudofermions  belonging to several $\alpha\nu\neq c0,\,s1$ branches, except 
that the number of equations defining the the set of rapidities $\{\Lambda_{\alpha\nu}\}$
is in general larger than above and each of these equations is more involved than the 
two equations given in Eq. (\ref{QIsnNV}). Importantly, such  $\alpha\nu$ 
pseudofermions are also invariant under the electron - rotated-electron unitary 
transformation. 

%%%%%%%%%%%%%%%%%%%%%%%%%%%%%%%%%%%%%%%%%%%%%%%%%%%%%%%%%%%%%%%%
\section{CONCLUDING REMARKS}

In this paper we have shown that the non-perturbative and strongly
correlated scattering problem described in terms of electrons by
the Hamiltonian (\ref{H}) considerably simplifies in terms of the pseudofermions
of Refs. \cite{I,II,IIIb,V,V-1,LE}: We have confirmed here that in terms of 
pseudofermion scattering the spectral and dynamical properties are controlled at all
energy scales and for all values of the on-site electronic
repulsion by two-pseudofermion zero-momentum forward
scattering only. This agrees with the preliminary analysis of
the problem of Ref. \cite{Car04}. The matrix elements 
between the ground state and excited energy eigenstates of the corresponding
spectral functions can be expressed in terms of the pseudofermion 
anticommutators \cite{V,V-1,LE}. Our results show that such anticommutators
are controlled by two-pseudofermion zero-momentum forward scattering through 
the associated pseudofermion and hole $S$ matrices. 

Our results have also clarified the relation between the pseudofermion scattering
properties and symmetry. Specifically, it was found for the metallic pase at
finite spin density that the invariance under the electron - rotated-electron unitary 
transformation of the $\alpha\nu\neq c0,\,s1$
pseudofermions with limiting canonical momentum $\bar{q}=\pm q^0_{\alpha\nu}$
implies that such composite objects separate into $2\nu$ independent holons
($\alpha =c$) or $2\nu$ independent spinons ($\alpha =s$) and a current
excitation. The latter excitation is felt by all $\alpha\nu$ active scatterers 
of arbitrary momentum as a shift of both $c0$ {\it Fermi} points and 
shifts of both $c0$ and $s1$ {\it Fermi} points, respectively. 
The effects of the invariance under both the electron - rotated-electron unitary 
transformation and pseudoparticle - pseudofermion unitary transformation
were also studied for the Mott-Hubbard insulator at zero spin density.
The property that only the holons and spinons that remain invariant under 
such a transformation are allowed to exist
as independent quantum objects which are not part of $2\nu$-holon and 
$2\nu$-spinon composite pseudofermions, respectively, is general and also
applies to the Yang holons and HL spinons. Such objects are neither
scatterers nor scattering centers. 

That our choice of scatterers and scattering centers profits from the transformation
laws under the electron - rotated-electron unitary transformation justifies that all 
``in" and ``out" states of the theory are excited energy eigenstates. All these
states can be written as a direct product of ``in" asymptote and ``out" asymptote
one-pseudofermion scattering states, respectively. 
Such a property combined with the simple form obtained 
for the pseudofermion and hole $S$ matrices is behind the suitability of the
pseudofermion representation for the study of the finite-energy spectral and dynamical
properties \cite{V,V-1,LE}. The studies of Ref. \cite{relation} clarify the relation
of the pseudofermion phase shifts and $S$ matrices introduced in this paper to the 
corresponding quantities of the conventional spinon-holon scattering theory 
of Refs. \cite{Natan,S0,S}. The choice 
of scatterers and scattering centers of the latter theory is
also based on associating the Bethe-ansatz quantum numbers with quantum
objects, yet it does not profit from the invariances under the electron - rotated-electron 
unitary transformation. The clarification in this paper of the relation between the pseudofermion 
scattering mechanisms and symmetry provides further useful information about
the PDT microscopic processes \cite{V-1,LE}.  

Since the transport and dynamical properties \cite{dynamical} and other properties
predicted by the 1D Hubbard model were observed in low-dimensional complex
materials \cite{properties} and the investigations presented in Refs.
\cite{spectral0,spectral,super} confirm that the PDT describes successfully the unusual
finite-energy spectral features observed by angle-resolved photoelectron spectroscopy 
in quasi-1D organic metals, our results also contribute to the further understanding of the
non-perturbative scattering mechanisms behind these properties. While the studies of this
paper considered the 1D Hubbard model, our results are of general nature for many
integrable interacting problems \cite{tj} and therefore have wide applicability.

%%%%%%%%%%%%%%%%%%%%%%%%%%%%%%%%%%%%%%%%%%%%%%%%%%%%%%%%%%%%%%%%%%%%%%%%%%
\begin{acknowledgments}
We thank Natan Andrei, Katrina E. Hibberd, Vladimir E. Korepin, Patrick A. Lee, Sung-Sik Lee, Karlo
Penc, and Tiago C. Ribeiro for stimulating discussions and the support of FCT under the
grant POCTI/FIS/58133/2004 and that of the ESF Science Programme INSTANS 2005-2010.
J.M.P.C. and D.B. thank the financial support of the Calouste Gulbenkian Foundation and
the hospitality and support of MIT where most of this research was performed, J.M.P.C. thanks
the support of the Fulbright Commission, and D.B. thanks the financial support of the
FCT grant SFRH/BD/6930/2001.
\end{acknowledgments}
%%%%%%%%%%%%%%%%%%%%%%%%%%%%%%%%%%%%%%%%%%%%%%%%%%%%%%%%%%%%%%%%%%%%%%%%%%
\appendix

\section{PSEUDOFERMION SCATTERING PROCESSES ENERGY CONSERVATION}

Both the scattering phase shift $Q^{\Phi}_{\alpha\nu} (q_j)/2$, Eq. (\ref{qcan1j}), and
the overall phase shift $Q_{\alpha\nu}(q_j)/2 = Q_{\alpha\nu}^0/2 + Q^{\Phi}_{\alpha\nu}
(q_j)/2$, Eq. (\ref{Qcan1j}), conserve the total energy.

Let us start by confirming that for $L>>1$ the above overall phase shift
conserves the total energy. The $c0$ and $c1$ pseudofermion momentum distribution
function deviations of Eq. (\ref{DE}) can be written as,
\begin{equation}
\Delta {\cal{N}}_{\alpha\nu} ({\bar{q}}_j) = \Delta {\cal{N}}^{(1)}_{\alpha\nu}
({\bar{q}}_j) + \Delta {\cal{N}}^{(2)}_{\alpha\nu} ({\bar{q}}_j) \, ; \hspace{1cm}
\alpha\nu = c0,\,s1 \, . \label{Nqcs1-12}
\end{equation}
Here $\Delta {\cal{N}}_{\alpha\nu} ({\bar{q}}_j)\equiv \Delta N_{\alpha\nu} (q_j)$ and
the deviations $\Delta {\cal{N}}^{(1)}_{\alpha\nu} ({\bar{q}}_j)$ and
$\Delta {\cal{N}}^{(2)}_{\alpha\nu} ({\bar{q}}_j)$ are associated 
with the ground-state - virtual-state transition and the virtual-state - ``out"-state 
transition, respectively. The latter deviation reads,
\begin{eqnarray}
\Delta {\cal{N}}^{(2)}_{\alpha\nu} ({\bar{q}}) & = & {\cal{N}}^0_{\alpha\nu} (q+
Q_{\alpha\nu}(q)/L)-{\cal{N}}^0_{\alpha\nu} (q) \nonumber
\\ & = & {1\over L}\,Q_{\alpha\nu}(q)\,{\partial
{\cal{N}}^0_{\alpha\nu} ({\bar{q}})\over \partial {\bar{q}}} = -{1\over L}\,{\rm sgn}
(\bar{q})\,\Bigl[\,{Q_{\alpha\nu}({\rm sgn}\, (\bar{q}) q^0_{F\alpha\nu})}\Bigr] \,\delta
(q^0_{F\alpha\nu}-\vert\,\bar{q}\vert)  \, ; \hspace{0.5cm} \alpha = c0,\,s1 \, .
\label{DevEt}
\end{eqnarray}
Here $q=\bar{q}$ is the virtual-state canonical momentum value, the {\it Fermi} 
momentum $q^0_{F\alpha\nu}$ of the $c0$ and $s1$ bands is given in Eq. (\ref{q0Fcs}),
and $Q_{\alpha\nu}(q)/2$ is the phase shift (\ref{Qcan1j}). Use of Eq. (\ref{DevEt}) in
the energy spectrum (\ref{DE}) leads to the following energy spectrum to first order in
the canonical-momentum distribution-function deviations,
\begin{eqnarray}
\Delta E^{(2)}_{c0,\,s1} & = & \sum_{\alpha\nu =c0,s1}\sum_{{{\bar{q}}_j}=
-q^0_{\alpha\nu }}^{+q^0_{\alpha\nu }} \,\Delta {\cal{N}}^{(2)}_{\alpha\nu} (\bar{q})\,
\epsilon_{\alpha\nu} (\bar{q})= -{1\over L}\,\sum_{\iota =\pm 1}\sum_{\alpha\nu
=c0,s1}\iota \,Q_{\alpha\nu}(\iota\, q^0_{F\alpha\nu})\,\epsilon_{\alpha\nu}
(q^0_{F\alpha\nu})
\nonumber \\
& = & -{1\over L}\,\sum_{\iota =\pm 1}\sum_{\alpha\nu =c0,s1}\iota
\,\Bigl[\,Q^0_{\alpha\nu}+Q^{\Phi}_{\alpha\nu}(\iota\,
q^0_{F\alpha\nu})\Bigr]\,\epsilon_{\alpha\nu} (q^0_{F\alpha\nu}) =0 \, .
\label{VanishDEt}
\end{eqnarray}
In order to confirm that the factor $-\sum_{\iota =\pm 1}\sum_{\alpha\nu =c0,s1}\iota
\,Q_{\alpha\nu}(\iota\, q^0_{F\alpha\nu})\,\epsilon_{\alpha\nu} (q^0_{F\alpha\nu})$,
which multiplies $1/L$ in the last term on the right-hand side of Eq. (\ref{VanishDEt}),
vanishes we have used the symmetry $\epsilon_{\alpha\nu} (q)=\epsilon_{\alpha\nu} (-q)$
and Eq. (\ref{eplev0}) such that $\epsilon_{\alpha\nu} (q^0_{F\alpha\nu})=0$ for
$\alpha\nu=c0,s1$. (We recall that the occupancies of other $\alpha\nu\neq c0,\,s1$
pseudofermion branches vanish for the ground state.) Thus, we have just found that the
energy contribution of first order in the canonical-momentum distribution-function
deviations originated by the collective excitation (2) vanishes.

Note that the energy (\ref{VanishDEt}) decouples into two contributions, corresponding to
the scatter-less phase shift $Q^0_{\alpha\nu}/2$ and scattering phase shift
$Q^{\Phi}_{\alpha\nu}(\iota\, q^0_{F\alpha\nu})/2$. This confirms that both these phase
shifts conserve the energy independently.

Since the collective excitation (2) involves all virtual-state $c0$ and $s1$
pseudofermions, we used similar procedures for the evaluation of the energy contributions
of order larger than one in the canonical-momentum distribution-function deviations. We
find that all such contributions also vanish and decouple into independent and vanishing
contributions corresponding to the scatter-less and scattering phase shifts. It
follows that both the overall phase shift $Q_{\alpha\nu}(q_j)/2$ and the phase shifts
$Q_{\alpha\nu}^0/2$ and $Q^{\Phi}_{\alpha\nu} (q_j)/2$ conserve the total energy.

%%%%%%%%%%%%%%%%%%%%%%%%%%%%%%%%%%%%%%%%%%%%%%%%%%%%%%%%%%%%%%%%
\section{ELEMENTARY TWO-PSEUDOFERMION PHASE SHIFT EXPRESSIONS FOR $m\rightarrow 0$
AND $n\leq 1$ AND FOR $m\rightarrow 0$ AND $n\rightarrow 1$}

We start by considering the limit $m\rightarrow 0$. The rapidity two-pseudofermion phase
shifts $\pi\,{\bar{\Phi }}_{\alpha\nu,\,\alpha'\nu'}(r,r')$ are defined by the integral
equations (A1)-(A13) of Ref. \cite{IIIb}. By Fourier transforming these equations after
considering that $B=\infty$ and thus $r^0_s = 4t\,B/U=\infty$ for finite values of $U/t$, 
we arrive to the following equations valid for $m\rightarrow0$ and $U/t$ finite,
\begin{equation}
\pi\,{\bar{\Phi }}_{c0,\,c0}(r,\,r') = - B (r-r') + \int_{-r_0}^{+r_0}dr''\,A
(r-r'')\,\pi\,{\bar{\Phi }}_{c0,\,c0}(r'',\,r') \, , \label{Phicc}
\end{equation}
\begin{equation}
\pi\,{\bar{\Phi }}_{c0,\,s1}(r,\,r') = -{1\over 2}\,{\rm arc}\tan\Bigl(\sinh
\Bigl({\pi\over 2}(r-r')\Bigr)\Bigr) + \int_{-r_0}^{+r_0}dr''\,A (r-r'')\,\pi\,{\bar{\Phi
}}_{c0,\,s1}(r'',\,r') \, , \label{Phics}
\end{equation}
\begin{eqnarray}
\pi\,{\bar{\Phi }}_{s1,\,c0}(r,\,r') & = & -{1\over 2}\,{\rm arc}\tan\Bigl(\sinh
\Bigl({\pi\over 2}(r-r')\Bigr)\Bigr) + {1\over 4}\int_{-r_0}^{+r_0}dr''\,{\pi\,{\bar{\Phi
}}_{c0,\,c0}(r'',\,r')\over \cosh \Bigl({\pi\over 2}(r-r'')\Bigr)} \, ;
\hspace{0.5cm} r \neq \pm \infty \nonumber \\
& = & -{{\rm sgn} (r)\pi\over 2\sqrt{2}} \, ; \hspace{0.5cm} r = \pm \infty \, ,
\label{Phisc}
\end{eqnarray}
\begin{eqnarray}
\pi\,{\bar{\Phi }}_{s1,\,s1}(r,\,r') & = & B (r-r') + {1\over
4}\int_{-r_0}^{+r_0}dr''\,{\pi\,{\bar{\Phi }}_{c0,\,s1}(r'',\,r')\over \cosh \Bigl({\pi\over
2}(r-r'')\Bigr)} \, ; \hspace{0.3cm} r\neq\pm\infty \nonumber
\\
& = & {{\rm sgn} (r)\pi\over 2\sqrt{2}} \, ; \hspace{0.5cm} r = \pm\infty \, ,
\hspace{0.5cm} r' \neq r \nonumber
\\
& = & [{\rm sgn} (r)]\Bigl({3\over 2\sqrt{2}}-1\Bigr)\pi \, ; \hspace{0.5cm} r = r' =
\pm\infty \, , \label{Phiss}
\end{eqnarray}
\begin{equation}
\pi\,{\bar{\Phi }}_{c0,\,c\nu}(r,\,r') = -{\rm arc}\tan\Bigl({r-r'\over
\nu}\Bigr) + \int_{-r_0}^{+r_0}dr''\,A (r-r'')\,\pi\,{\bar{\Phi }}_{c0,\,c\nu}(r'',\,r') \, ;
\hspace{0.5cm} \nu
>0 \, , \label{Phiccn}
\end{equation}
\begin{equation}
\pi\,{\bar{\Phi }}_{c0,\,s\nu}(r,\,r') = 0 \, ; \hspace{0.5cm} \nu
>1 \, , \label{Phicsn}
\end{equation}
\begin{equation}
\pi\,{\bar{\Phi }}_{s1,\,c\nu}(r,\,r') =  {1\over 4}\int_{-r_0}^{+r_0}dr''\,{\pi\,{\bar{\Phi
}}_{c0,\,c\nu}(r'',\,r')\over \cosh \Bigl({\pi\over 2}(r-r'')\Bigr)} \, ; \hspace{0.5cm}
\nu >0 \, , \label{Phiscn}
\end{equation}
\begin{eqnarray}
\pi\,{\bar{\Phi }}_{s1,\,s\nu}(r,\,r') & = & {\rm arc}\tan\Bigl({r-r'\over \nu
-1}\Bigr) \, ; \hspace{0.5cm} r\neq \pm\infty \, ; \hspace{0.5cm} \nu
>1 \, , \nonumber \\
& = & \pm {\pi\over\sqrt{2}} \, ; \hspace{0.5cm} r= \pm\infty \, ; \hspace{0.5cm} \nu
>1 \, , \label{Phissn}
\end{eqnarray}
\begin{equation}
\pi\,{\bar{\Phi }}_{s\nu,\,c0}\left(r,r'\right) = \pi\,\bar{\Phi }_{s\nu,\,c\nu'}\left(r,r'\right)
= 0 \, ; \hspace{0.5cm} \nu
> 1 \, , \label{Phisnc}
\end{equation}
\begin{equation}
\pi\,{\bar{\Phi }}_{s\nu,\,s1}\left(r,r'\right) = {\rm arc}{\rm
tan}\Bigl({r-r'\over {\nu -1}}\Bigr) \, ; \hspace{0.5cm} \nu > 1 \, , \label{Phisns1}
\end{equation}
\begin{equation}
\pi\,\bar{\Phi }_{s\nu,\,s\nu'}\left(r,r'\right) = {1\over{2}}\,\Theta_{\nu,\,\nu'}(r-r') -
{\rm arc}{\rm tan}\Bigl({r-r'\over {\nu +\nu' -2}}\Bigr) -
{\rm arc}{\rm tan}\Bigl({r-r'\over {\nu +\nu'}}\Bigr) \, , \hspace{0.5cm}
\nu ,\,\nu' > 1  \, . \label{Phisnsn}
\end{equation}
In the above expressions the function $\Theta_{\nu,\,\nu'}(x)$ is defined in Eq. (B.5) of
Ref. \cite{I},
\begin{equation}
B (r) = \int_{0}^{\infty} d\omega{\sin (\omega\,r)\over \omega
(1+e^{2\omega})} = {i\over 2} \ln {\Gamma \Bigl({1\over 2}+i{r\over 4}\Bigr)\,\Gamma
\Bigl(1-i{r\over 4}\Bigr)\over \Gamma \Bigl({1\over 2}-i{r\over 4}\Bigr)\,\Gamma
\Bigl(1+i{r\over 4}\Bigr)} \, , \label{Br}
\end{equation}
\begin{equation}
A (r) = {1\over\pi}{d B (r)\over dr} = {1\over\pi}\int_{0}^{\infty} d\omega{\cos (\omega\,r)\over
1+e^{2\omega}} \, , \label{Ar}
\end{equation}
and
\begin{equation}
r_0 = {4t\sin Q \over U} \, , \label{r0}
\end{equation}
where $Q$ is the parameter defined by Eq. (A.5) of Ref. \cite{V-1} and $\Gamma (x)$ is the usual
$\Gamma$ function.

Moreover, the two-pseudofermion phase shifts,
\begin{equation}
\pi\,{\bar{\Phi }}_{c\nu,\,c0}\left(r,r'\right) = {\rm arc}{\rm
tan}\Bigl({r-r'\over {\nu}}\Bigr) - {1\over{\pi}}\int_{-r_0}^{+r_0} dr''{\pi\,{\bar{\Phi
}}_{c0,\,c0}\left(r'',r'\right) \over {\nu[1+({r-r''\over {\nu}})^2]}} \, ;
\hspace{0.5cm} \nu
>0 \, , \label{Phicnc}
\end{equation}
\begin{equation}
\pi\,\bar{\Phi }_{c\nu,\,c\nu'}\left(r,r'\right) = {1\over{2}}\,\Theta_{\nu,\,\nu'}(r-r') -
{1\over{\pi}}\int_{-r_0}^{+r_0} dr''{\pi\,\bar{\Phi }_{c0,\,c\nu'}\left(r'',r'\right) \over
{\nu[1+({r-r''\over\nu})^2]}} \, ; \hspace{0.5cm} \nu ,\,\nu'
>0 \, , \label{Phicncn}
\end{equation}
\begin{equation}
\pi\,\bar{\Phi }_{c\nu,\,s\nu'}\left(r,r'\right) = - {1\over{\pi}}\,\int_{-r_0}^{+r_0}
dr''{\pi\,\bar{\Phi }_{c0,\,s\nu'}\left(r'',r'\right) \over {\nu[1+({r-r''\over\nu})^2]}} \, ;
\hspace{0.5cm} \nu >0 \, , \label{Phicnsn}
\end{equation}
remain as in Ref. \cite{IIIb}. Furthermore, we note that the four expressions
(\ref{Phicc})-(\ref{Phiss}) with $c0 =c$, $s1 =s$, $r =x$, and $r_0 =x_0$ are equivalent
to expressions (A9)-(A12) of Ref. \cite{Carmelo92}. (For the phase shifts (\ref{Phisc})
and (\ref{Phiss}) this equality refers to values of $r$ such that $r\neq\infty$.)
The rapidity phase-shift expressions given here for $m\rightarrow 0$ correspond
to some of the two-pseudofermion phase shifts plotted in units of $\pi$
in Figs. 1-6.

Let us now consider the limit $m\rightarrow 0$ and $n\rightarrow 1$. Note that the
expressions (\ref{Phicsn}) and (\ref{Phissn})-(\ref{Phisnsn}) are valid for $m\rightarrow
0$ and all values of $n$ such that $n\leq 1$. Thus, they also apply to the limit
$n\rightarrow 1$. On the other hand, note that for all other rapidity two-pseudofermion
phase-shifts the expressions for both $m\rightarrow 0$ and $n\rightarrow 1$ are obtained
by considering $r_0 =0$ in the above integral equations. Such a procedure leads to closed
form analytical expressions for all rapidity two-pseudofermion phase shifts. Thus, for
$m\rightarrow 0$ and $n\rightarrow 1$ we find,
\begin{equation}
\pi\,{\bar{\Phi }}_{c0,\,c0}(r,\,r') = - B (r-r') \, ; \hspace{0.5cm} \pi\,{\bar{\Phi
}}_{c0,\,s1}(r,\,r') = -{1\over 2}\,{\rm arc}\tan\Bigl(\sinh \Bigl({\pi\over
2}(r-r')\Bigr)\Bigr) \, , \label{Phicc-cs1}
\end{equation}
\begin{eqnarray}
\pi\,{\bar{\Phi }}_{s1,\,c0}(r,\,r') & = & -{1\over 2}\,{\rm arc}\tan\Bigl(\sinh
\Bigl({\pi\over 2}(r-r')\Bigr)\Bigr) \, ;
\hspace{0.5cm} r \neq \pm \infty  \nonumber \\
& = & -{{\rm sgn} (r)\pi\over 2\sqrt{2}} \, ; \hspace{0.5cm} r = \pm \infty \, ,
\label{Phisc1}
\end{eqnarray}
\begin{eqnarray}
\pi\,{\bar{\Phi }}_{s1,\,s1}(r,\,r') & = & B (r-r') \, ; \hspace{0.3cm} r\neq\pm\infty  
\nonumber
\\
& = & {{\rm sgn} (r)\pi\over 2\sqrt{2}} \, ; \hspace{0.5cm} r = \pm\infty \, ,
\hspace{0.5cm} r' \neq r \nonumber
\\
& = & [{\rm sgn} (r)]\Bigl({3\over 2\sqrt{2}}-1\Bigr)\pi \, ; \hspace{0.5cm} r = r' =
\pm\infty \, , \label{Phiss1}
\end{eqnarray}
\begin{equation}
\pi\,{\bar{\Phi }}_{c0,\,c\nu}(r,\,r') = -{\rm arc}\tan\Bigl({r-r'\over
\nu}\Bigr) \, , \hspace{0.5cm} 
\pi\,{\bar{\Phi }}_{s1,\,c\nu}(r,\,r') = \pi\,{\bar{\Phi }}_{c0,\,s\nu'}(r,\,r') = 0 
\, , \hspace{0.5cm} \nu >0 \, , \hspace{0.15cm} \nu' >1 \, , 
\label{Phiccn-csn1}
\end{equation}
\begin{eqnarray}
\pi\,{\bar{\Phi }}_{s1,\,s\nu}(r,\,r') & = & {\rm arc}\tan\Bigl({r-r'\over \nu
-1}\Bigr) \, ; \hspace{0.5cm} r\neq \pm\infty \, ; \hspace{0.5cm} \nu
>1 \, , \nonumber \\
& = & \pm {\pi\over\sqrt{2}} \, ; \hspace{0.5cm} r= \pm\infty \, ; \hspace{0.5cm} \nu
>1 \, , \label{Phicn-ssn1}
\end{eqnarray}
\begin{equation}
\pi\,{\bar{\Phi }}_{c\nu,\,c0}\left(r,r'\right) = {\rm arc}{\rm
tan}\Bigl({r-r'\over {\nu}}\Bigr) \, ; \hspace{0.5cm} \pi\,\bar{\Phi
}_{c\nu,\,c\nu'}\left(r,r'\right) = {1\over{2}}\,\Theta_{\nu,\,\nu'}(r-r') \, ;
\hspace{0.5cm} \pi\,\bar{\Phi }_{c\nu,\,s\nu'}\left(r,r'\right) = 0 \, , \hspace{0.5cm} \nu >
0 \, , \label{Phicnc-cncn-cnsn1}
\end{equation}
and the phase-shift expressions (\ref{Phisnc})-(\ref{Phisnsn}) are $n$ independent and
then have the same expressions as for $n<1$. The two-pseudofermion phase shift
expressions derived here are used in the derivation of the equalities of Eq.
(\ref{QIsnNV}).

%%%%%%%%%%%%%%%%%%%%%%%%%%%%%%%%%%%%%%%%%%%%%%%%%%%%%%%%%%%%%%%%%%%%%%%%%%
\section{THE TWO-PSEUDOFERMION PHASE SHIFTS FOR $U/t\rightarrow 0$}

Here we derive expressions for the $m\rightarrow 0$ two-pseudofermion
phase shifts given in Appendix B that are plotted in Figs. 1-6 in units of
$\pi$, $\pi\,\Phi_{c0,\,c0}(q,\,q')$, $\pi\,\Phi_{c0,\,s1}(q\,q')$, $\pi\,\Phi_{s1,\,c0}(q,\,q')$, 
$\pi\,\Phi_{s1,\,s1}(q,\,q')$, $\pi\,\Phi_{c0,\,c1}(q,\,q')$, and $\pi\,\Phi_{s1,\,c1}(q,\,q')$, 
for the limit $U/t\rightarrow 0$. Moreover, we also provide large-$U/t$ 
expressions for the rapidity functions involved
in the evaluation of the two-pseudofermion phase-shift expansions 
(\ref{PhiccUinf})-(\ref{Phisc1Uinf}). 

The evaluation of the two-pdseudofermion expressions provided here and 
of the expansions given in Eqs. (\ref{PhiccUinf})-(\ref{Phisc1Uinf})
involves the use of Eq. (\ref{Phi-barPhi}) where the 
ground-state rapidity functions $\Lambda^{0}_{\alpha\nu}(q')$
are defined in terms of their inverse functions, whose expressions are given in 
Eqs. (A.1) and (A.2) of Ref. \cite{V-1}. First, we use the latter expressions to derive the following
closed-form expressions for the ground-state functions $k^0 (q)$, $\Lambda_{c0}^0 (q)$,
$\Lambda_{c\nu}^0 (q)$, and $\Lambda_{s1}^0 (q)$, valid for zero spin density
$m\rightarrow 0$, values of the electronic density $0\leq n\leq 1$, and limiting on-site
repulsion values $U/t\rightarrow 0$ and $U/t\gg 1$,
\begin{eqnarray}
k^0 (q) & = & {q\over 2} \, ; \hspace{0.5cm} \vert q\vert \leq 2k_F \, , \hspace{0.5cm}
U/t\rightarrow 0 \nonumber \\ & = & {\rm sgn} (q)\,[\vert q\vert - k_F] \, ;
\hspace{1cm} 2k_F \leq \vert q\vert < \pi/a \, , \hspace{0.5cm} U/t\rightarrow 0 
\nonumber \\ & = & {\rm sgn} (q)\,\pi \, ; \hspace{1cm} \vert q\vert = \pi \, ,
\hspace{0.5cm} U/t\rightarrow 0 \nonumber
\\ & = & q - {4tn\over U}\,\ln (2)\,\sin (q) \, ;
\hspace{0.5cm} \vert q\vert \leq \pi \, , \hspace{0.2cm} U/t \gg 1 \, , \label{k0lim}
\end{eqnarray}
\begin{eqnarray}
\Lambda^0_{c0} (q) & = & \sin\Bigl({q\over 2}\Bigr) \, ; \hspace{0.5cm} \vert q\vert \leq
2k_F \, , \hspace{0.5cm} U/t\rightarrow 0 \nonumber \\ & = & {\rm sgn}
(q)\,\sin\Bigl((\vert q\vert - k_F)\Bigr) \, ; \hspace{1cm} 2k_F \leq \vert q\vert < \pi
\, , \hspace{0.5cm} U/t\rightarrow 0 \nonumber \\ & = & 0 \, ; \hspace{1cm} \vert
q\vert = \pi \, , \hspace{0.5cm} U/t\rightarrow 0 \nonumber \\ & = & \sin (q) -
{2tn\over U}\,\ln (2)\,\sin (2q) \, ; \hspace{0.5cm} \vert q\vert \leq \pi \, ,
\hspace{0.2cm} U/t \gg 1 \, , \label{Gc0lim}
\end{eqnarray}
\begin{eqnarray}
\Lambda^0_{c\nu} (q) & = & {\rm sgn} (q)\,\sin\Bigl({(\vert q\vert + \pi n)\over 2}\Bigr)
\, ; \hspace{0.5cm}  0<\vert q\vert< (\pi -2k_F) \, , \hspace{0.5cm} U/t\rightarrow 0
\nonumber \\ & = & 0 \, ; \hspace{0.5cm} q = 0 \, , \hspace{0.5cm} U/t\rightarrow 0
\nonumber \\ & = & \pm\infty \, ; \hspace{0.5cm} q = \pm (\pi -2k_F) \, , \hspace{0.5cm}
U/t\rightarrow 0 \nonumber \\ & = & {\nu U\over 4t}\,\tan \bigl({q\over 2\delta}\Bigr) \,
; \hspace{0.5cm} 0\leq \vert q\vert \leq (\pi -2k_F) \, , \hspace{0.2cm} U/t \gg 1 \, ,
\label{Gcnlim}
\end{eqnarray}
for $\nu >0$, and
\begin{eqnarray}
\Lambda^0_{s1} (q) & = & \sin (q) \, ; \hspace{0.5cm} \vert q\vert < k_F \, ,
\hspace{0.5cm} U/t\rightarrow 0  \nonumber \\ & = & \pm\infty \, ; \hspace{0.5cm} q =\pm
k_F \, , \hspace{0.5cm} U/t\rightarrow 0 \nonumber \\ & = & {U\over 2\pi t}\, {\rm
arcsinh} \Bigl(\tan ({q\over n})\Bigr) \, ; \hspace{0.5cm} \vert q\vert \leq k_F \, ,
\hspace{0.2cm} U/t \gg 1 \, , \label{Gs1lim}
\end{eqnarray}
respectively.

Next, we use Eqs. (\ref{k0lim})-(\ref{Gs1lim}) in the integral equations given in the
Appendix A of Ref. \cite{IIIb}, which define the two-pseudofermion phase shifts. (Such
equations were also presented in Ref. \cite{97} with a slightly different notation.) By
manipulation of these equations for the limit $U/t\rightarrow 0$,
we find the expressions for the above bare-momentum two-pseudofermion phase shifts in
units of $\pi$ given below. For the phase shifts $\pi\,\Phi_{s1,\,\alpha\nu}(q,\,q')$ we
provide the values for $U/t= 0$ and $U/t\rightarrow 0$, when different. The expressions read,
\begin{equation}
\pi\,\Phi_{c0,\,c0}(q,\,q') = - {\rm sgn} \Bigl(\sin k^0_{c0} (q)-\sin k^0_{c0}
(q')\Bigr){\pi\over C_c (q)} + \delta_{\vert q\vert ,2k_F}\delta_{q,q'}[{\rm sgn}
(q)]\Bigl({3\over 2\sqrt{2}}-1\Bigr)\pi \, , \label{PhiccU0}
\end{equation}
\begin{equation}
\pi\,\Phi_{c0,\,s1}(q,\,q') = - {\rm sgn} \Bigl(\sin k^0_{c0} (q)-c_{s1}(q')\sin
(q')\Bigr){\pi\over C_c (q)} \, , \label{PhicsU0}
\end{equation}
\begin{eqnarray}
\pi\,\Phi_{s1,\,c0}(q,\,q') & = & - {\rm sgn} \Bigl(\sin (q)-\sin k^0_{c0} (q')\Bigr){\pi\over 2}
\, ;
\hspace{0.5cm} q \neq \pm k_F \nonumber \\
& = & -{{\rm sgn} (q)\pi\over 2\sqrt{2}} \, ; \hspace{0.5cm} q = \pm k_F
\, ; \hspace{0.5cm} U/t\rightarrow 0  \nonumber \\
& = & 0 \, ; \hspace{0.5cm} q = \pm k_F \, ; \hspace{0.5cm} U/t= 0 \, , \label{Phisc0}
\end{eqnarray}
\begin{eqnarray}
\pi\,\Phi_{s1,\,s1}(q,\,q') & = & 0 \, ;
\hspace{0.5cm} q \neq \pm k_F \nonumber \\
& = & {{\rm sgn} (q)\pi\over 2\sqrt{2}}\Bigl[1 + \delta_{q,q'}2(1-\sqrt{2})\Bigr] \, ;
\hspace{0.5cm} q = \pm k_F
\, ; \hspace{0.5cm} U/t\rightarrow 0  \nonumber \\
& = & 0 \, ; \hspace{0.5cm} q = \pm k_F \, ; \hspace{0.5cm} U/t= 0 \, , \label{PhissU0}
\end{eqnarray}
\begin{equation}
\pi\,\Phi_{c0,\,c1}(q,\,q') = - {\rm sgn} \Bigl(\sin k^0_{c0} (q)-c_{c1}(q')\sin k^0_{c1}
(q')\Bigr){2\pi\over C_c (q)} \, , \label{PhictU0}
\end{equation}
and
\begin{equation}
\pi\,\Phi_{s1,\,c1}(q,\,q') = - \theta (k_F -\vert q\vert )\,{\rm sgn} \Bigl(\sin
q-\sin k^0_{c1} (q')\Bigr){\pi\over 2} \, , \label{PhistU0}
\end{equation}
respectively. Here the sign function is such that ${\rm sgn} (0)=0$ and $\theta (x) = 1$
for $x> 0$ and $\theta (x) = 0$ for $x\leq 0$ and thus $\pi\,\Phi_{s1,\,c1}(\pm k_F,\,q')=0$.
In the above equations, $k^0_{c0} (q)=\lim_{U/t\rightarrow 0}k^0 (q)$ where the
$U/t\rightarrow 0$ value of $k^0 (q)$ is given in Eq. (\ref{k0lim}),
\begin{eqnarray}
k^0_{c1} (q) & = & {q\over 2} +{\rm sgn} (q)\,k_F \, ; \hspace{0.5cm} 0<\vert q\vert
\leq [\pi -2k_F] \nonumber \\
& = & 0 \, ; \hspace{0.5cm} q=0 \, , \label{k0c1}
\end{eqnarray}
\begin{equation}
C_c (q) = 2\Bigl[\theta (2k_F -\vert q\vert ) + \sqrt{2}\,\delta_{\vert q\vert ,2k_F}
+2\,\theta (\pi -\vert q\vert )\theta (\vert q\vert -2k_F) + \delta_{\vert q\vert
,\pi}\Bigr] \, , \label{Ccsq}
\end{equation}
and
\begin{eqnarray}
c_{s1} (q) & = & 1 \, , \hspace{0.5cm} \vert q\vert <k_F \, ; \hspace{0.5cm} c_{s1} (q) =
\infty \, , \hspace{0.5cm} q=\pm k_F \nonumber \\ c_{c1} (q) & = & 1 \, , \hspace{0.5cm}
\vert q\vert <[\pi -2k_F] \, ; \hspace{0.5cm} c_{c1} (q) = \infty \, , \hspace{0.5cm}
q=\pm [\pi -2k_F] \, . \label{ccsq}
\end{eqnarray}

%%%%%%%%%%%%%%%%%%%%%%%%%%%%%%%%%%%%%%%%%%%%%%%%%%%%%%%%%%%%%%%%%%%%%%%%%%


\begin{references}
\bibitem[1]{Lieb}
        Elliott H. Lieb, F. Y. Wu, Phys. Rev. Lett. 20 (1968)
        1445.
\bibitem[2]{Takahashi}
        M. Takahashi, Prog. Theor. Phys. 47 (1972) 69.
\bibitem[3]{Martins98}
        P. B. Ramos, M. J. Martins, J. Phys. A 30 (1997) L195;
        M. J. Martins, P.B. Ramos, Nucl. Phys. B 522 (1998) 413.
\bibitem[4]{I}
        J. M. P. Carmelo, J. M. Rom\'an, K. Penc, Nucl. Phys. B
        683 (2004) 387.
\bibitem[5]{II}
        J. M. P. Carmelo, P. D. Sacramento, Phys. Rev. B 68
        (2003) 085104.
\bibitem[6]{IIIb}
        J. M. P. Carmelo,  cond-mat/0405411.
\bibitem[7]{Woy}
        F. Woynarovich, J. Phys. A: Math. Gen. 22 (1989) 4243.
\bibitem[8]{Ogata}
        M. Ogata, H. Shiba, Phys. Rev. B 41(1990) 2326;
        M. Ogata, T. Sugiyama, H. Shiba, Phys. Rev. B 43 (1991) 8401.
\bibitem[9]{Kawakami}
        N. Kawakami, S. K. Yang, Phys. Lett. A 148 (1990) 359.
\bibitem[10]{Frahm}
        H. Frahm, V. E. Korepin, Phys. Rev. B 42 (1990) 10 553;
        H. Frahm, V. E. Korepin, Phys. Rev. B 43 (1991) 5653.
\bibitem[11]{Brech}
        M. Brech, J. Voit, H. Buttner, Europhys. Lett. 12 (1990) 289.
\bibitem[12]{93-94}
    	J. M. P. Carmelo, A. H. Castro Neto,
        Phys. Rev. Lett. 70 (1993) 1904; J. M. P. Carmelo, A. H. Castro Neto,
        D. K. Campbell, Phys. Rev. B 50 (1994) 3667; J. M. P. Carmelo, A. H.
        Castro Neto, D. K. Campbell, Phys. Rev. B 50 (1994) 3683.    
\bibitem[13]{Karlo}
        K. Penc, J. S\'olyom, Phys. Rev. B 47 (1993) 6273 (1993). 
\bibitem[14]{V}
        J. M. P. Carmelo, K. Penc, cond-mat/0311075.
\bibitem[15]{V-1}
        J. M. P. Carmelo, K. Penc, D. Bozi, Nucl. Phys. B 725
        (2005) 421; J. M. P. Carmelo, K. Penc, D. Bozi, Nucl. Phys. B 737 (2006) 351,
        Erratum.
\bibitem[16]{LE}
        J. M. P. Carmelo, L. M. Martelo, K. Penc, Nucl. Phys. B 737 (2006) 237;
        J. M. P. Carmelo, K. Penc, Phys. Rev. B 73 (2006) at press.
\bibitem[17]{Harris}
        A. Brooks Harris, Robert V. Lange, Phys. Rev. 157
        (1967) 295; A. H. MacDonald, S. M. Girvin, D. Yoshioka,
        Phys. Rev. B 37 (1988) 9753.
\bibitem[18]{HL}
        O. J. Heilmann, E. H. Lieb, Ann. N. Y. Acad. Sci. 172
        (1971) 583; E. H. Lieb, Phys. Rev. Lett. 62 (1989) 1201.
\bibitem[19]{Yang89}
        C. N. Yang, Phys. Rev. Lett. 63 (1989) 2144.
\bibitem[20]{Penc}
        Karlo Penc, Karen Hallberg, Fr\'ed\'eric Mila, Hiroyuki Shiba,
        Phys. Rev. Lett. 77 (1996) 1390; Karlo Penc, Karen Hallberg, 
        Fr\'ed\'eric Mila, Hiroyuki Shiba, Phys. Rev. B 55 (1997) 15 475.
\bibitem[21]{CFT}
        A. A. Belavin, A. M. Polyakov, A. B.
        Zamolodchikov, Nucl. Phys. B 241 (1984) 333.
\bibitem[22]{Bozo}
        H. J. Schulz, Phys. Rev. Lett. 64 (1990) 2831;
        J. M. P. Carmelo, A. H. Castro Neto, D. K. Campbell,
        Phys. Rev. Lett. 73 (1994) 926 and Erratum 74 (1995) 3089.
\bibitem[23]{spectral0}
        M. Sing, U. Schwingenschl\"ogl, R. Claessen, P.
        Blaha, J. M. P. Carmelo, L. M. Martelo, P. D. Sacramento, M.
        Dressel, C. S. Jacobsen, Phys. Rev. B 68 (2003) 125111.
\bibitem[24]{spectral}
        J. M. P. Carmelo, K. Penc, L. M. Martelo, P. D. Sacramento,
        J. M. B. Lopes dos Santos, R. Claessen, M. Sing,
        U. Schwingenschl\"ogl, Europhys. Lett. 67 (2004) 233;
        J. M. P. Carmelo, D. Bozi, P. D. Sacramento, K. Penc,
        submitted for publication.
\bibitem[25]{super}
        J. M. P. Carmelo, F. Guinea, K. Penc, P. D. Sacramento,
        Europhys. Lett. 68 (2004) 839.
\bibitem[26]{Eric}
        H. Benthien, F. Gebhard, E. Jeckelmann, Phys. Rev.
        Lett. 92 (2004) 256401.
\bibitem[27]{Zoller}
        D. Jaksch, P. Zoller, Ann. Phys. 315 (2005) 52.
\bibitem[28]{Chen}
	   Y. Chen, private communication.
\bibitem[29]{Car04}
        J. M. P. Carmelo, J. Phys.: Cond. Mat. 17 (2005) 5517.
\bibitem[30]{Natan}
        N. Andrei, {\it Series on Modern Condensed Matter Physics -
        Vol. 6}, 458, World Scientific, Lecture Notes of ICTP Summer
        Course, Editors: S. Lundquist, G. Morandi, Yu Lu [cond-mat/9408101].
\bibitem[31]{S0}
        Fabian H. L. Essler, Vladimir E. Korepin,
        Phys. Rev. Lett. 72 (1994) 908.
\bibitem[32]{S}
        Fabian H. L. Essler, Vladimir E. Korepin,
        Nucl. Phys. B 426 (1994) 505.
\bibitem[33]{relation}
        J. M. P. Carmelo, K. E. Hibberd, N. Andrei, cond-mat/0603446.
\bibitem[34]{Taylor} John R. Taylor, {\em Scattering theory: the quantum theory of
        nonrelativistic collisions} (Robert E. Krieger Publishing Company, Malabar,
        Florida, 1987).
\bibitem[35]{Martinus} For a basic introduction to chromodynamics see Martinus Veltman, 
	  {\em Facts and Mysteries in Elementary Particle Physics} (World Scientific, New
	  Jersey, 2003).
\bibitem[36]{Impurity}
        J. M. P. Carmelo, D. Bozi, unpublished.
\bibitem[37]{Mahan} Gerald D. Mahan, {\em Many-particle physics} (Kluwer Academic/Plenum
        Publishers, New York, 2000), Chapter 4; F. G. Fumi, Philos. Mag. 46 (1955) 1007.
\bibitem[38]{Carmelo91}
        J. Carmelo and A. A. Ovchinnikov, J. Phys.: Condens.
        Matter {\bf 3}, 757 (1991).
\bibitem[39]{Carmelo92}
        J. M. P. Carmelo, P. Horsch, A. A. Ovchinnikov,
        Phys. Rev. B 45 (1992) 7899.
\bibitem[40]{Ohanian} H. C. Ohanian, {\em Principles of quantum mechanics}
        (Prentice Hall, Englewood Cliffs, New Jersey, 1993).
\bibitem[41]{dynamical}
        N. M. R. Peres, J. M. P. Carmelo, D. K. Campbell, A. W. Sandvik, Z. Phys. B
        103 (1997) 217; J. M. P. Carmelo, P. Horsch, D. K. Campbell, A. H. Castro Neto,
        Phys. Rev. B (RC) 48 (1993) 4200.
\bibitem[42]{properties}
        Dionys Baeriswyl, Jos\'e Carmelo, Kazumi Maki, Synth. Met. 21 (1987) 271.
        J. M. P. Carmelo, N. M. R. Peres, P. D.
        Sacramento, Phys. Rev. Lett. 84 (2000) 4673;
        D. Controzzi, F.H.L. Essler, A.M. Tsvelik,
        Phys. Rev. Lett. 86 (2001) 680; 
        H. Kishida, M. Ono, K. Miura, H. Okamoto, M. Izumi, T. Manako,
        M. Kawasaki, Y. Taguchi, Y. Tokura, T. Tohyama, K. Tsutsui, 
        S. Maekawa, Phys. Rev. Lett. 87 (2001) 177401.
\bibitem[43]{tj}
        H. Bethe, Z. Phys. 71 (1931) 205; Elliott H. Lieb, Werner Liniger, Phys. Rev.
        130 (1963) 1605;  N. Andrei, J. Lowenstein, Phys. Rev. Lett. 43 (1979) 1698;
        N. Andrei, Phys. Rev. Lett. 45 (1980) 379; P.- A. Bares, J. M. P. Carmelo, J. Ferrer,
        P. Horsch, Phys. Rev. B 46 (1992) 14624.
\bibitem[44]{97}
        J. M. P. Carmelo, N. M. R. Peres, Phys. Rev. B 56 (1997) 3717.
\end{references}
\end{document}